\documentclass[longauth]{aa}

\usepackage{graphicx}
\usepackage{natbib}
\usepackage{scalerel}
\usepackage[table]{xcolor}
\usepackage{txfonts}
\usepackage[pdfencoding=auto,psdextra]{hyperref}
\hypersetup{
    colorlinks=true,
    linkcolor=blue,
    filecolor=magenta,      
    urlcolor=blue,
    citecolor=blue
}
\makeatletter
\renewcommand*\aa@pageof{, page \thepage{} of \pageref*{LastPage}}
\makeatother

\usepackage[utf8]{inputenc}

\usepackage[switch, modulo]{lineno}
\linenumbers

\newcommand{\hstfull}{\HST\xspace}
\newcommand{\hst}{HST\xspace}

\newcommand{\jwst}{JWST\xspace}
\newcommand{\gaia}{\textit{Gaia}\xspace}
\newcommand{\euclid}{\textit{Euclid}\xspace}
\newcommand{\qfit}{\texttt{QFIT}\xspace}

\newcommand{\hpass}{\texttt{hst1pass}\xspace}

\newcommand{\epass}{\texttt{euclid1pass}\xspace}

\newcommand{\egpm}{\mbox{\textit{E}-\textit{G}}\xspace}

\usepackage{euclid}

\begin{document} 
\nolinenumbers

\title{\euclid: High-precision imaging astrometry and photometry from Early Release Observations\thanks{This paper is published on behalf of the Euclid Consortium.}}
\subtitle{I. Internal kinematics of NGC\,6397 by combining \euclid and \gaia data}

\titlerunning{\euclid: High-precision imaging astrometry and photometry from Early Release Observations. I.}
\authorrunning{Libralato et al.}

\author{M.~Libralato\orcid{0000-0001-9673-7397}\thanks{\email{mattia.libralato@inaf.it}}\inst{\ref{aff1}}
\and L.~R.~Bedin\orcid{0000-0003-4080-6466}\inst{\ref{aff1}}
\and M.~Griggio\orcid{0000-0002-5060-1379}\inst{\ref{aff1},\ref{aff2}}
\and D.~Massari\orcid{0000-0001-8892-4301}\inst{\ref{aff3}}
\and J.~Anderson\orcid{0000-0003-2861-3995}\inst{\ref{aff2}}
\and J.-C.~Cuillandre\orcid{0000-0002-3263-8645}\inst{\ref{aff4}}
\and A.~M.~N.~Ferguson\inst{\ref{aff5}}
\and A.~Lan\c{c}on\orcid{0000-0002-7214-8296}\inst{\ref{aff6}}
\and S.~S.~Larsen\orcid{0000-0003-0069-1203}\inst{\ref{aff7}}
\and M.~Schirmer\orcid{0000-0003-2568-9994}\inst{\ref{aff8}}
\and F.~Annibali\inst{\ref{aff3}}
\and E.~Balbinot\orcid{0000-0002-1322-3153}\inst{\ref{aff9},\ref{aff10}}
\and E.~Dalessandro\orcid{0000-0003-4237-4601}\inst{\ref{aff3}}
\and D.~Erkal\orcid{0000-0002-8448-5505}\inst{\ref{aff11}}
\and P.~B.~Kuzma\orcid{0000-0003-1980-8838}\inst{\ref{aff5},\ref{aff12}}
\and T.~Saifollahi\orcid{0000-0002-9554-7660}\inst{\ref{aff6}}
\and G.~Verdoes~Kleijn\orcid{0000-0001-5803-2580}\inst{\ref{aff9}}
\and M.~K\"ummel\orcid{0000-0003-2791-2117}\inst{\ref{aff13}}
\and R.~Nakajima\inst{\ref{aff14}}
\and M.~Correnti\orcid{0000-0001-6464-3257}\inst{\ref{aff15},\ref{aff16}}
\and G.~Battaglia\orcid{0000-0002-6551-4294}\inst{\ref{aff17}}
\and B.~Altieri\orcid{0000-0003-3936-0284}\inst{\ref{aff18}}
\and A.~Amara\inst{\ref{aff11}}
\and S.~Andreon\orcid{0000-0002-2041-8784}\inst{\ref{aff19}}
\and C.~Baccigalupi\orcid{0000-0002-8211-1630}\inst{\ref{aff20},\ref{aff21},\ref{aff22},\ref{aff23}}
\and M.~Baldi\orcid{0000-0003-4145-1943}\inst{\ref{aff24},\ref{aff3},\ref{aff25}}
\and A.~Balestra\orcid{0000-0002-6967-261X}\inst{\ref{aff1}}
\and S.~Bardelli\orcid{0000-0002-8900-0298}\inst{\ref{aff3}}
\and A.~Basset\inst{\ref{aff26}}
\and P.~Battaglia\orcid{0000-0002-7337-5909}\inst{\ref{aff3}}
\and D.~Bonino\orcid{0000-0002-3336-9977}\inst{\ref{aff27}}
\and E.~Branchini\orcid{0000-0002-0808-6908}\inst{\ref{aff28},\ref{aff29},\ref{aff19}}
\and M.~Brescia\orcid{0000-0001-9506-5680}\inst{\ref{aff30},\ref{aff31},\ref{aff32}}
\and J.~Brinchmann\orcid{0000-0003-4359-8797}\inst{\ref{aff33},\ref{aff34}}
\and A.~Caillat\inst{\ref{aff35}}
\and S.~Camera\orcid{0000-0003-3399-3574}\inst{\ref{aff36},\ref{aff37},\ref{aff27}}
\and V.~Capobianco\orcid{0000-0002-3309-7692}\inst{\ref{aff27}}
\and C.~Carbone\orcid{0000-0003-0125-3563}\inst{\ref{aff38}}
\and J.~Carretero\orcid{0000-0002-3130-0204}\inst{\ref{aff39},\ref{aff40}}
\and S.~Casas\orcid{0000-0002-4751-5138}\inst{\ref{aff41},\ref{aff42}}
\and M.~Castellano\orcid{0000-0001-9875-8263}\inst{\ref{aff15}}
\and G.~Castignani\orcid{0000-0001-6831-0687}\inst{\ref{aff3}}
\and S.~Cavuoti\orcid{0000-0002-3787-4196}\inst{\ref{aff31},\ref{aff32}}
\and A.~Cimatti\inst{\ref{aff43}}
\and C.~Colodro-Conde\inst{\ref{aff44}}
\and G.~Congedo\orcid{0000-0003-2508-0046}\inst{\ref{aff5}}
\and C.~J.~Conselice\orcid{0000-0003-1949-7638}\inst{\ref{aff45}}
\and L.~Conversi\orcid{0000-0002-6710-8476}\inst{\ref{aff46},\ref{aff18}}
\and Y.~Copin\orcid{0000-0002-5317-7518}\inst{\ref{aff47}}
\and F.~Courbin\orcid{0000-0003-0758-6510}\inst{\ref{aff48},\ref{aff49},\ref{aff50}}
\and H.~M.~Courtois\orcid{0000-0003-0509-1776}\inst{\ref{aff51}}
\and M.~Cropper\orcid{0000-0003-4571-9468}\inst{\ref{aff52}}
\and A.~Da~Silva\orcid{0000-0002-6385-1609}\inst{\ref{aff53},\ref{aff54}}
\and H.~Degaudenzi\orcid{0000-0002-5887-6799}\inst{\ref{aff55}}
\and G.~De~Lucia\orcid{0000-0002-6220-9104}\inst{\ref{aff21}}
\and J.~Dinis\orcid{0000-0001-5075-1601}\inst{\ref{aff53},\ref{aff54}}
\and F.~Dubath\orcid{0000-0002-6533-2810}\inst{\ref{aff55}}
\and X.~Dupac\inst{\ref{aff18}}
\and S.~Dusini\orcid{0000-0002-1128-0664}\inst{\ref{aff56}}
\and M.~Fabricius\orcid{0000-0002-7025-6058}\inst{\ref{aff57},\ref{aff13}}
\and M.~Farina\orcid{0000-0002-3089-7846}\inst{\ref{aff58}}
\and S.~Farrens\orcid{0000-0002-9594-9387}\inst{\ref{aff4}}
\and F.~Faustini\orcid{0000-0001-6274-5145}\inst{\ref{aff16},\ref{aff15}}
\and S.~Ferriol\inst{\ref{aff47}}
\and P.~Fosalba\orcid{0000-0002-1510-5214}\inst{\ref{aff59},\ref{aff60}}
\and M.~Frailis\orcid{0000-0002-7400-2135}\inst{\ref{aff21}}
\and E.~Franceschi\orcid{0000-0002-0585-6591}\inst{\ref{aff3}}
\and M.~Fumana\orcid{0000-0001-6787-5950}\inst{\ref{aff38}}
\and S.~Galeotta\orcid{0000-0002-3748-5115}\inst{\ref{aff21}}
\and B.~Garilli\orcid{0000-0001-7455-8750}\inst{\ref{aff38}}
\and K.~George\orcid{0000-0002-1734-8455}\inst{\ref{aff13}}
\and W.~Gillard\orcid{0000-0003-4744-9748}\inst{\ref{aff61}}
\and B.~Gillis\orcid{0000-0002-4478-1270}\inst{\ref{aff5}}
\and C.~Giocoli\orcid{0000-0002-9590-7961}\inst{\ref{aff3},\ref{aff62}}
\and P.~G\'omez-Alvarez\orcid{0000-0002-8594-5358}\inst{\ref{aff63},\ref{aff18}}
\and A.~Grazian\orcid{0000-0002-5688-0663}\inst{\ref{aff1}}
\and F.~Grupp\inst{\ref{aff57},\ref{aff13}}
\and L.~Guzzo\orcid{0000-0001-8264-5192}\inst{\ref{aff64},\ref{aff19}}
\and S.~V.~H.~Haugan\orcid{0000-0001-9648-7260}\inst{\ref{aff65}}
\and J.~Hoar\inst{\ref{aff18}}
\and H.~Hoekstra\orcid{0000-0002-0641-3231}\inst{\ref{aff10}}
\and W.~Holmes\inst{\ref{aff66}}
\and F.~Hormuth\inst{\ref{aff67}}
\and A.~Hornstrup\orcid{0000-0002-3363-0936}\inst{\ref{aff68},\ref{aff69}}
\and P.~Hudelot\inst{\ref{aff70}}
\and K.~Jahnke\orcid{0000-0003-3804-2137}\inst{\ref{aff8}}
\and M.~Jhabvala\inst{\ref{aff71}}
\and E.~Keih\"anen\orcid{0000-0003-1804-7715}\inst{\ref{aff72}}
\and S.~Kermiche\orcid{0000-0002-0302-5735}\inst{\ref{aff61}}
\and A.~Kiessling\orcid{0000-0002-2590-1273}\inst{\ref{aff66}}
\and M.~Kilbinger\orcid{0000-0001-9513-7138}\inst{\ref{aff4}}
\and B.~Kubik\orcid{0009-0006-5823-4880}\inst{\ref{aff47}}
\and M.~Kunz\orcid{0000-0002-3052-7394}\inst{\ref{aff73}}
\and H.~Kurki-Suonio\orcid{0000-0002-4618-3063}\inst{\ref{aff74},\ref{aff75}}
\and R.~Laureijs\inst{\ref{aff76}}
\and D.~Le~Mignant\orcid{0000-0002-5339-5515}\inst{\ref{aff35}}
\and S.~Ligori\orcid{0000-0003-4172-4606}\inst{\ref{aff27}}
\and P.~B.~Lilje\orcid{0000-0003-4324-7794}\inst{\ref{aff65}}
\and V.~Lindholm\orcid{0000-0003-2317-5471}\inst{\ref{aff74},\ref{aff75}}
\and I.~Lloro\inst{\ref{aff77}}
\and E.~Maiorano\orcid{0000-0003-2593-4355}\inst{\ref{aff3}}
\and O.~Mansutti\orcid{0000-0001-5758-4658}\inst{\ref{aff21}}
\and O.~Marggraf\orcid{0000-0001-7242-3852}\inst{\ref{aff14}}
\and K.~Markovic\orcid{0000-0001-6764-073X}\inst{\ref{aff66}}
\and M.~Martinelli\orcid{0000-0002-6943-7732}\inst{\ref{aff15},\ref{aff78}}
\and N.~Martinet\orcid{0000-0003-2786-7790}\inst{\ref{aff35}}
\and F.~Marulli\orcid{0000-0002-8850-0303}\inst{\ref{aff79},\ref{aff3},\ref{aff25}}
\and R.~Massey\orcid{0000-0002-6085-3780}\inst{\ref{aff80}}
\and E.~Medinaceli\orcid{0000-0002-4040-7783}\inst{\ref{aff3}}
\and S.~Mei\orcid{0000-0002-2849-559X}\inst{\ref{aff81}}
\and M.~Melchior\inst{\ref{aff82}}
\and Y.~Mellier\inst{\ref{aff83},\ref{aff70}}
\and M.~Meneghetti\orcid{0000-0003-1225-7084}\inst{\ref{aff3},\ref{aff25}}
\and E.~Merlin\orcid{0000-0001-6870-8900}\inst{\ref{aff15}}
\and G.~Meylan\inst{\ref{aff48}}
\and M.~Moresco\orcid{0000-0002-7616-7136}\inst{\ref{aff79},\ref{aff3}}
\and L.~Moscardini\orcid{0000-0002-3473-6716}\inst{\ref{aff79},\ref{aff3},\ref{aff25}}
\and C.~Neissner\orcid{0000-0001-8524-4968}\inst{\ref{aff84},\ref{aff40}}
\and R.~C.~Nichol\orcid{0000-0003-0939-6518}\inst{\ref{aff11}}
\and S.-M.~Niemi\inst{\ref{aff76}}
\and J.~W.~Nightingale\orcid{0000-0002-8987-7401}\inst{\ref{aff85}}
\and C.~Padilla\orcid{0000-0001-7951-0166}\inst{\ref{aff84}}
\and S.~Paltani\orcid{0000-0002-8108-9179}\inst{\ref{aff55}}
\and F.~Pasian\orcid{0000-0002-4869-3227}\inst{\ref{aff21}}
\and K.~Pedersen\inst{\ref{aff86}}
\and W.~J.~Percival\orcid{0000-0002-0644-5727}\inst{\ref{aff87},\ref{aff88},\ref{aff89}}
\and V.~Pettorino\inst{\ref{aff76}}
\and S.~Pires\orcid{0000-0002-0249-2104}\inst{\ref{aff4}}
\and G.~Polenta\orcid{0000-0003-4067-9196}\inst{\ref{aff16}}
\and M.~Poncet\inst{\ref{aff26}}
\and L.~A.~Popa\inst{\ref{aff90}}
\and L.~Pozzetti\orcid{0000-0001-7085-0412}\inst{\ref{aff3}}
\and F.~Raison\orcid{0000-0002-7819-6918}\inst{\ref{aff57}}
\and R.~Rebolo\inst{\ref{aff44},\ref{aff91},\ref{aff92}}
\and A.~Refregier\inst{\ref{aff93}}
\and A.~Renzi\orcid{0000-0001-9856-1970}\inst{\ref{aff94},\ref{aff56}}
\and J.~Rhodes\orcid{0000-0002-4485-8549}\inst{\ref{aff66}}
\and G.~Riccio\inst{\ref{aff31}}
\and E.~Romelli\orcid{0000-0003-3069-9222}\inst{\ref{aff21}}
\and M.~Roncarelli\orcid{0000-0001-9587-7822}\inst{\ref{aff3}}
\and E.~Rossetti\orcid{0000-0003-0238-4047}\inst{\ref{aff24}}
\and R.~Saglia\orcid{0000-0003-0378-7032}\inst{\ref{aff13},\ref{aff57}}
\and Z.~Sakr\orcid{0000-0002-4823-3757}\inst{\ref{aff95},\ref{aff96},\ref{aff97}}
\and A.~G.~S\'anchez\orcid{0000-0003-1198-831X}\inst{\ref{aff57}}
\and D.~Sapone\orcid{0000-0001-7089-4503}\inst{\ref{aff98}}
\and B.~Sartoris\orcid{0000-0003-1337-5269}\inst{\ref{aff13},\ref{aff21}}
\and M.~Sauvage\orcid{0000-0002-0809-2574}\inst{\ref{aff4}}
\and P.~Schneider\orcid{0000-0001-8561-2679}\inst{\ref{aff14}}
\and T.~Schrabback\orcid{0000-0002-6987-7834}\inst{\ref{aff99}}
\and A.~Secroun\orcid{0000-0003-0505-3710}\inst{\ref{aff61}}
\and E.~Sefusatti\orcid{0000-0003-0473-1567}\inst{\ref{aff21},\ref{aff20},\ref{aff22}}
\and G.~Seidel\orcid{0000-0003-2907-353X}\inst{\ref{aff8}}
\and M.~Seiffert\orcid{0000-0002-7536-9393}\inst{\ref{aff66}}
\and S.~Serrano\orcid{0000-0002-0211-2861}\inst{\ref{aff59},\ref{aff100},\ref{aff60}}
\and C.~Sirignano\orcid{0000-0002-0995-7146}\inst{\ref{aff94},\ref{aff56}}
\and G.~Sirri\orcid{0000-0003-2626-2853}\inst{\ref{aff25}}
\and J.~Skottfelt\orcid{0000-0003-1310-8283}\inst{\ref{aff101}}
\and L.~Stanco\orcid{0000-0002-9706-5104}\inst{\ref{aff56}}
\and J.~Steinwagner\orcid{0000-0001-7443-1047}\inst{\ref{aff57}}
\and P.~Tallada-Cresp\'{i}\orcid{0000-0002-1336-8328}\inst{\ref{aff39},\ref{aff40}}
\and A.~N.~Taylor\inst{\ref{aff5}}
\and H.~I.~Teplitz\orcid{0000-0002-7064-5424}\inst{\ref{aff102}}
\and I.~Tereno\inst{\ref{aff53},\ref{aff103}}
\and R.~Toledo-Moreo\orcid{0000-0002-2997-4859}\inst{\ref{aff104}}
\and F.~Torradeflot\orcid{0000-0003-1160-1517}\inst{\ref{aff40},\ref{aff39}}
\and A.~Tsyganov\inst{\ref{aff105}}
\and I.~Tutusaus\orcid{0000-0002-3199-0399}\inst{\ref{aff96}}
\and L.~Valenziano\orcid{0000-0002-1170-0104}\inst{\ref{aff3},\ref{aff106}}
\and T.~Vassallo\orcid{0000-0001-6512-6358}\inst{\ref{aff13},\ref{aff21}}
\and A.~Veropalumbo\orcid{0000-0003-2387-1194}\inst{\ref{aff19},\ref{aff29},\ref{aff107}}
\and Y.~Wang\orcid{0000-0002-4749-2984}\inst{\ref{aff102}}
\and J.~Weller\orcid{0000-0002-8282-2010}\inst{\ref{aff13},\ref{aff57}}
\and G.~Zamorani\orcid{0000-0002-2318-301X}\inst{\ref{aff3}}
\and E.~Zucca\orcid{0000-0002-5845-8132}\inst{\ref{aff3}}
\and C.~Burigana\orcid{0000-0002-3005-5796}\inst{\ref{aff108},\ref{aff106}}
\and V.~Scottez\inst{\ref{aff83},\ref{aff109}}
\and D.~Scott\orcid{0000-0002-6878-9840}\inst{\ref{aff110}}
\and R.~L.~Smart\orcid{0000-0002-4424-4766}\inst{\ref{aff27},\ref{aff111}}}
										   
\institute{INAF-Osservatorio Astronomico di Padova, Via dell'Osservatorio 5, 35122 Padova, Italy\label{aff1}
\and
Space Telescope Science Institute, 3700 San Martin Dr, Baltimore, MD 21218, USA\label{aff2}
\and
INAF-Osservatorio di Astrofisica e Scienza dello Spazio di Bologna, Via Piero Gobetti 93/3, 40129 Bologna, Italy\label{aff3}
\and
Universit\'e Paris-Saclay, Universit\'e Paris Cit\'e, CEA, CNRS, AIM, 91191, Gif-sur-Yvette, France\label{aff4}
\and
Institute for Astronomy, University of Edinburgh, Royal Observatory, Blackford Hill, Edinburgh EH9 3HJ, UK\label{aff5}
\and
Universit\'e de Strasbourg, CNRS, Observatoire astronomique de Strasbourg, UMR 7550, 67000 Strasbourg, France\label{aff6}
\and
Department of Astrophysics/IMAPP, Radboud University, PO Box 9010, 6500 GL Nijmegen, The Netherlands\label{aff7}
\and
Max-Planck-Institut f\"ur Astronomie, K\"onigstuhl 17, 69117 Heidelberg, Germany\label{aff8}
\and
Kapteyn Astronomical Institute, University of Groningen, PO Box 800, 9700 AV Groningen, The Netherlands\label{aff9}
\and
Leiden Observatory, Leiden University, Einsteinweg 55, 2333 CC Leiden, The Netherlands\label{aff10}
\and
School of Mathematics and Physics, University of Surrey, Guildford, Surrey, GU2 7XH, UK\label{aff11}
\and
National Astronomical Observatory of Japan, 2-21-1 Osawa, Mitaka, Tokyo 181-8588, Japan\label{aff12}
\and
Universit\"ats-Sternwarte M\"unchen, Fakult\"at f\"ur Physik, Ludwig-Maximilians-Universit\"at M\"unchen, Scheinerstrasse 1, 81679 M\"unchen, Germany\label{aff13}
\and
Universit\"at Bonn, Argelander-Institut f\"ur Astronomie, Auf dem H\"ugel 71, 53121 Bonn, Germany\label{aff14}
\and
INAF-Osservatorio Astronomico di Roma, Via Frascati 33, 00078 Monteporzio Catone, Italy\label{aff15}
\and
Space Science Data Center, Italian Space Agency, via del Politecnico snc, 00133 Roma, Italy\label{aff16}
\and
Instituto de Astrof\'isica de Canarias (IAC); Departamento de Astrof\'isica, Universidad de La Laguna (ULL), 38200, La Laguna, Tenerife, Spain\label{aff17}
\and
ESAC/ESA, Camino Bajo del Castillo, s/n., Urb. Villafranca del Castillo, 28692 Villanueva de la Ca\~nada, Madrid, Spain\label{aff18}
\and
INAF-Osservatorio Astronomico di Brera, Via Brera 28, 20122 Milano, Italy\label{aff19}
\and
IFPU, Institute for Fundamental Physics of the Universe, via Beirut 2, 34151 Trieste, Italy\label{aff20}
\and
INAF-Osservatorio Astronomico di Trieste, Via G. B. Tiepolo 11, 34143 Trieste, Italy\label{aff21}
\and
INFN, Sezione di Trieste, Via Valerio 2, 34127 Trieste TS, Italy\label{aff22}
\and
SISSA, International School for Advanced Studies, Via Bonomea 265, 34136 Trieste TS, Italy\label{aff23}
\and
Dipartimento di Fisica e Astronomia, Universit\`a di Bologna, Via Gobetti 93/2, 40129 Bologna, Italy\label{aff24}
\and
INFN-Sezione di Bologna, Viale Berti Pichat 6/2, 40127 Bologna, Italy\label{aff25}
\and
Centre National d'Etudes Spatiales -- Centre spatial de Toulouse, 18 avenue Edouard Belin, 31401 Toulouse Cedex 9, France\label{aff26}
\and
INAF-Osservatorio Astrofisico di Torino, Via Osservatorio 20, 10025 Pino Torinese (TO), Italy\label{aff27}
\and
Dipartimento di Fisica, Universit\`a di Genova, Via Dodecaneso 33, 16146, Genova, Italy\label{aff28}
\and
INFN-Sezione di Genova, Via Dodecaneso 33, 16146, Genova, Italy\label{aff29}
\and
Department of Physics "E. Pancini", University Federico II, Via Cinthia 6, 80126, Napoli, Italy\label{aff30}
\and
INAF-Osservatorio Astronomico di Capodimonte, Via Moiariello 16, 80131 Napoli, Italy\label{aff31}
\and
INFN section of Naples, Via Cinthia 6, 80126, Napoli, Italy\label{aff32}
\and
Instituto de Astrof\'isica e Ci\^encias do Espa\c{c}o, Universidade do Porto, CAUP, Rua das Estrelas, PT4150-762 Porto, Portugal\label{aff33}
\and
Faculdade de Ci\^encias da Universidade do Porto, Rua do Campo de Alegre, 4150-007 Porto, Portugal\label{aff34}
\and
Aix-Marseille Universit\'e, CNRS, CNES, LAM, Marseille, France\label{aff35}
\and
Dipartimento di Fisica, Universit\`a degli Studi di Torino, Via P. Giuria 1, 10125 Torino, Italy\label{aff36}
\and
INFN-Sezione di Torino, Via P. Giuria 1, 10125 Torino, Italy\label{aff37}
\and
INAF-IASF Milano, Via Alfonso Corti 12, 20133 Milano, Italy\label{aff38}
\and
Centro de Investigaciones Energ\'eticas, Medioambientales y Tecnol\'ogicas (CIEMAT), Avenida Complutense 40, 28040 Madrid, Spain\label{aff39}
\and
Port d'Informaci\'{o} Cient\'{i}fica, Campus UAB, C. Albareda s/n, 08193 Bellaterra (Barcelona), Spain\label{aff40}
\and
Institute for Theoretical Particle Physics and Cosmology (TTK), RWTH Aachen University, 52056 Aachen, Germany\label{aff41}
\and
Institute of Cosmology and Gravitation, University of Portsmouth, Portsmouth PO1 3FX, UK\label{aff42}
\and
Dipartimento di Fisica e Astronomia "Augusto Righi" - Alma Mater Studiorum Universit\`a di Bologna, Viale Berti Pichat 6/2, 40127 Bologna, Italy\label{aff43}
\and
Instituto de Astrof\'isica de Canarias, Calle V\'ia L\'actea s/n, 38204, San Crist\'obal de La Laguna, Tenerife, Spain\label{aff44}
\and
Jodrell Bank Centre for Astrophysics, Department of Physics and Astronomy, University of Manchester, Oxford Road, Manchester M13 9PL, UK\label{aff45}
\and
European Space Agency/ESRIN, Largo Galileo Galilei 1, 00044 Frascati, Roma, Italy\label{aff46}
\and
Universit\'e Claude Bernard Lyon 1, CNRS/IN2P3, IP2I Lyon, UMR 5822, Villeurbanne, F-69100, France\label{aff47}
\and
Institute of Physics, Laboratory of Astrophysics, Ecole Polytechnique F\'ed\'erale de Lausanne (EPFL), Observatoire de Sauverny, 1290 Versoix, Switzerland\label{aff48}
\and
Institut de Ci\`{e}ncies del Cosmos (ICCUB), Universitat de Barcelona (IEEC-UB), Mart\'{i} i Franqu\`{e}s 1, 08028 Barcelona, Spain\label{aff49}
\and
Instituci\'o Catalana de Recerca i Estudis Avan\c{c}ats (ICREA), Passeig de Llu\'{\i}s Companys 23, 08010 Barcelona, Spain\label{aff50}
\and
UCB Lyon 1, CNRS/IN2P3, IUF, IP2I Lyon, 4 rue Enrico Fermi, 69622 Villeurbanne, France\label{aff51}
\and
Mullard Space Science Laboratory, University College London, Holmbury St Mary, Dorking, Surrey RH5 6NT, UK\label{aff52}
\and
Departamento de F\'isica, Faculdade de Ci\^encias, Universidade de Lisboa, Edif\'icio C8, Campo Grande, PT1749-016 Lisboa, Portugal\label{aff53}
\and
Instituto de Astrof\'isica e Ci\^encias do Espa\c{c}o, Faculdade de Ci\^encias, Universidade de Lisboa, Campo Grande, 1749-016 Lisboa, Portugal\label{aff54}
\and
Department of Astronomy, University of Geneva, ch. d'Ecogia 16, 1290 Versoix, Switzerland\label{aff55}
\and
INFN-Padova, Via Marzolo 8, 35131 Padova, Italy\label{aff56}
\and
Max Planck Institute for Extraterrestrial Physics, Giessenbachstr. 1, 85748 Garching, Germany\label{aff57}
\and
INAF-Istituto di Astrofisica e Planetologia Spaziali, via del Fosso del Cavaliere, 100, 00100 Roma, Italy\label{aff58}
\and
Institut d'Estudis Espacials de Catalunya (IEEC),  Edifici RDIT, Campus UPC, 08860 Castelldefels, Barcelona, Spain\label{aff59}
\and
Institute of Space Sciences (ICE, CSIC), Campus UAB, Carrer de Can Magrans, s/n, 08193 Barcelona, Spain\label{aff60}
\and
Aix-Marseille Universit\'e, CNRS/IN2P3, CPPM, Marseille, France\label{aff61}
\and
Istituto Nazionale di Fisica Nucleare, Sezione di Bologna, Via Irnerio 46, 40126 Bologna, Italy\label{aff62}
\and
FRACTAL S.L.N.E., calle Tulip\'an 2, Portal 13 1A, 28231, Las Rozas de Madrid, Spain\label{aff63}
\and
Dipartimento di Fisica "Aldo Pontremoli", Universit\`a degli Studi di Milano, Via Celoria 16, 20133 Milano, Italy\label{aff64}
\and
Institute of Theoretical Astrophysics, University of Oslo, P.O. Box 1029 Blindern, 0315 Oslo, Norway\label{aff65}
\and
Jet Propulsion Laboratory, California Institute of Technology, 4800 Oak Grove Drive, Pasadena, CA, 91109, USA\label{aff66}
\and
Felix Hormuth Engineering, Goethestr. 17, 69181 Leimen, Germany\label{aff67}
\and
Technical University of Denmark, Elektrovej 327, 2800 Kgs. Lyngby, Denmark\label{aff68}
\and
Cosmic Dawn Center (DAWN), Denmark\label{aff69}
\and
Institut d'Astrophysique de Paris, UMR 7095, CNRS, and Sorbonne Universit\'e, 98 bis boulevard Arago, 75014 Paris, France\label{aff70}
\and
NASA Goddard Space Flight Center, Greenbelt, MD 20771, USA\label{aff71}
\and
Department of Physics and Helsinki Institute of Physics, Gustaf H\"allstr\"omin katu 2, 00014 University of Helsinki, Finland\label{aff72}
\and
Universit\'e de Gen\`eve, D\'epartement de Physique Th\'eorique and Centre for Astroparticle Physics, 24 quai Ernest-Ansermet, CH-1211 Gen\`eve 4, Switzerland\label{aff73}
\and
Department of Physics, P.O. Box 64, 00014 University of Helsinki, Finland\label{aff74}
\and
Helsinki Institute of Physics, Gustaf H{\"a}llstr{\"o}min katu 2, University of Helsinki, Helsinki, Finland\label{aff75}
\and
European Space Agency/ESTEC, Keplerlaan 1, 2201 AZ Noordwijk, The Netherlands\label{aff76}
\and
NOVA optical infrared instrumentation group at ASTRON, Oude Hoogeveensedijk 4, 7991PD, Dwingeloo, The Netherlands\label{aff77}
\and
INFN-Sezione di Roma, Piazzale Aldo Moro, 2 - c/o Dipartimento di Fisica, Edificio G. Marconi, 00185 Roma, Italy\label{aff78}
\and
Dipartimento di Fisica e Astronomia "Augusto Righi" - Alma Mater Studiorum Universit\`a di Bologna, via Piero Gobetti 93/2, 40129 Bologna, Italy\label{aff79}
\and
Department of Physics, Institute for Computational Cosmology, Durham University, South Road, DH1 3LE, UK\label{aff80}
\and
Universit\'e Paris Cit\'e, CNRS, Astroparticule et Cosmologie, 75013 Paris, France\label{aff81}
\and
University of Applied Sciences and Arts of Northwestern Switzerland, School of Engineering, 5210 Windisch, Switzerland\label{aff82}
\and
Institut d'Astrophysique de Paris, 98bis Boulevard Arago, 75014, Paris, France\label{aff83}
\and
Institut de F\'{i}sica d'Altes Energies (IFAE), The Barcelona Institute of Science and Technology, Campus UAB, 08193 Bellaterra (Barcelona), Spain\label{aff84}
\and
School of Mathematics, Statistics and Physics, Newcastle University, Herschel Building, Newcastle-upon-Tyne, NE1 7RU, UK\label{aff85}
\and
DARK, Niels Bohr Institute, University of Copenhagen, Jagtvej 155, 2200 Copenhagen, Denmark\label{aff86}
\and
Waterloo Centre for Astrophysics, University of Waterloo, Waterloo, Ontario N2L 3G1, Canada\label{aff87}
\and
Department of Physics and Astronomy, University of Waterloo, Waterloo, Ontario N2L 3G1, Canada\label{aff88}
\and
Perimeter Institute for Theoretical Physics, Waterloo, Ontario N2L 2Y5, Canada\label{aff89}
\and
Institute of Space Science, Str. Atomistilor, nr. 409 M\u{a}gurele, Ilfov, 077125, Romania\label{aff90}
\and
Departamento de Astrof\'isica, Universidad de La Laguna, 38206, La Laguna, Tenerife, Spain\label{aff91}
\and
Consejo Superior de Investigaciones Cientificas, Calle Serrano 117, 28006 Madrid, Spain\label{aff92}
\and
Institute for Particle Physics and Astrophysics, Dept. of Physics, ETH Zurich, Wolfgang-Pauli-Strasse 27, 8093 Zurich, Switzerland\label{aff93}
\and
Dipartimento di Fisica e Astronomia "G. Galilei", Universit\`a di Padova, Via Marzolo 8, 35131 Padova, Italy\label{aff94}
\and
Institut f\"ur Theoretische Physik, University of Heidelberg, Philosophenweg 16, 69120 Heidelberg, Germany\label{aff95}
\and
Institut de Recherche en Astrophysique et Plan\'etologie (IRAP), Universit\'e de Toulouse, CNRS, UPS, CNES, 14 Av. Edouard Belin, 31400 Toulouse, France\label{aff96}
\and
Universit\'e St Joseph; Faculty of Sciences, Beirut, Lebanon\label{aff97}
\and
Departamento de F\'isica, FCFM, Universidad de Chile, Blanco Encalada 2008, Santiago, Chile\label{aff98}
\and
Universit\"at Innsbruck, Institut f\"ur Astro- und Teilchenphysik, Technikerstr. 25/8, 6020 Innsbruck, Austria\label{aff99}
\and
Satlantis, University Science Park, Sede Bld 48940, Leioa-Bilbao, Spain\label{aff100}
\and
Centre for Electronic Imaging, Open University, Walton Hall, Milton Keynes, MK7~6AA, UK\label{aff101}
\and
Infrared Processing and Analysis Center, California Institute of Technology, Pasadena, CA 91125, USA\label{aff102}
\and
Instituto de Astrof\'isica e Ci\^encias do Espa\c{c}o, Faculdade de Ci\^encias, Universidade de Lisboa, Tapada da Ajuda, 1349-018 Lisboa, Portugal\label{aff103}
\and
Universidad Polit\'ecnica de Cartagena, Departamento de Electr\'onica y Tecnolog\'ia de Computadoras,  Plaza del Hospital 1, 30202 Cartagena, Spain\label{aff104}
\and
Centre for Information Technology, University of Groningen, P.O. Box 11044, 9700 CA Groningen, The Netherlands\label{aff105}
\and
INFN-Bologna, Via Irnerio 46, 40126 Bologna, Italy\label{aff106}
\and
Dipartimento di Fisica, Universit\`a degli studi di Genova, and INFN-Sezione di Genova, via Dodecaneso 33, 16146, Genova, Italy\label{aff107}
\and
INAF, Istituto di Radioastronomia, Via Piero Gobetti 101, 40129 Bologna, Italy\label{aff108}
\and
Junia, EPA department, 41 Bd Vauban, 59800 Lille, France\label{aff109}
\and
Department of Physics and Astronomy, University of British Columbia, Vancouver, BC V6T 1Z1, Canada\label{aff110}
\and
School of Physics, Astronomy and Mathematics, University of Hertfordshire, College Lane, Hatfield AL10 9AB, UK\label{aff111}}    

\date{Received 18 September 2024; Accepted 24 October 2024}

\abstract{
The instruments at the focus of the \euclid space observatory offer superb, diffraction-limited imaging over an unprecedented (from space) wide field of view of 0.57 deg$^2$. This exquisite image quality has the potential to produce high-precision astrometry for point sources once the undersampling of \euclid's cameras is taken into account by means of accurate, effective point spread function (ePSF) modelling. We present a complex, detailed workflow to simultaneously solve for the geometric distortion (GD) and model the undersampled ePSFs of the \euclid detectors. Our procedure was successfully developed and tested with data from the Early Release Observations (ERO) programme focused on the nearby globular cluster NGC~6397.  Our final one-dimensional astrometric precision for a well-measured star just below saturation is 0.7\,mas (0.007\,pixel) for the Visible Instrument (VIS) and 3\,mas (0.01\,pixel) for the Near-Infrared Spectrometer and Photometer (NISP). Finally, we present a specific scientific application of this high-precision astrometry: the combination of \euclid and \gaia data to compute proper motions and study the internal kinematics of NGC~6397. Future work, when more data become available, will allow for a better characterisation of the ePSFs and GD corrections that are derived here, along with assessment of their temporal stability, and their dependencies on the spectral energy distribution of the sources as seen through the wide-band filters of \euclid.
}


\maketitle

\section{Introduction}

Astrometry is the classical example of a branch of astronomy that cyclically comes back to the fore when new, more advanced telescopes and instruments with higher spatial resolution and better stability than their predecessors begin their operations. Since the past decade, the astrometric scene has been dominated by the \hstfull (\hst) and \gaia. The proper motions (PMs) obtained thanks to the data from these observatories have provided critical pieces of information on a large variety of investigations, such as on the Galaxy's structure and kinematics \citep{2023A&A...674A..37G,2024ApJ...967...89I,2024NewAR..9901696C}), the internal motions of stellar clusters \citep[e.g.][]{2014BelliniPMcat,2018ComAC...5....2V,2021VasilievGCkin,2022LibralatoPMcat,2024HaberlewCenIMBH}, and the confirmation and characterisation of (sub-)stellar or exoplanetary systems in proximity to the Sun \citep{2024AN....34530158B}. Astronomers have become familiar with the pros and cons of the two observatories: \hst works best in crowded regions and can observe objects over a wide range of magnitudes, but its limited field of view hampers investigations of large-scale structures; \gaia has provided high-precision PMs for the entire sky, but it has a limited depth and it struggles in extreme regions, such as the cores of globular clusters or towards the Galactic centre. In 2023, a new facility that potentially combines the strong points of these two telescopes has finally begun its operation: \euclid \citep{EuclidSkyOverview}.

\euclid's instruments offer superb image quality -- comparable to that of the \hst\footnote{The angular resolution of the \euclid telescope at 700\,nm (near the centre of the wavelength range covered by the VIS \IE filter) is twice that of \hst at 814\,nm and about the same as \hst at 1.25\,$\mu$m.} -- over an unprecedentedly wide field of view of 0.57 deg$^2$ for a 1-m class, diffraction-limited space telescope. The exquisite image quality immediately translates into high-precision astrometry for point sources, even in crowded environments. The wide field of view, the high resolving power, and the depth of \euclid's overall survey, when compared to most ground- and space-based surveys, provide an immense astrometric potential for the \euclid mission. However, in order to achieve their full astrometric capability, the undersampled \euclid instruments require more detailed knowledge of the point-spread function (PSF) than is necessary in fully sampled photometry.

In this study, we illustrate the well-established `state-of-the-art' techniques adopted to obtain high-precision imaging astrometry and photometry of point sources with undersampled detectors in space. We describe in detail the procedures specifically developed to analyse the data of the globular cluster NGC~6397 taken as part of the \euclid Early Release Observations (ERO) programme \citep{EROcite}. In particular, we show how we derive for both of \euclid's instruments: (i) high-accuracy geometric distortion (GD) corrections and, (ii) high-accuracy models of the cores of the PSFs. 

Finally, we combine \euclid and \gaia Data Release 3 \citep[DR3;][]{2016A&A...595A...1G,2023A&A...674A...1G} data to compute PMs for all stars in common. Thanks to the high-precision PMs we derive, we are able to measure the internal motions of the globular cluster NGC~6397 out to half of its tidal radius ($r_{\rm t}$).  Future papers will focus on accurate photometry, and on the characterisation of the stability of the derived PSFs and GD corrections over time and as a function of the colours of the sources. As supplementary online material of this work, we release to the community a high-accuracy astrometric catalogue and high-resolution multi-filter atlases of sources within the field of NGC~6397.

\begin{figure*}
\includegraphics[width=\textwidth]{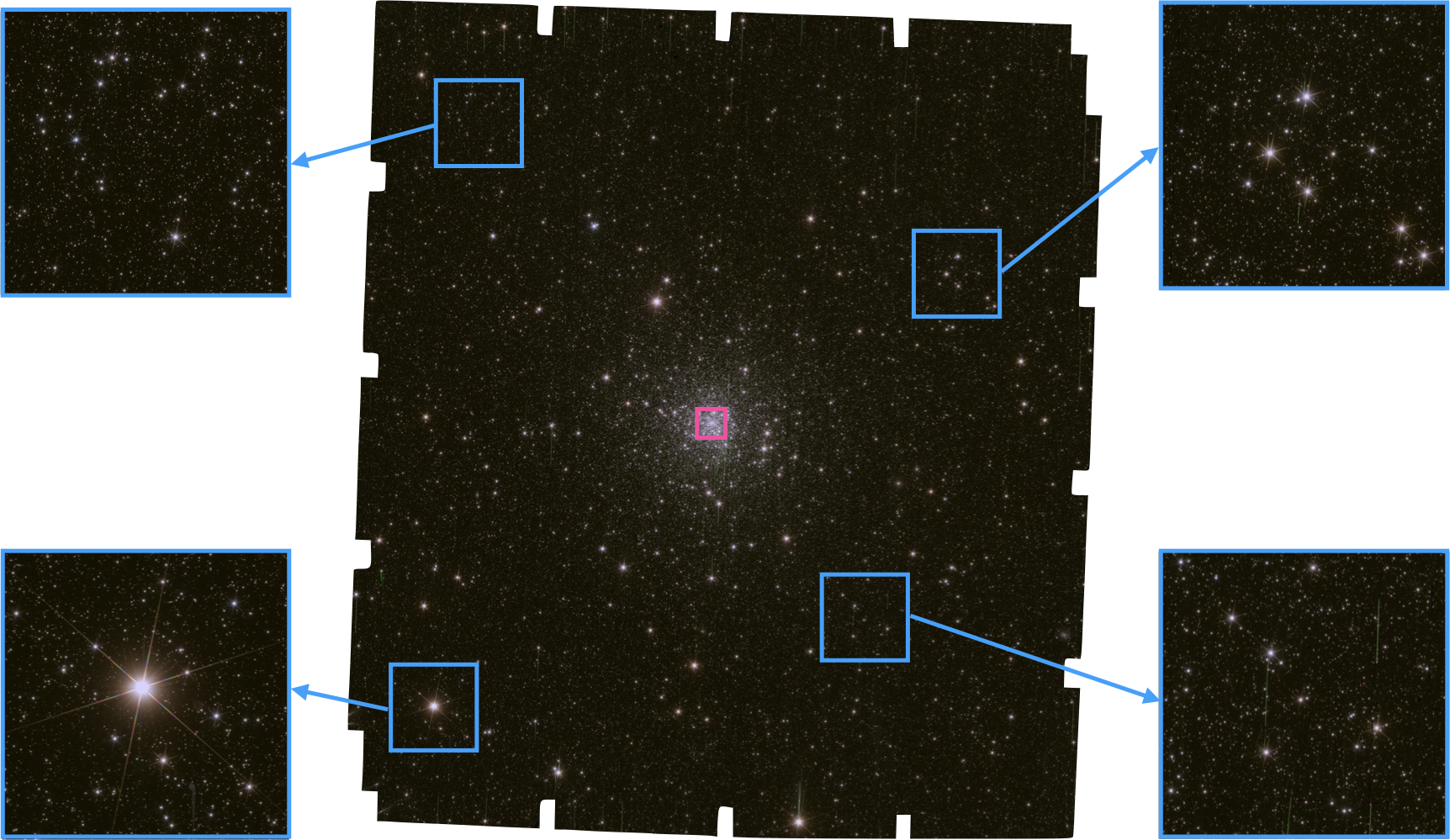}\\~\\~\\
\includegraphics[width=\textwidth]{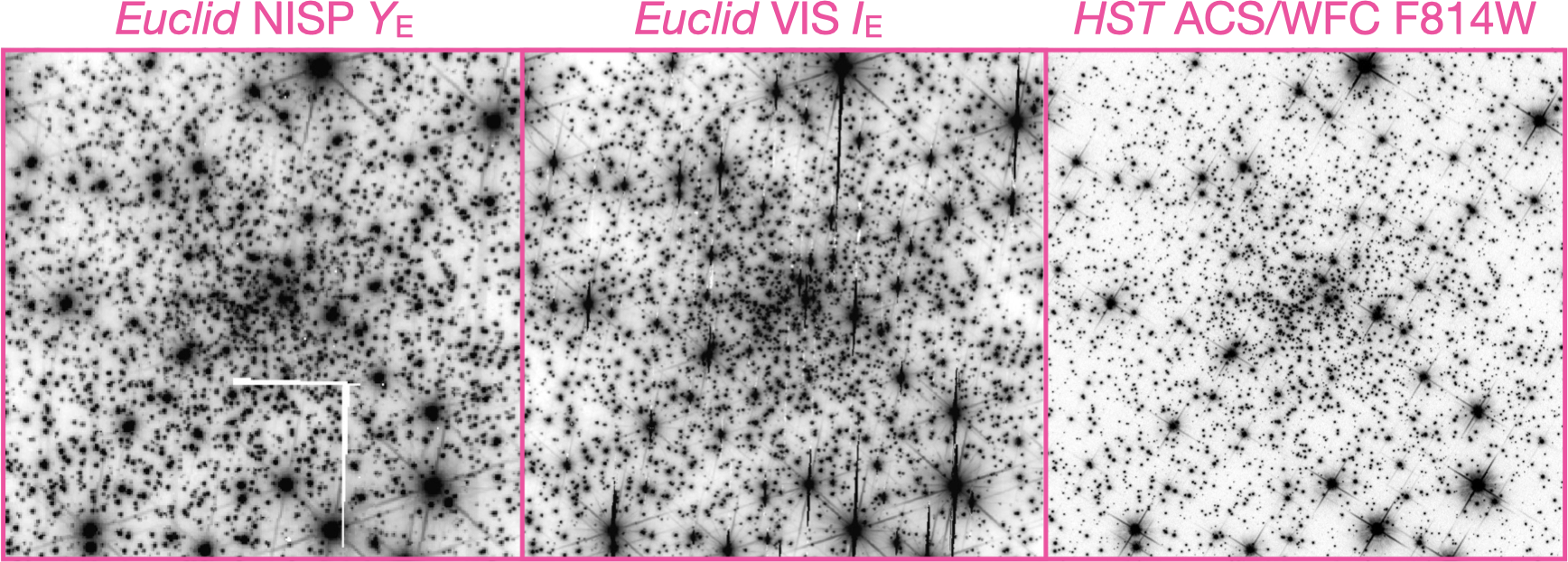}
\caption{\euclid view of NGC~6397 and comparison with \hst. The top-centre panel presents a colour stacked image (red$=$NISP $\HE$, green$=$NISP $\JE$, and blue$=$NISP $\YE$) of the entire field of NGC~6397 covered by the ERO data. The four zoom-ins (blue squares) highlight various regions in the outskirts of the field. The magenta box is centred on the core of the cluster. A zoom-in on this region is shown in the three bottom panels. The leftmost panel is taken from the NISP $\YE$-filter stacked image, while the middle panel refers to the VIS $\IE$-filter data. The empty space in the leftmost panel is an artefact of the stacking process. As a comparison, a view of the same region taken with \hst's Wide Field Camera of the Advanced Cameras for Surveys in the F814W filter \citep{2018MNRAS.481.3382N} is shown in the rightmost panel. In all these panels, north is up and east is to the left. All plots are on a logarithmic grey scale, with minimum and maximum values tailored in each image to enhance the faint details.}
\label{fig:EFoV}
\end{figure*}

\section{Observations}\label{S:obs}

The globular cluster NGC~6397 was observed on 22 September 2023 as one of the targets of the \euclid ERO programme with the Visible Instrument \citep[VIS;][]{EuclidSkyVIS} and the Near-Infrared Spectrometer and Photometer \citep[NISP;][]{EuclidSkyNISP} instruments. VIS is an imager composed of a 6$\times$6 array of 4k$\times$4k CCDs, with a pixel scale of 100 mas pixel$^{-1}$ and an instantaneous field of view of 0.54 deg$^2$. Each CCD has four quadrants that, because of the charge injection lines, are analysed independently. VIS is equipped with a single broadband (550--900 nm) filter (\IE).  NISP is designed to provide both spectroscopy and photometry, but in this work we make use only of its photometric capabilities. NISP is composed of 16 2k$\times$2k near-infrared detectors with a pixel scale of 300 mas pixel$^{-1}$ and a total field of view of $\sim$0.57 deg$^2$. NISP is equipped with three filters (\YE, \JE, \HE), which allow a spectral coverage from 950 to 2021 nm \citep{Schirmer-EP18}.

The ERO data set of NGC~6397 comprises four dithered exposures \citep{EROData} with VIS in the $\IE$ filter (of 560\,s  each) and with NISP in the $\YE$, $\JE$, and $\HE$ filters (of 87.2\,s each)\footnote{The NISP detectors use a non-destructive readout. Each of these NISP ERO exposures was obtained with four groups, 16 reads, and four drops. Each of the four groups was obtained by averaging 16 consecutive reads. Between each group, four frames were dropped (i.e. not read).}. Additional short exposures were also taken as part of the ERO programme, but they are not considered in our investigation. Figure~\ref{fig:EFoV} shows an overview of the \euclid data set analysed here and a comparison with \hst.

The VIS and NISP raw images were corrected for instrumental effects using the ERO pipeline as described in \citet{EROData}, to which we refer for a detailed overview of the process. The only change with respect to \citet{EROData} is that we did not correct the VIS data for the effects of the cosmic rays by the \texttt{deepCR} code \citep{2020ApJ...889...24Z} through pixel in-painting \citep[see][]{EROData}. We empirically found a small, yet sizeable, improvement in the astrometric and photometric precision in our analysis with respect to what is obtained with cosmic-ray corrected images (see Appendix~\ref{appendix:crphot}). In contrast to the approach in \citet{EROGalGCs}, here we analyse only the calibrated but not resampled images, which are the images best suited for high-precision PSF modelling. Hence, for analysing the astrometry of sources, we do not use the stacked images of the ERO public data release, but instead use the individual exposures (still not publicly available). In the following, we consider each VIS quadrant as a stand-alone detector for which to model PSFs and GD corrections.

\begin{figure*}
    \centering
    \includegraphics[height=5.3cm]{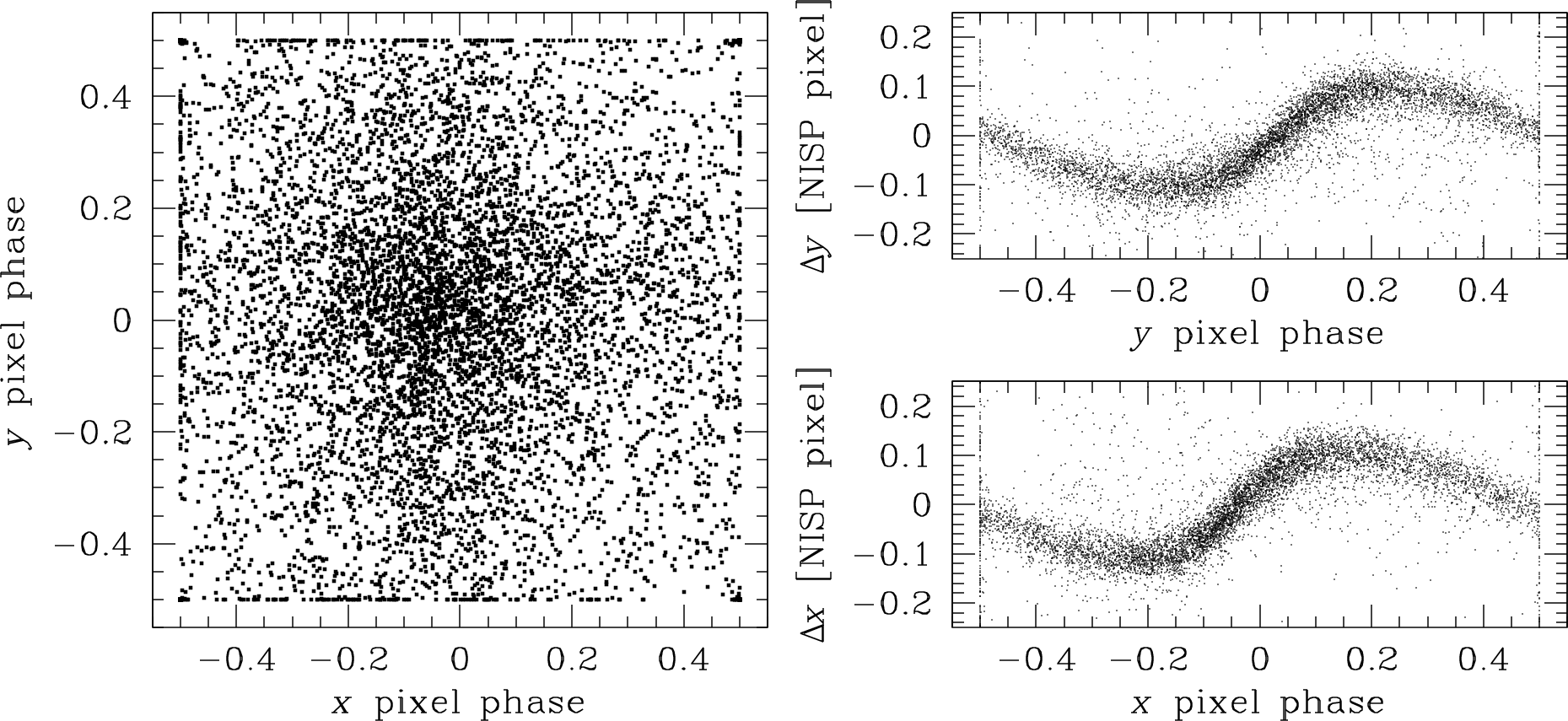}\hspace{0.15 cm}
    \includegraphics[height=5.3cm]{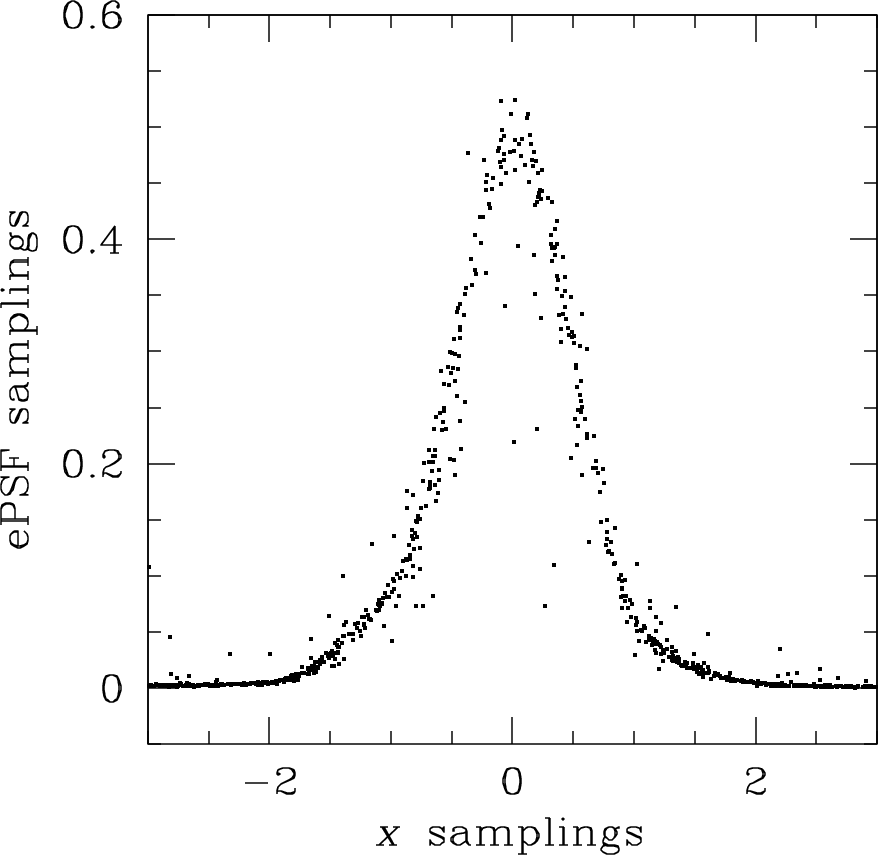}\\~\\~\\
    \includegraphics[height=5.3cm]{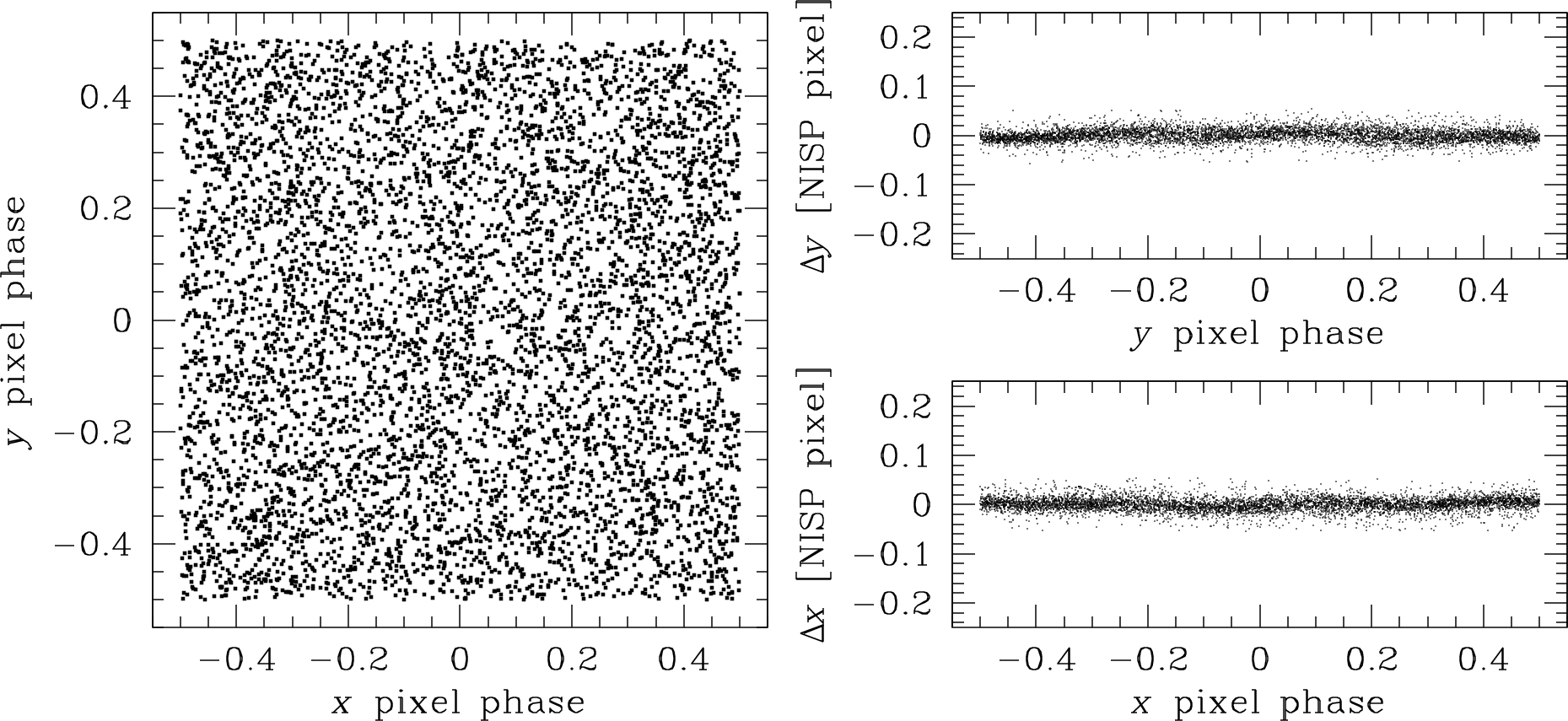}\hspace{0.15 cm}
    \includegraphics[height=5.3cm]{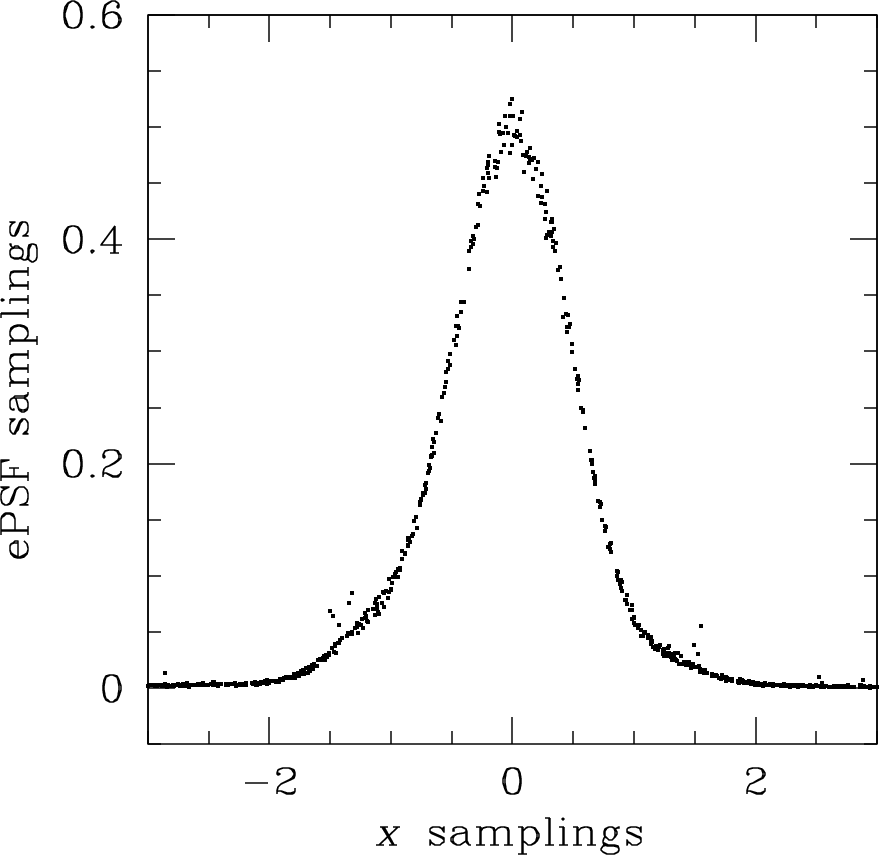}
    \caption{Example of the iterative derivation of the ePSF for the most undersampled instrument/filter combination (i.e. the NISP $\YE$ filter). The top row shows the first iteration, while the bottom row is the last iteration. In each row, we show from left to right: (i) where the centres of the stars are measured to be with respect to the pixel boundaries (the origin of the plot is the centre of the pixel); (ii) pixel-phase errors for the positional residuals (defined as the difference between the single \euclid exposure positions and the \gaia positions transformed back onto the \euclid frame) along the $x$ (top) and $y$ (bottom) axes; and (iii) ePSF samplings with respect to the centre of the ePSF placed at 0 along the $x$ axis in a small strip ($\Delta y < 0.01$ pixel) around the centre of the ePSF. In the rightmost panels, a given star contributes to the plot with different points (one per sampled pixel), at odds with the panels in the first two columns from the left where each point corresponds the position of a star.}
    \label{fig:ePSF_Y}
\end{figure*}

\section{ePSF and GD modelling}

The \euclid VIS PSFs are moderately undersampled, and the NISP PSFs are extremely undersampled, as is discussed in \citet{EuclidSkyVIS} and \citet{EuclidSkyNISP}, respectively (we quantify the level of undersampling in the following section). Thus, the modelling of these PSFs requires particular attention. We followed the prescriptions of \citet[later described in \citealt{2016LibralatoK2i,2023LibralatoNIRISS,2022NardielloNIRCam}]{2000AKWFPC2} to create effective PSFs (ePSFs). The ePSF is the convolution between the instrumental PSF (that due to the telescope optics) and the pixel response, and it is what one actually measures from an image. Because the ePSF modelling requires an estimate of the GD correction of the detector, we iterated the entire procedure three times, each time improving the GD correction and ePSF models with the products from the previous iteration. Below we briefly describe the two parts separately to simplify the presentation.

In the following, positions are estimated using the \texttt{FORTRAN} code \epass, which is the \euclid-tailored version of the code \hpass \citep{2022acs..rept....2A,2022wfc..rept....5A} that we developed for our analysis. The \epass code allows the users to measure positions and fluxes of all detectable sources in a given exposure in a single pass and provides useful diagnostic parameters to assess the quality of the results. In our iterative process discussed below, we initially used \epass to perform aperture photometry and then, when the ePSF models had been constructed, we fit the ePSF models to obtain more accurate positions and fluxes.

\subsection{ePSF modelling}

As described in \citet[see, in particular, the overview diagram in their Fig.~8]{2000AKWFPC2}, \citet{2016LibralatoK2i,2023LibralatoNIRISS,2024LibralatoMIRI}, and \citet{2022NardielloNIRCam}, every pixel in the vicinity of a star can provide one constraint on the ePSF model at a particular location in its domain. Collecting many samplings from many stars in different images allowed us to map the pixel-phase space and build accurate ePSF models. However, this is a complex procedure when dealing with undersampled PSFs.

\begin{figure*}
    \centering
    \includegraphics[width=\textwidth]{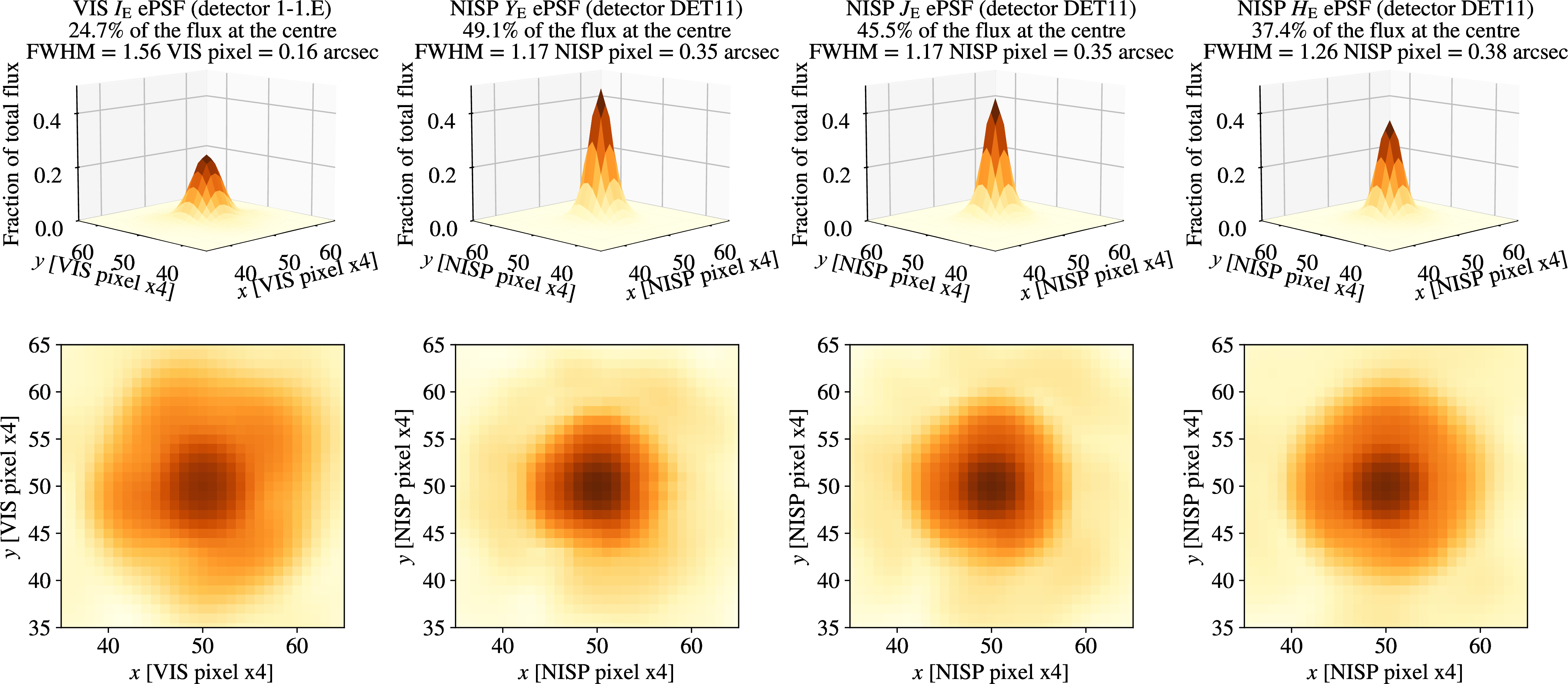}
    \caption{Overview of the \euclid's ePSFs modelled in our analysis. The top row shows a 3D representation of our ePSFs for the VIS quadrant 1-1.E and for the NISP detector DET11 in each of the three NISP filters. At the top of each panel, we provide the percentage of flux of the star within the centremost pixel, and the full width at half maximum (FWHM) in pixel and in arcsec. The bottom row contains the 2D view of the ePSFs. The scale is logarithmic and the same in all these four panels.}
    \label{fig:ePSFs}
\end{figure*}

The first steps of the modelling of undersampled PSFs present an important hurdle that needs to be overcome: PSFs can be obtained when accurate stellar positions are available, but without accurate PSF models it is not possible to measure positions accurately. To break this degeneracy, we took advantage of an external catalogue to provide the accurate positions we need. We chose the \gaia DR3 catalogue for this task since it provides hundreds of stars that are bright enough in the VIS or NISP images to be used for the ePSF modelling.

The \gaia DR3 positions were moved to the epoch of the \euclid ERO observations by means of the PMs provided in the catalogue. The equatorial coordinates in the \gaia catalogue were projected onto a tangent plane centred at the centre of the VIS or NISP mosaic (taken from the header of each image). The pixel scale was set to be the nominal pixel scale of VIS or NISP (100 and 300 mas pixel$^{-1}$, respectively), and the $x$ and $y$ axes were oriented west and north, respectively. We refer to these positions as $(x^{\rm Gaia},y^{\rm Gaia})$. These were transformed into the raw reference frame of each VIS quadrant or NISP detector by means of: (1) six-parameter linear transformations (to go from the \gaia plane to the GD-corrected plane); and (2) the inverse GD correction (to go from the GD-corrected to the raw plane). These positions then provide a bias-free estimate of the true location of the stars on the pixel grid. We refer to these positions as $(x^{\rm Gaia}_{\rm RAW, 6p},y^{\rm Gaia}_{\rm RAW, 6p})$.

We iteratively measured position and flux of bright\footnote{Our ePSFs were made using stars in a specific magnitude range and, as such, these ePSFs resemble the flux distribution of objects with flux similar to that of the sources used in the modelling. Stars brighter or fainter than those used for the ePSF modelling show small deviations from what predicted by the ePSF. This is what we would expect from the so-called `brighter-fatter effect', as discussed in, for example, \citet{2023LibralatoNIRISS,2024LibralatoMIRI}. For now, we chose to make our ePSF models flux-independent and postpone more detailed analyses to future works.}, isolated sources in each VIS quadrant or NISP detector. At the first iteration, positions were just defined as the photocentre, while fluxes came from aperture photometry. In the subsequent iterations, when the ePSFs were available, these values were updated with the result of the ePSF fit.

The bright and isolated stars in the \euclid and \gaia catalogues are cross-identified and  transformations between the two frames are computed. The \gaia positions are transformed back to the \euclid frame and then the inverse GD correction is applied to provide positions in the raw chip-based system $(x^{\rm Gaia}_{\rm RAW, 6p},y^{\rm Gaia}_{\rm RAW, 6p})$, where ePSFs are modelled. Once a star has been pin-pointed on the pixel grid, its pixels are used to sample the ePSF. At each iteration, new, improved \euclid positions are available and are used to compute more accurate transformations between frames. As in \citet{2023LibralatoNIRISS}, we allow the ePSF fitting procedure to also fit source positions instead of imposing positions from the \gaia catalogue in the last iterations. An example of the iterative process for the detector DET11 of NISP in the $\YE$ filter is presented in Fig.~\ref{fig:ePSF_Y}. At the beginning, the ePSF samplings are preferentially clustered at the centre of the pixel and a clear trend in the positional errors as a function of the pixel boundaries is present (top row). When the ePSF modelling is complete (bottom row), the ePSF samplings are now homogeneously distributed across the pixel and the pixel-phase errors have significantly improved. Small residual pixel-phase errors are still present in ePSF models for some of the VIS quadrants or NISP detectors. The use of \gaia positions, for which uncertainties are larger than the nominal DR3 positions (because they were propagated to the epoch of the \euclid observations), is likely the major limitation to our modelling. As shown in, for example, \citet{2000AKWFPC2}, a self-calibration approach could lead to more accurate ePSFs but the \euclid ERO data set is not designed for this purpose (few images and only large dithers). Although not perfect because it relies on limited data, our procedures already allow us to obtain accurate ePSFs for the \euclid instruments for high-precision and accuracy astrometry, which are a significant improvement over existing products (as described in the next sections).

The available data sets for VIS and NISP allowed us to model oversampled ePSFs by a factor of 4. Various constraints were applied to ensure the ePSF models are smooth and continuous \citep{2000AKWFPC2}. We normalised our ePSFs to sum up to 1 within a radius of 5 pixels (chosen arbitrarily). The ePSF modelling started with one ePSF for the entire detector and then we let the ePSF model vary spatially, dividing the array into regions and computing an ePSF for each of them. For the VIS quadrants, we designed a $3 \times 3$ array of ePSF models, while for the NISP detectors the 10 times more stars made it possible to obtain a $5 \times 5$ array of ePSFs. We quantified the spatial variation of the ePSFs by computing the difference between the centremost pixel of each ePSF and the average ePSF. The median and maximum differences among all chips for a given detector and filter are the following:
\begin{itemize}
    \item VIS $\IE$ -- median value: 1.2\%, maximum value: 4.4\%;
    \item NISP $\YE$ -- median value: 2.6\%, maximum value: 3.6\%;
    \item NISP $\JE$ -- median value: 2.4\%, maximum value: 3.1\%;
    \item NISP $\HE$ -- median value: 1.6\%, maximum value: 2.0\%.
\end{itemize}

Figure~\ref{fig:ePSFs} shows a 3D view of the ePSFs for the bottom-left quadrant and detector of the VIS and NISP instruments, respectively. All ePSFs are undersampled, with the most extreme case being that of the NISP $\YE$-filter data. The sampling parameter $r$ is a pure number defined as the ratio between the PSF full-width at half maximum (FWHM) and the pixel size in the same units (mas or pixel). For an undersampled PSF, $r$ is smaller than the `Nyquist' sampling 2.3. Using the values in Fig.~\ref{fig:ePSFs}, we conclude that the VIS instrument is moderately undersampled ($r\simeq1.5$), while the NISP instrument is severely undersampled ($r \simeq 1.2$--1.3).

\subsection{GD polynomial corrections}

The GD correction for each mosaic was calibrated by leveraging the \gaia DR3 catalogue. The large FoV covered by VIS and NISP requires a specific GD correction that takes into account that every dithered image is in a different tangent plane. To perform this, we followed the prescriptions of \citet{2015LibralatoVIRCAM} and  \citet{2022MNRAS.515.1841G} for wide-field imagers. For every image, the \gaia positions were propagated to the epoch of the observations, thus removing the contribution of the PMs that could significantly increase the noise in the mapping of the GD. As described in the previous section, we projected the \gaia positions onto a tangent plane centred at the centre of each chip, fixing the pixel scale to be the nominal pixel scale of the instrument analysed, and imposing the $x$ and $y$ axes to be oriented west and north, respectively. This step allowed us to disentangle the projection effects form the GD and use the information from different images to solve for the GD.

The first step of the solution consists of two third-order polynomial functions, one for each coordinate, for every quadrant of the VIS or detector of the NISP mosaic. For NISP, we obtained a GD correction for each filter, so as to take into account the filter-dependent terms in the solution. We used the best-available ePSF models to measure positions and fluxes of bright, well-measured, unsaturated objects. We excluded from the \gaia catalogue stars brighter than $G$$=$13 (saturation makes \gaia astrometry worse for stars brighter than this threshold) and fainter than 20.5 (to exclude the few stars with very large PM errors). Then, we cross-identified the same stars in the \euclid and \gaia catalogues. For VIS, the number of stars in common with \gaia varied between about 1000 and 4300, whereas for NISP the number ranged between about 9000 and 25\,000, thanks to the shallower NISP exposures. The \gaia positions were transformed onto the raw reference system of the \euclid VIS quadrant or NISP detector by means of four-parameter linear transformations (rigid shifts in the two coordinates, one rotation, and one change of scale) and the available inverse GD correction to obtain $(x^{\rm Gaia}_{\rm RAW, 4p},y^{\rm Gaia}_{\rm RAW, 4p})$. Finally, positional residuals were defined \citep[e.g. Sect. 4.1.3 of][]{2010BelliniLBT} as the difference between these transformed \gaia positions and the raw \euclid positions $(x_{\rm RAW},y_{\rm RAW})$:
\begin{equation}
    \left\{
    \begin{array}{c}
        \Delta x = x^{\rm Gaia}_{\rm RAW, 4p} - x_{\rm RAW} \, , \\
        \Delta y = y^{\rm Gaia}_{\rm RAW, 4p} - y_{\rm RAW} \, .
    \end{array}
    \right.
\end{equation}

\begin{figure*}[th!]
    \sidecaption
    \centering
    \includegraphics[width=6.35cm]{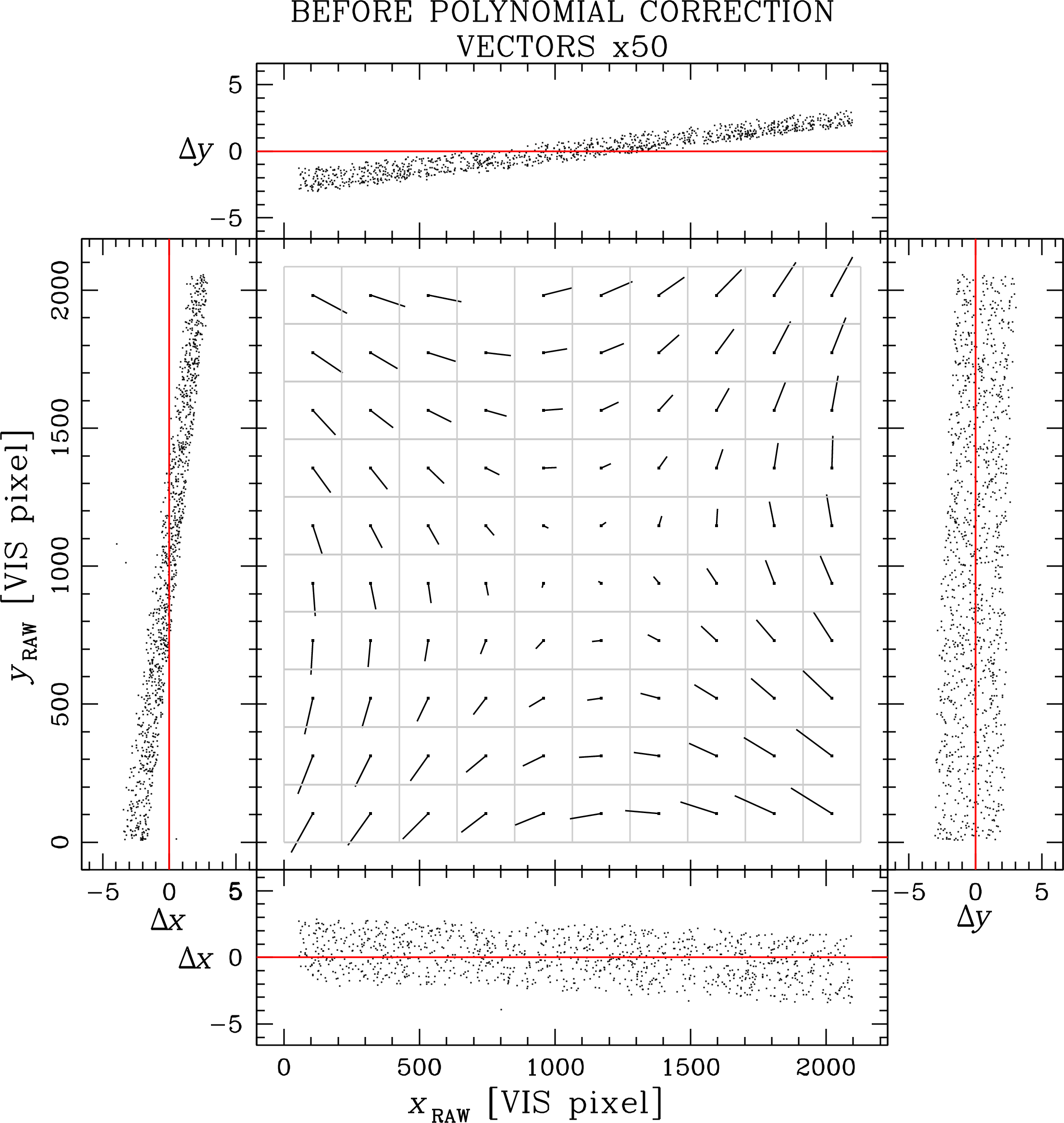}
    \includegraphics[width=6.35cm]{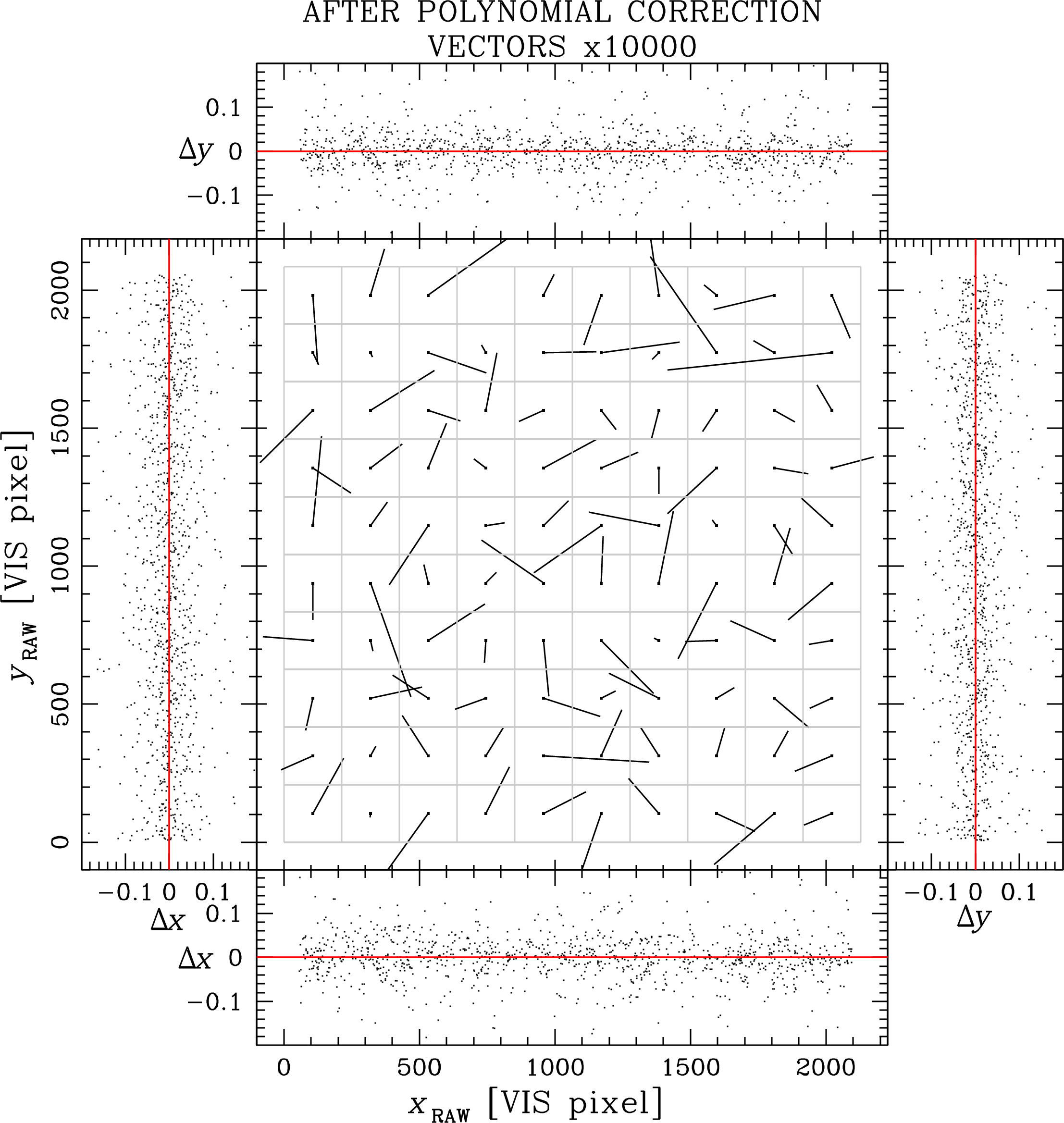}
    \caption{GD maps for the VIS quadrant 1-1.E before (left panel) and after (right panel) applying the polynomial correction. Vectors are magnified by a factor of 50 (left) and 10\,000 (right) to enhance the details. The positional $x$ and $y$ positional residuals as a function of $x$ and $y$ raw VIS positions are shown in the side panels.}
    \label{fig:VISGDdet1}
\end{figure*}

\begin{figure*}[th!]
    \sidecaption
    \centering
    \includegraphics[width=6.35cm]{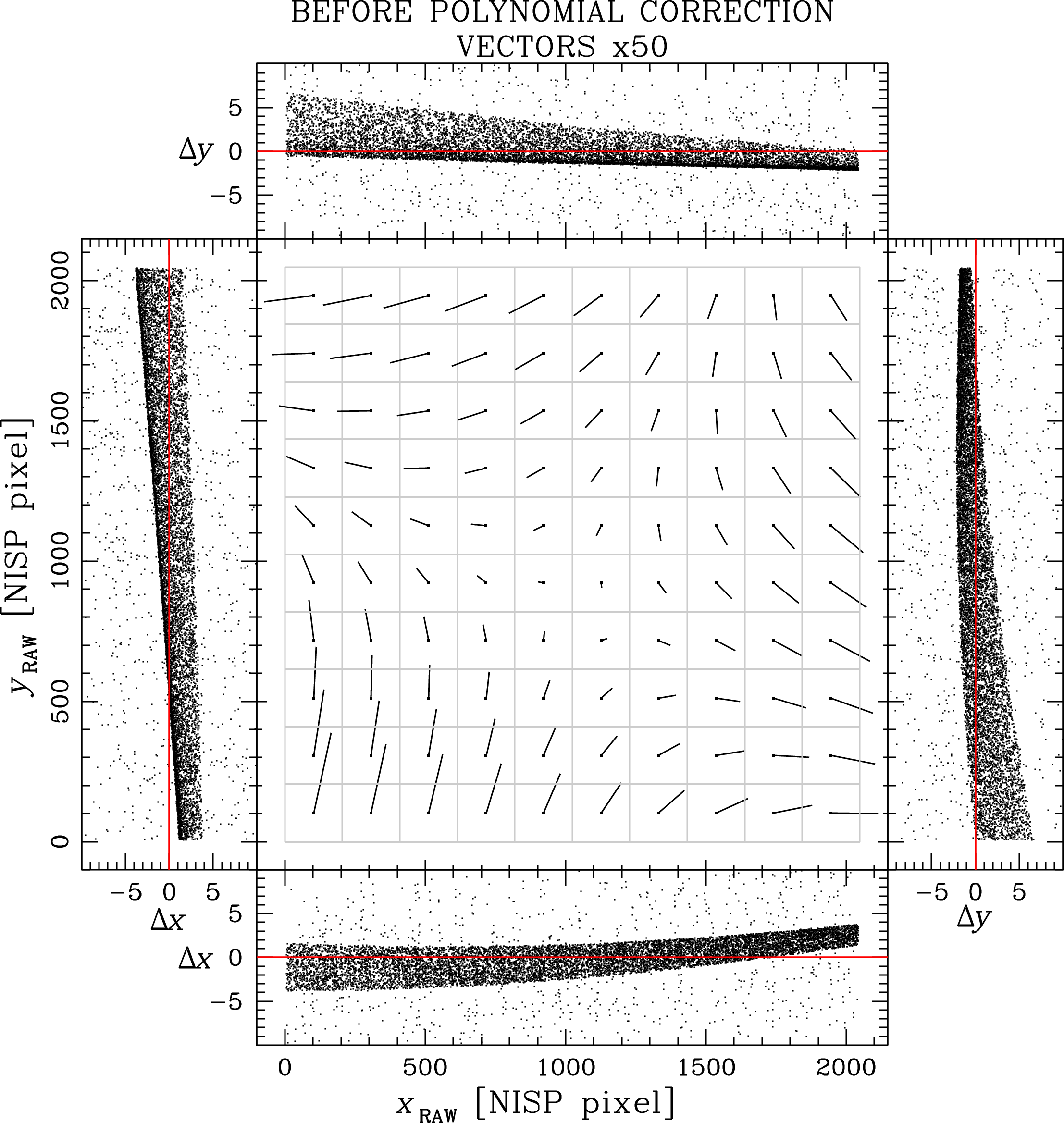}
    \includegraphics[width=6.35cm]{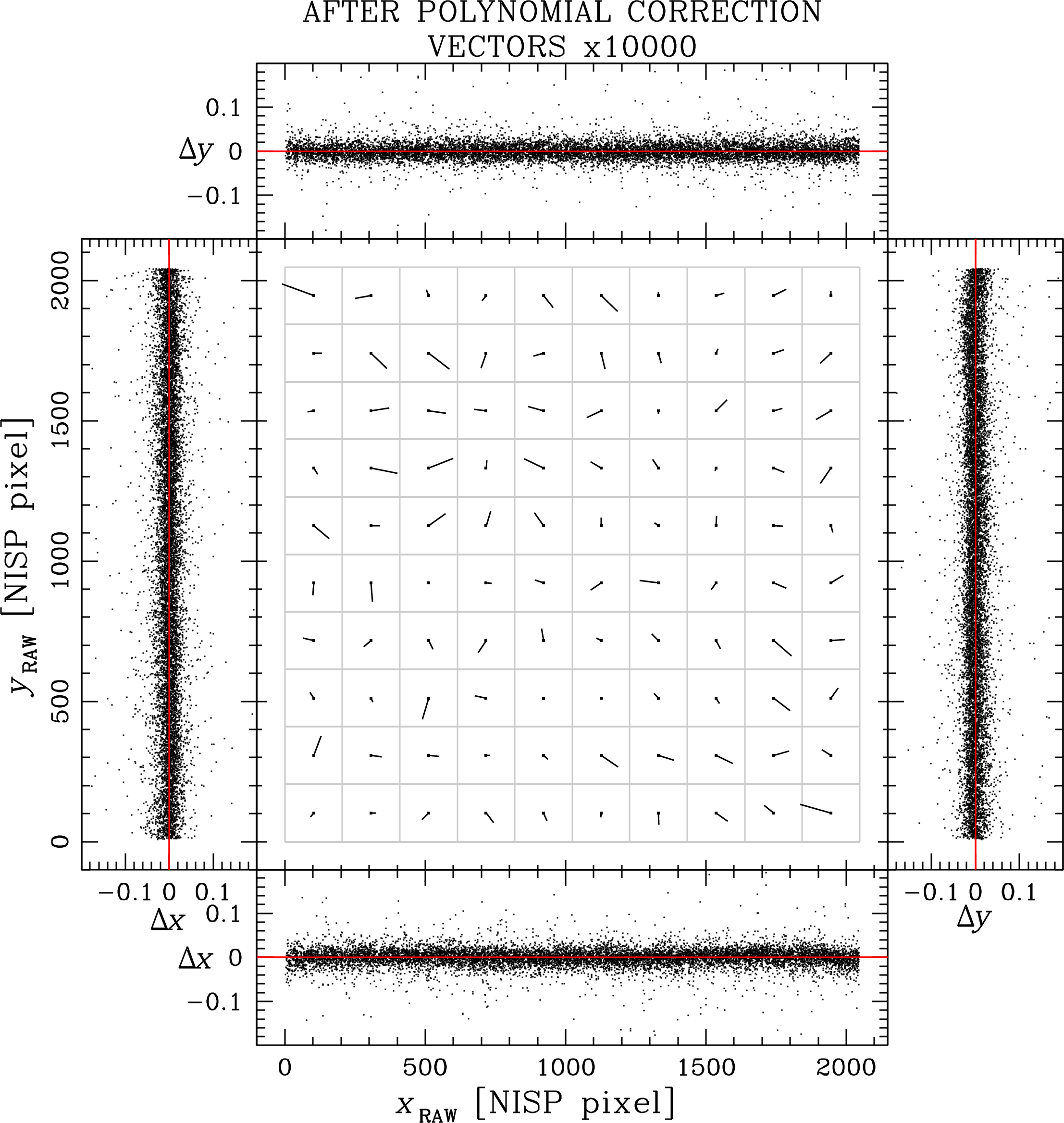}
    \caption{Similar to Fig.~\ref{fig:VISGDdet1}, but for the NISP DET11 detector with $\YE$-filter data. Units are now NISP pixels.}
    \label{fig:NISPYGDdet1}
\end{figure*}

We collected all residuals and fit them with two third-order polynomial functions. The coefficients of the polynomial functions were obtained via a least-square fit of all positional residuals. As in other analyses, \citep[e.g.][]{2006AndersonWFI,2010BelliniLBT,2014LibralatoHAWKI}, we selected the centre of the detector $(x_{\textrm{ref}},y_{\textrm{ref}})$ as the reference pixel with respect to which we solve for the GD. This choice makes the interpretation of the different terms of the GD straightforward. We selected $(x_{\textrm{ref}},y_{\textrm{ref}}) = (1064,1043)$ and $(1024,1024)$ for the VIS quadrants and NISP detectors, respectively. The GD correction is defined as \citep[see also Sect. 4.1.2 of][]{2010BelliniLBT}
\begin{equation}
    \left\{
    \begin{array}{c}
        \delta x = \sum_{i = 1, 3}\sum_{j = 1, 3-i} a_{ij} \left(\frac{x-x_{\textrm{ref}}}{x_{\textrm{ref}}}\right)^i\left(\frac{y-y_{\textrm{ref}}}{y_{\textrm{ref}}}\right)^j \, , \\
        \delta y = \sum_{i = 1, 3}\sum_{j = 1, 3-i} b_{ij} \left(\frac{x-x_{\textrm{ref}}}{x_{\textrm{ref}}}\right)^i\left(\frac{y-y_{\textrm{ref}}}{y_{\textrm{ref}}}\right)^j \, ,
    \end{array}
    \right.
\end{equation}
\begin{equation}
    \left\{
    \begin{array}{c}
        x_{\rm CORR} = x_{\rm RAW} + \delta x \, , \\
        y_{\rm CORR} = y_{\rm RAW} + \delta y \, .
    \end{array}
    \right.
\end{equation}
We iterated the process 200 times, each time adding only 75\% of the coefficient values to the previous estimates to ensure the convergence of the solution. Figures~\ref{fig:VISGDdet1} and \ref{fig:NISPYGDdet1} show the GD maps and the positional residuals against the \gaia catalogue as a function of $x$ or $y$ coordinates before (left) and after (right) applying the polynomial correction in the case of the VIS and NISP $\YE$-filter data. The GD maps for all detectors are presented in Appendix~\ref{appendix:gdmaps}.

The distortion maps and residuals for both VIS quadrants and NISP detectors show that our corrections effectively remove most of the GD. For VIS, some fine structures ($<$0.1 pixel) are still present in the positional residuals. An additional look-up table of residuals would remove these systematics, but the number of sources in common with \gaia is not high enough for this task \citep[e.g.][]{2023LibralatoNIRISS}. Alternatively, one could compute such look-up tables by means of a self-calibration approach \citep[e.g.][]{2010BelliniLBT,2014LibralatoHAWKI}, but the limited number of images and the particular dither pattern used for the \euclid ERO observations of NGC~6397 is not appropriate for this approach \citep[see discussion in][]{2021HaberleNACO}. For NISP, the number of sources in common with \gaia is higher than for VIS, and we are able to obtain residuals with a better dispersion than for VIS (see below) and comparable with other studies in the literature on the topic of the GD \citep[e.g.][]{2011BelliniWFC3}.

We note that the residuals shown in the distortion maps in Figs.~\ref{fig:VISGDdet1} and \ref{fig:NISPYGDdet1} include the contributions of the ePSF-fit errors, the residual uncorrected GD and the positional errors of \gaia. For the latter part, the dominant source of error is that related to the PM propagation from 2016.0 (the reference epoch of the \gaia DR3) to the epoch of the \euclid observations. We measured the RMS of the positional residuals obtained by comparing the \euclid and \gaia positions (i.e. like in the side panels of Figs.~\ref{fig:VISGDdet1} and \ref{fig:NISPYGDdet1}). For all 144 VIS quadrants, we find that the RMS ranges from 0.032 and 0.053 VIS pixel, which corresponds to 3.2--5.3 mas. For the 16 NISP detectors, we find instead 0.014--0.020 pixel, which corresponds to 4.2--6.0 mas. These values in mas are strikingly similar and could suggest that the \gaia PM errors dominate the error budget in the positional residuals and that our \euclid astrometry is much better than what is shown in the two figures. We will further investigate this in Sect.~\ref{sec:precision}, where we present an analysis to assess the astrometric precision of our data reduction using \euclid data alone, and in Sect.~\ref{sec:pms} where we compare the \gaia DR3 PMs with the PMs obtained by combining the \euclid and \gaia positions.

Solving for the GD of a detector by means of an external reference catalogue strongly depends on the quality of the reference catalogue itself \citep{2019WFIRST}. Any systematic error present in the reference catalogue can propagate in the GD solution. The situation worsen when PMs are included in the process. In our work, the contribution of the PM propagation of the \gaia positions dominates the error budget, as we have shown above. However, for relative astrometry, like what we present in Sect.~\ref{sec:precision}, the \gaia systematics do not explicitly manifest themselves in the positions. Combining dithered observations, we can compare positions of stars measured in different parts of the VIS or NISP mosaic and randomize any systematic errors, thus improving our reference frame.

\subsection{GD meta frame}

The second step of the GD correction consists of putting all chips on to the same distortion-free reference frame. Again, we followed the prescriptions of \citet{2003PASP..115..113A}, \citet{2010BelliniLBT}, \citet{2015LibralatoVIRCAM} and \citet{2023GriggioNIRCam}. First, we applied the GD correction to all bright stars in our \euclid single-chip catalogues found in common with the \gaia DR3 catalogue. Then, we found the four-parameter linear transformations to transform the GD-corrected positions of each chip $k$ on to the \gaia DR3 catalogue, and then back to the reference system of the bottom-left VIS quadrant (1-1.E) or NISP detector (DET11). The relations between the positions of a star in the chip-$k$ system $(x_{k}^{\rm corr},y_{k}^{\rm corr})$ and that in the chip-1 system $(x_{1}^{\rm corr},y_{1}^{\rm corr})$ are \citep[see also the detailed description of the process in Sect.~5 of][]{2010BelliniLBT}:
\begin{equation}
  \begin{array}{ll}
  \left(
  \begin{array}{c}
    x_{1}^{\rm corr}\\
    y_{1}^{\rm corr}\\
  \end{array}
  \right)&
  \hspace{-0.3 cm}= \frac{\alpha_k}{\alpha_{1}}
  \left[
    \begin{array}{rr}
      \cos(\theta_k-\theta_{1}) & \sin(\theta_k-\theta_{1}) \\
     -\sin(\theta_k-\theta_{1}) & \cos(\theta_k-\theta_{1}) \\
    \end{array}
    \right]
  \left(
  \begin{array}{c}
    x_{\it k}^{\rm corr}-x_{\rm ref} \\
    y_{\it k}^{\rm corr}-y_{\rm ref} \\
  \end{array}
  \right)
  \\ & +
  \left(
  \begin{array}{c}
    (x^{\rm offset})_{\it k} \\
    (y^{\rm offset})_{\it k} \\
  \end{array}
  \right) \, ,
  \end{array}
\end{equation}
where $(x_{\textrm{ref}},y_{\textrm{ref}})$ refer to the centre of the detector defined above. The scale factor is indicated as $\alpha_{\it k}$ and the angle as $\theta_{\it k}$, while $(x_{\textrm{offset}},y_{\textrm{offset}})$ are two offsets that include the relative position of the centre of the chip $k$ in the reference frame of chip 1 and place the centre of the image in the pixel (0,0). This meta-frame solution places all detectors onto a GD-corrected system and in the same tangent plane. Hereafter, we refer to this system simply as the `meta-frame catalogue'. The advantage of using the meta frame solution is that all 144 VIS quadrants or 16 NISP channels, respectively, are collected in the same catalogue, which makes the handling of the data easier.

\begin{figure}[t!]
    \centering
    \includegraphics[width=\columnwidth]{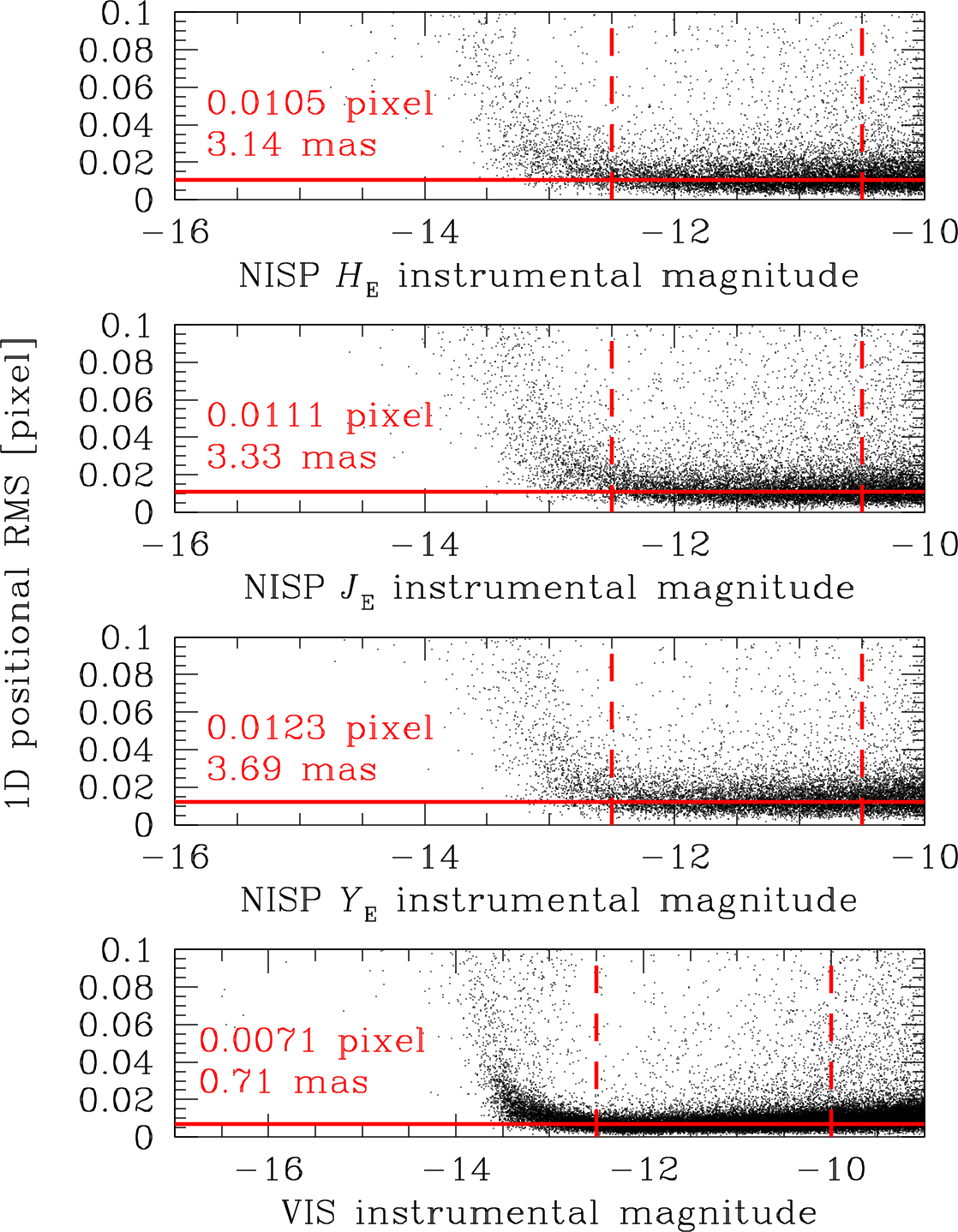}
    \caption{Oned dimensional (1D) positional RMS (expressed in units of VIS or NISP pixel depending on the panel) as a function of VIS or NISP instrumental magnitude. The red horizontal line is set at the median value of bright, well-measured (\qfit $<$ 0.05) stars that lie within the two red, dashed, vertical lines. The median values in pixels and mas are reported in each panel. Only 20\% of the points are shown for clarity.}
    \label{fig:1drms}
\end{figure}

\begin{figure}[th!]
    \centering
    \includegraphics[width=\columnwidth]{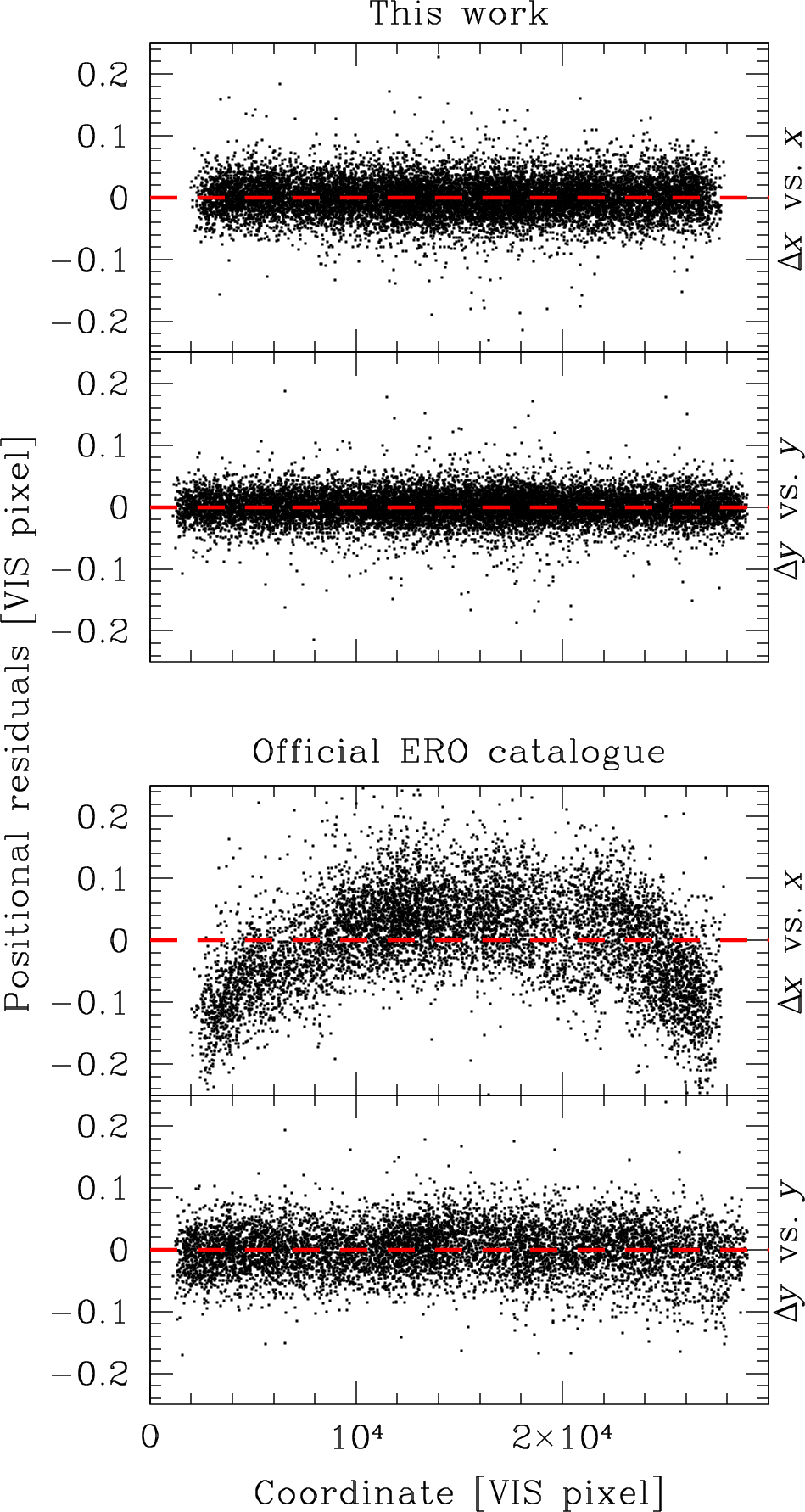}
    \caption{Comparison of the astrometry in our new and the ERO official VIS catalogues. In each panel, we show the positional residuals as a function of position in the \gaia catalogue. The positional residuals are computed as the difference between the \gaia DR3 positions and our or the ERO positions transformed on to the reference frame of the \gaia catalogue. All units are VIS pixels. The labels on the right side of each plot describe the quantities depicted in each panel. The red, dashed horizontal line is set to 0 as a reference.}
    \label{fig:astrocomp}
\end{figure}

\subsection{Astrometric precision}\label{sec:precision}

At the end of our iterative process, we have a final set of ePSF models and GD corrections for each detector, instrument, or filter. We re-measured all detectable sources in each \euclid image via an ePSF fit using the code \epass. Then, we applied the GD correction and put all chips in the same meta frame.

We estimated the astrometric precision of our data reduction by combining all images together for a given instrument and filter. First, we placed all meta catalogues on to the same tangent plane. As a tangent point, we arbitrarily chose the centre of NGC~6397\footnote{(R.A.,Dec.) $=$ $(265.175385,-53.674335)$ from \href{https://people.smp.uq.edu.au/HolgerBaumgardt/globular/}{the GC online database of Holger Baumgardt}.}. None of the images is centred exactly on the cluster, so this choice allowed us to test the accuracy of our meta-frame solution. Because the tangent point of each image taken from the header is set to be the pixel (0,0) in the meta frame, this process is straightforward. Once all catalogues were projected onto the same tangent plane, we cross-identified the same stars in multiple meta catalogues, and averaged their positions and fluxes after they were transformed onto a common reference frame system, hereafter `master frame', by means of six-parameter linear transformations. The scale and orientation of this master frame were defined by means of the \gaia DR3 catalogue (similarly to what described in the previous sections). We projected the \gaia positions onto a tangent plane centred on the cluster, fixing the pixel scale and axis directions. We kept only stars measured in at least three VIS images. We defined the 1D positional RMS as the sum in quadrature of the positional RMS along the $x$ and $y$ axes divided by $\sqrt{2}$. The result for both VIS and NISP instruments is shown in Fig.~\ref{fig:1drms}. 

We selected well-measured stars from saturation to two magnitudes fainter and computed their median 1D positional RMS values. Well-measured objects were selected by means of the `Quality of PSF fit' or \qfit parameter \citep{2006AndersonWFI,2014LibralatoHAWKI} provided by \epass. The \qfit is defined as the absolute fractional error in the ePSF fit of a source. It is close to 0 for a well-measured source and it gets increasingly higher for poorly measured objects. The median 1D positional RMS for bright stars with \qfit $<$ 0.05 is reported in each panel in Fig.~\ref{fig:1drms}. Thanks to our accurate ePSF and GD modelling, we can measure stars with a precision better than about 0.7\,mas and 3\,mas with VIS and NISP, respectively. Pixel-wise, these values are consistent with what has been achieved with the same state-of-the-art techniques for cameras onboard \hst \citep{2006AndersonWFI,2009BelliniWFC3,2011BelliniWFC3} and \jwst \citep{2023GriggioNIRCam,2023LibralatoNIRISS,2024LibralatoMIRI}.

\begin{figure*}[th!]
    \sidecaption
    \centering
    \includegraphics[width=0.7\textwidth]{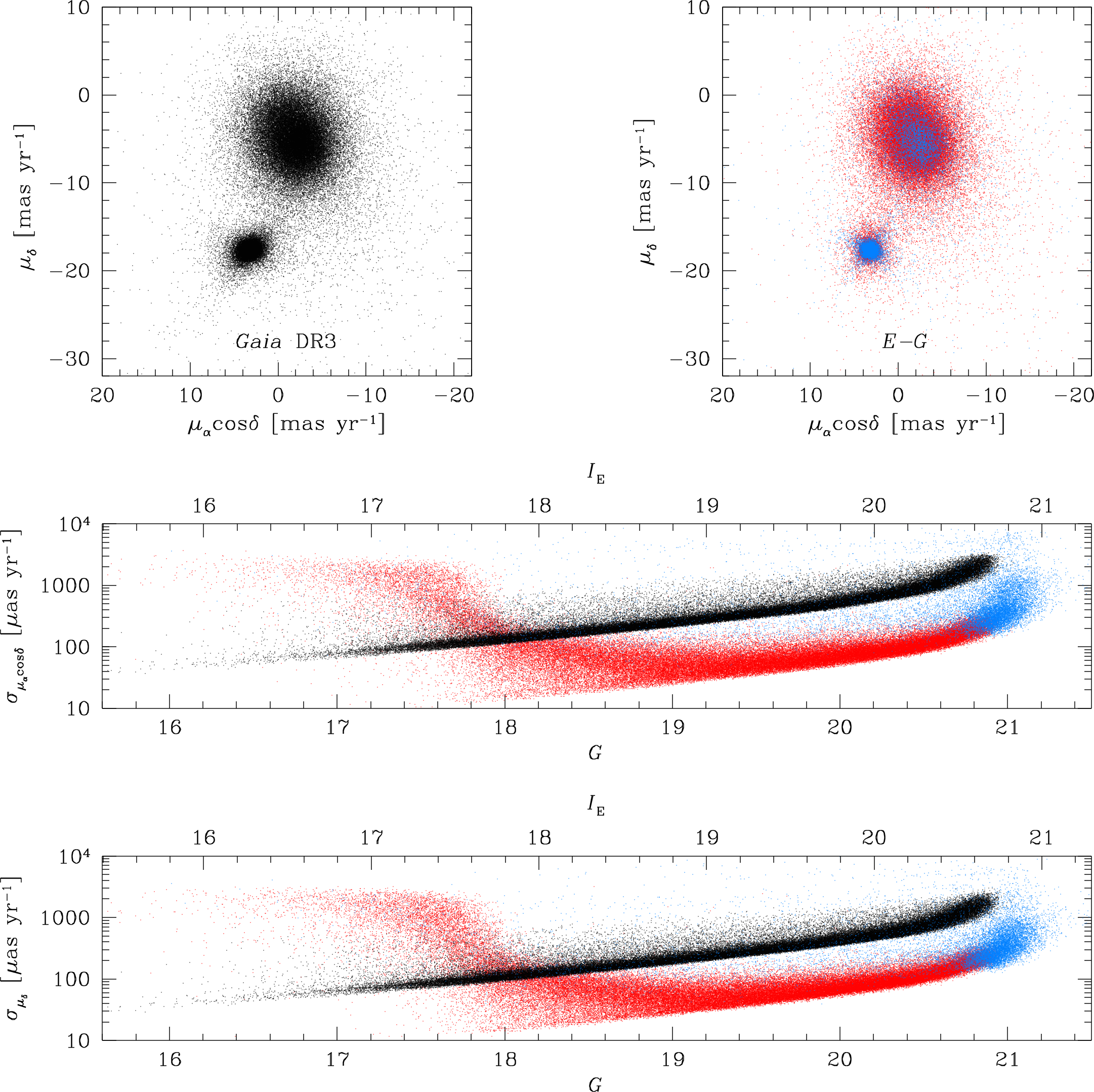}
    \caption{Comparison between the \gaia and \euclid-\gaia, hereafter \egpm, astrometry. The two VPDs at the top present the \gaia DR3 (left panel) and \egpm (right panel) PMs. Objects in common between the two catalogues are shown in black or red, depending on the panel. Blue points are stars with a two-parameter astrometric solution in \gaia (i.e. no PMs in the DR3) and an \egpm PM estimate. The middle and bottom panels show the PM errors along $\alpha \cos\delta$ and $\delta$ as a function of $G$ magnitude in the \gaia DR3 (black points) and \egpm (red and blue points) catalogues. The increase of the \egpm PM errors from \IE$\lesssim$18.8 is related to saturation and poorly measured objects in the VIS exposures.}
    \label{fig:EuclidVsGaiaDR3}
\end{figure*}

\begin{figure}[t!]
    \centering
    \includegraphics[width=0.8\columnwidth]{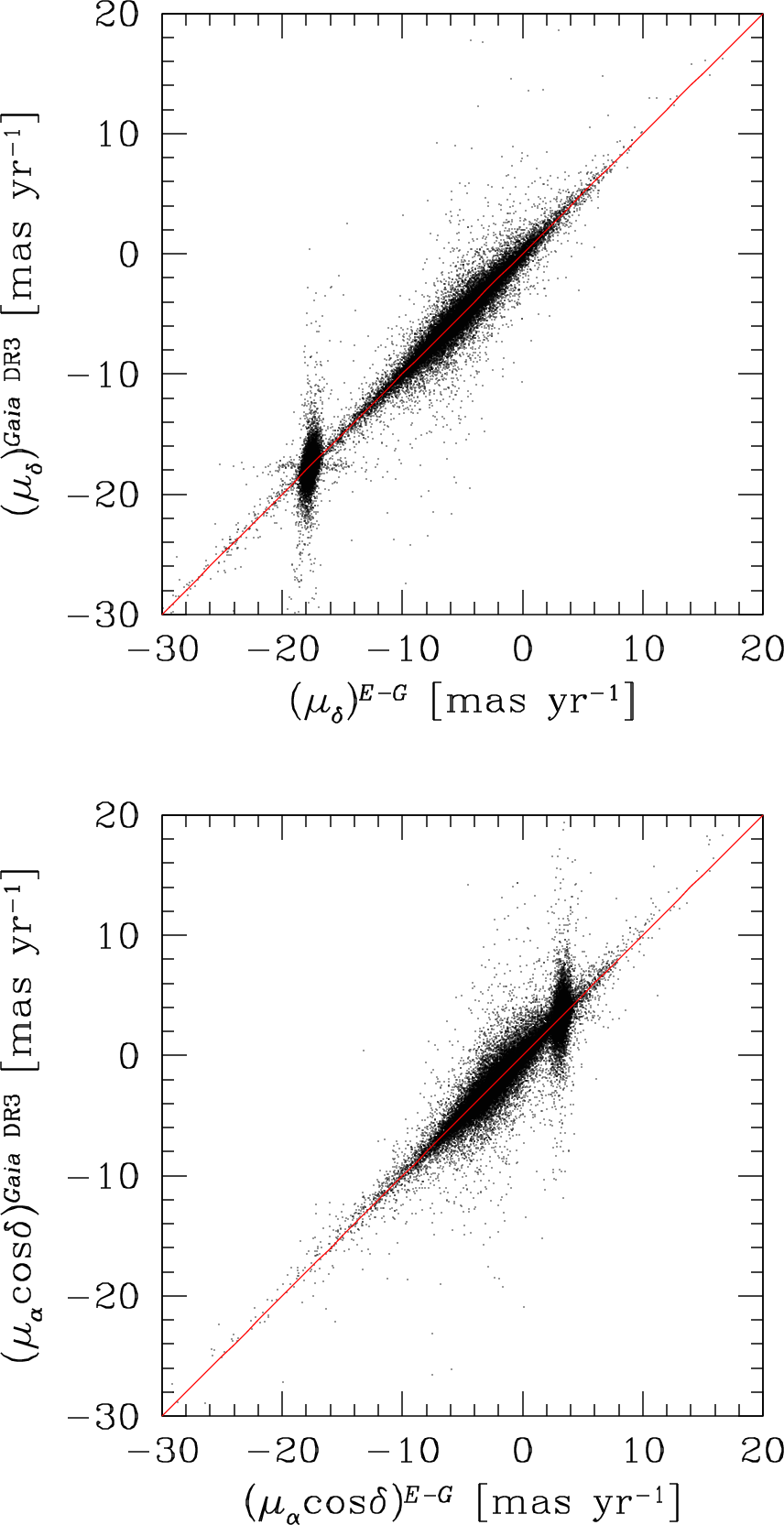}
    \caption{Comparison between the two components of the PMs in the \gaia DR3 and our \egpm catalogue. All unsaturated stars in common with a PM measurement are shown. The red line in each panel is the fit to the points (and not the bisector of the plot) and highlights the nice agreement between the two sets of PMs. The broad distribution along the $y$ axis is a proxy of the systematics errors in the \gaia DR3 PMs for the faint stars in common with the \euclid data.}
    \label{fig:pmcomp}
\end{figure}

\subsection{Comparison with the official ERO catalogue}\label{sec:astrocomp}

We compared our VIS astrometry and that of the \euclid ERO catalogue of NGC~6397 used in \citet{EROGalGCs}\footnote{Available at \href{https://euclid.esac.esa.int/dr/ero/ERO-NGC6397}{this \euclid ERO page}.}. The official ERO catalogues provide astrometry and photometry with multiple techniques. Here we used that from PSF fit for a fair comparison \citep[see][for a description of the PSF extraction and fit]{EROData}. Specifically, we took the equatorial coordinates obtained from the PSF fit (columns ALPHAPSF\_J2000 and DELTAPSF\_J2000), projected them onto the same tangent plane of our VIS catalogue (i.e. using as tangent point the centre of NGC~6397), and transformed these positions onto a tangent plane (in units of pixels) in the same way as we described for the \gaia catalogue in the previous sections. This step is necessary to avoid including projection effects in the comparison.

We cross-matched both catalogues with the \gaia DR3 catalogue (with positions that were PM-propagated to the epoch of the ERO observations) and considered only bright, well-measured stars in common between all three catalogues. We transformed the positions as measured in our or ERO catalogues on to the reference frame of the \gaia DR3 catalogue using six-parameter linear transformations and computed the difference between the two sets of positions. Figure~\ref{fig:astrocomp} shows the residuals along the $x$ or $y$ axis as a function of $x$ or $y$ positions for our catalogues (two panels from the top) and the ERO (two panels from the bottom) catalogues. The comparison shows that, although good, there are still some systematic trends at the 0.2-pixel (20-mas) level visible in the official ERO catalogue, while our astrometry provides an astrometrically flat reference frame comparable to that of \gaia. This comparison highlights once again the astrometric potential of the \euclid data with an optimised data reduction.

\section{Testing the astrometry: Proper motions from \gaia DR3 positions}\label{sec:pms}

We tested our astrometry and computed PMs for all stars in common between our \euclid and the \gaia DR3 catalogues as described in \citet{2021Libralato2DkinGC} and \citet{2024GriggioVVV} -- see also \citet{2022delPinoGaiaHub} for a similar application with \hst. In this test, we used only the VIS exposures since they provide the best astrometric precision (see Sect.~\ref{sec:precision}). 

First, we set up an absolute reference system using the \gaia catalogue as described in the previous section that is by projecting the \gaia positions propagated using PMs, but not parallaxes, to the epoch of the \euclid ERO on to a tangent plane centred on the centre of NGC~6397, fixing the pixel scale (100\,mas pixel$^{-1}$) and the orientation. Then, for each star in the meta catalogues, we used six-parameter linear transformations to transform its position onto the \gaia-based frame. We also applied a local offset to fine-tune the transformed positions and mitigate the effects of residual uncorrected GD or ePSF-related systematics.\footnote{This is the so-called `boresight' correction described in \citet{2010JvdMwCen} and it is defined as the average of the positional residuals between the transformed \euclid positions and the \gaia positions of the closest 25 stars to the target.} Finally, the \euclid position of each star was computed as the sigma-clipped average of the transformed positions of the four meta catalogues at our disposal. PMs were then defined as the difference between these \euclid and the \gaia DR3 (at epoch 2016.0) positions divided by the temporal baseline (7.724 yr) and multiplied by the pixel scale:
\begin{equation}
    \left\{
    \begin{array}{c}
        \mu_\alpha \cos\delta \, [{\rm mas \, yr^{-1}}] = -\paren{x^{Euclid, 2023.7} - x^{\rm Gaia, 2016.0}} \, [\rm VIS \, pixel] \\ \times \, \frac{100 \, [{\rm \,mas\,pixel^{-1}}]}{7.724 \, [{\rm yr}]} \, ,\\
        \mu_\delta \, [{\rm mas \, yr^{-1}}] = \paren{y^{Euclid, 2023.7} - y^{\rm Gaia, 2016.0}} \, [\rm VIS \, pixel] \\ \times \, \frac{100 \, [{\rm \,mas\,pixel^{-1}}]}{7.724 \, [{\rm yr}]} \, .
    \end{array}
    \right.
\end{equation}
We note that, by construction, our master frame is oriented with the $x$ axis increasing towards the west, thus the minus sign in the equation of $\mu_\alpha \cos\delta$.
The PM errors were computed as the sum in quadrature of the \euclid VIS and \gaia DR3 positional errors, again divided by the temporal baseline and multiplied by the pixel scale.

\begin{figure*}[th!]
    \centering
    \includegraphics[width=\textwidth]{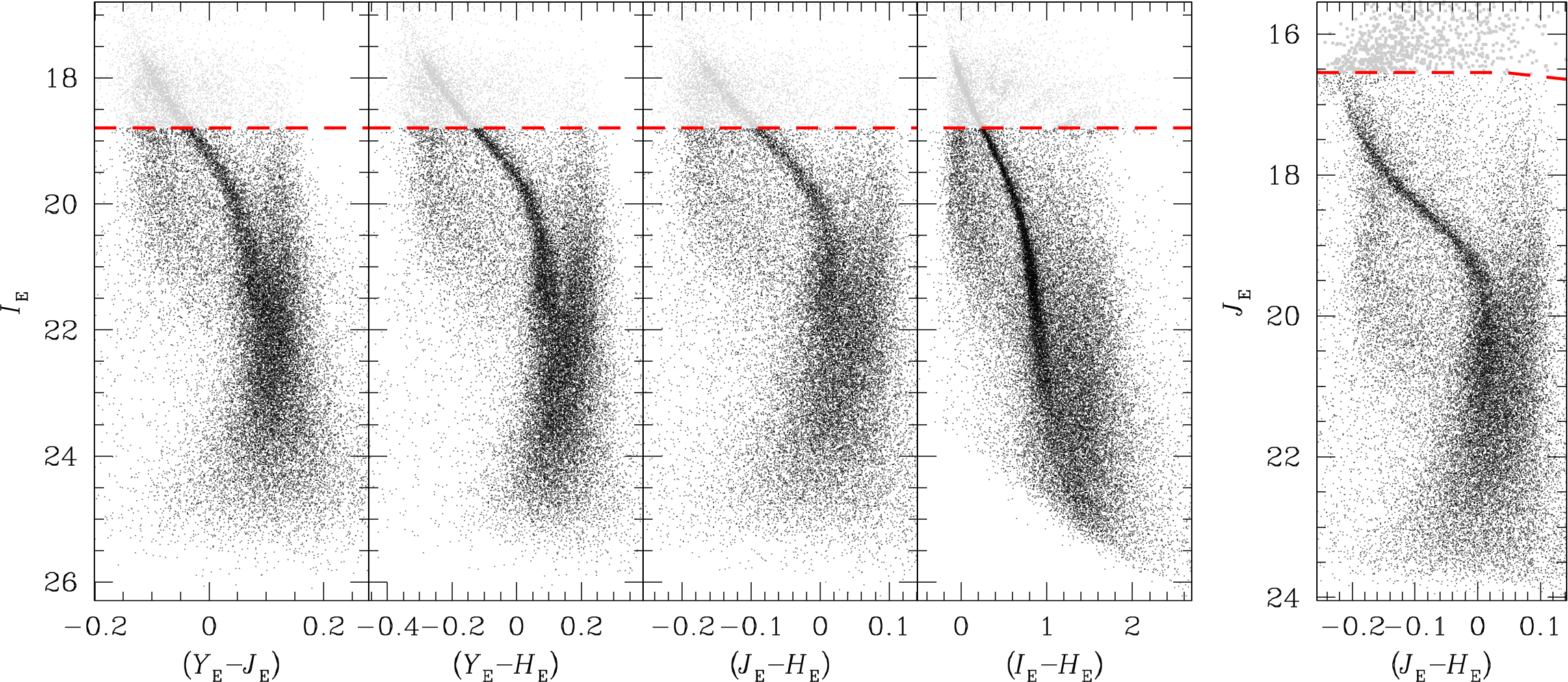}
    \caption{Collection of CMDs made using our VIS and NISP catalogues. The red dashed lines mark the saturation threshold in each plot. Unsaturated objects are shown in black, while saturated sources are depicted in grey. No quality selections are applied. Only 20\% of the points are shown for clarity.}
    \label{fig:EuclidCMDs}
\end{figure*}

\begin{figure*}[t!]
    \centering
    \includegraphics[width=\textwidth]{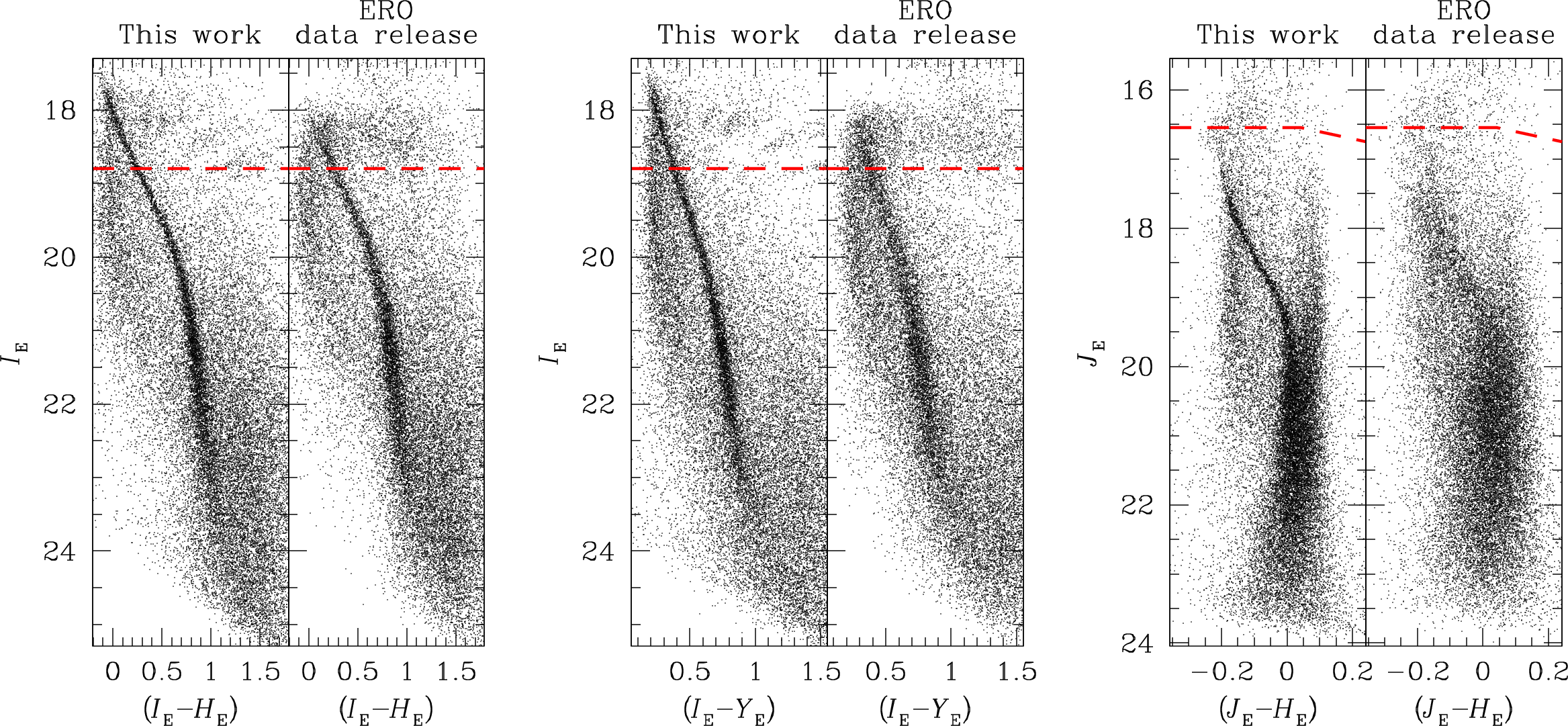}
    \caption{Comparison between three sets of CMDs made with the photometry in our work and in the official ERO data release. For clarity, only 20\% of the stars (randomly selected) among those in common between the two catalogues are shown.} 
    \label{fig:EuclidCMDcomp}
\end{figure*}

Figure~\ref{fig:EuclidVsGaiaDR3} compares the two PM catalogues. The top panels illustrate the vector-point diagrams (VPDs) for the two catalogues. We note that our computed PMs are in the absolute ICRS reference system of \gaia by design. The sources with no PM in \gaia DR3 but with an estimate of the position in both \gaia DR3 and \euclid images are plotted in blue in the rightmost VPD. The tight clump of sources at the bottom left of each VPD is made up by members of NGC~6397, while the other much broader distribution is composed of the Galactic field objects, mainly Bulge stars. The visual comparison between the two VPDs shows again the high precision reached with our \euclid-\gaia (hereafter \egpm) PMs. Indeed, the dispersion of the PMs among cluster members is much tighter using our PMs than with those of the \gaia DR3 catalogue. The broader and elliptical distribution of the cluster stars in the \gaia-only VPD is related to systematic errors in the \gaia DR3 PMs at the faint-end of the catalogue ($18 < G < 21$). Indeed, restricting the \gaia DR3 sample to stars with $G < 19$ and PM error better than 0.25 mas yr$^{-1}$ makes the distribution of NGC~6397 in the \gaia DR3 VPD tighter and comparable to that of our \egpm PMs.

The two bottom panels show the PM errors as a function of $G$ magnitude. For reference, we also report on the top $x$ axes the corresponding VIS $\IE$ magnitude levels. We note that this is a crude transformation (i.e. a zero point with no colour terms considered) and is just meant as a zero-order reference. Two features are immediately clear in these panels:
\begin{enumerate} 
    \item the PM errors of unsaturated stars (\IE $\gtrsim 18.8$) in the \egpm catalogue are about 10 times smaller than those in \gaia DR3;
    \item the stars without a PM estimate in the \gaia DR3 catalogue (i.e. sources with only a 2-parameter astrometric solution) now have a PM measurement and it is much better than that of sources of similar magnitudes in \gaia.    
\end{enumerate}
There are approximately 13300 \gaia sources without a PM in the DR3 catalogue within the \euclid ERO field studied here that now have an \egpm PM (blue points); 11000 of them with $G>20$ and 3000 with $G>21$. For reference, in this ERO field there are about 27\,000 objects with a PM measurement in the \gaia DR3 catalogue with $G>20$ and only five with $G>21$. The large dispersion in the distribution of the PM errors of bright sources for the \egpm sample (red points) is due to the saturation of the VIS exposures (\IE $\lesssim 18.8$) and poorly measured (i.e. large \qfit values) objects. In the following, unless declared otherwise, we excluded these objects from all our analyses.

\begin{figure*}[t!]
    \sidecaption
    \centering
    \includegraphics[width=12.975cm]{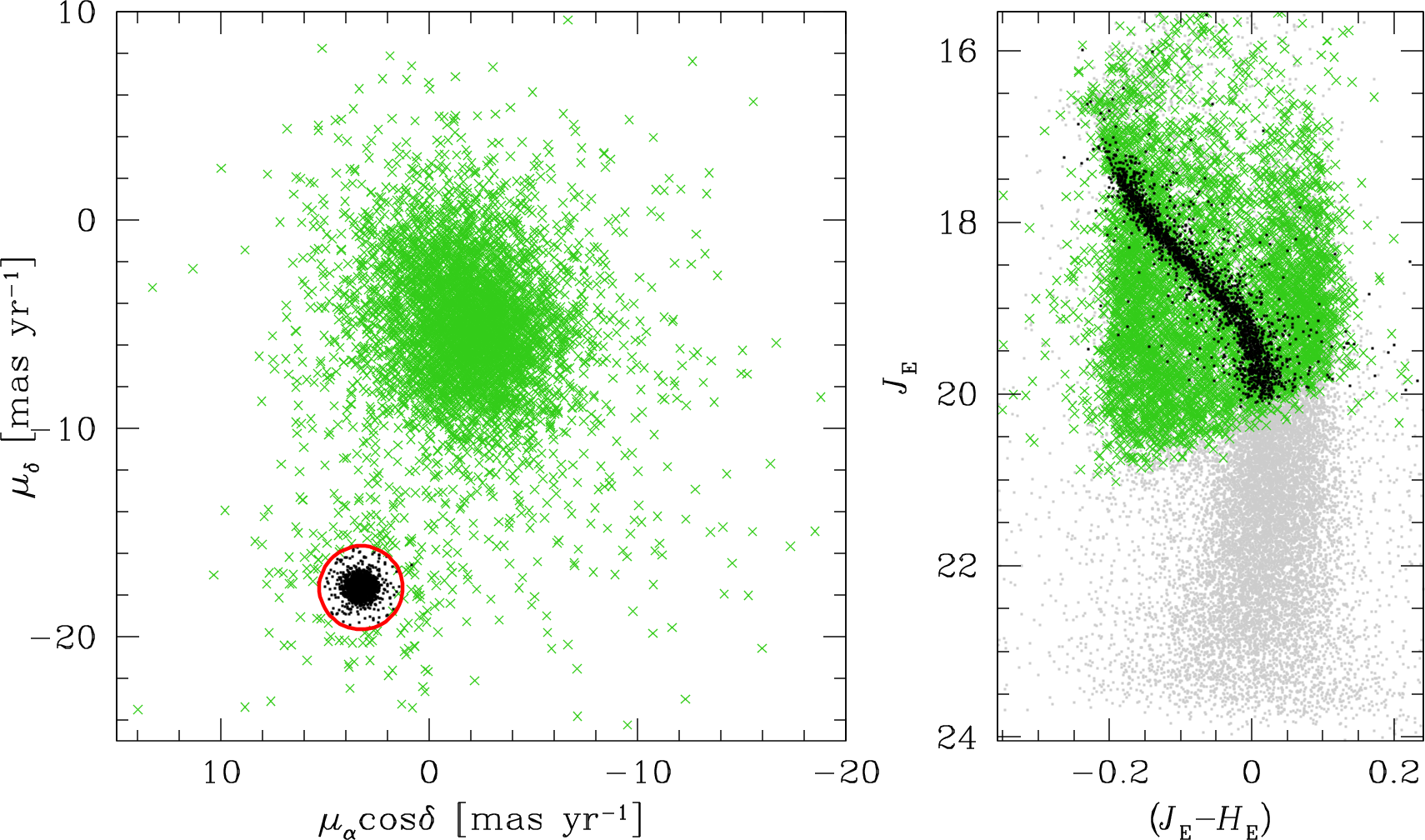}
    \caption{Overview of the VPD and CMD of stars with \egpm PMs. In the VPD on the left, we selected members of NGC~6397 (black points) as those objects having a PM within 2 mas yr$^{-1}$ of the bulk motion of the cluster (red circle). All other stars are considered background or foreground field populations and are shown in green. All objects, including those saturated, are shown for completeness. The two groups of stars are then plotted in $\JE$ as a function of ($\JE$$-$$\HE$) on the right. Our PMs are effective in discriminating between the various populations in the field. As a reference, we show in grey all other stars in the catalogue that do not have a PM measurement because they are fainter than the \gaia limit. The main limitation of our investigation is the shallow first-epoch \gaia catalogue used to compute PMs. Only 10\% of points are shown for clarity.}
    \label{fig:EuclidCMDVPD}
\end{figure*}

Figure~\ref{fig:pmcomp} presents a direct comparison between the \egpm and \gaia DR3 PMs. The red line in each panel is the fit to the point and not the plot bisector. The agreement between the two sets of PMs is clear. For cluster stars (those with the tighter distribution), the dispersion along the $y$ axis is larger than along the $x$ axis. This is an indicator of systematic uncertainties affecting the \gaia DR3 PMs, likely related to the inhomogeneous scanning pattern of \gaia \citep[clearly shown as an `X' shape pattern in the VPDs in Fig.~\ref{fig:EuclidVsGaiaDR3}; see, for example,][]{2018BianchiniRot} and partially to the crowding towards the centremost region of NGC~6397. Appendix~\ref{appendix:pmcomp} discusses a more detailed comparison between the \egpm and \gaia DR3 PMs. We found a clear systematic difference as a function of colour that we empirically corrected (see Appendix~\ref{appendix:pmcomp}). The VIS filter is very wide and as a consequence the PSFs of blue and red sources can be different. Our ePSF models were obtained using stars in a specific magnitude range (from saturation to about two magnitudes fainter) regardless of their colour, but the majority of the sources are confined within a specific colour range as shown in Fig.~\ref{fig:EuclidCMDVPD}. Thus, it is not surprising that very blue and red objects would behave differently from the others. Also, depending on the position in the field of view, the typical colour of the stars used in the modelling can be different. For example, towards the centremost regions the stars belonging to NGC~6397 dominate the sample, whereas at the edges of the field there is a mix of cluster and field stars. All these factors can contribute to creating colour-dependent systematic effects. Future studies will focus on the chromatic dependency of the ePSF models. The scientific application in Sect.~\ref{sec:kin} makes use of these corrected PMs.

\section{Testing the photometry: VIS and NISP colour-magnitude diagrams}\label{sec:cmd}

We cross-identified the same stars in all our ePSF-based VIS and NISP master catalogues. Only stars measured in at least three images per filter were considered. Our photometry was registered on to the official \euclid system by computing the zero-point difference with the corresponding photometry in the \euclid ERO data release. As in Sect.~\ref{sec:astrocomp}, we considered the PSF-based photometry in the official ERO catalogue (column MAG\_PSF).

Figure~\ref{fig:EuclidCMDs} presents a collection of colour-magnitude diagrams (CMDs) made using various combinations of VIS and NISP filters. Black and grey points represent unsaturated and saturated stars, respectively. The main sequence of NGC~6397 is clearly visible in each CMD. The rightmost $\JE$ versus ($\JE$$-$$\HE$) CMD shows a pure near-infrared diagram where it is even possible to distinguish features such as the kink of the main sequence at about (19.5,0.0). The kink is visible in near-infrared CMDs and arises from the effect of the collision-induced absorption of hydrogen molecules on the surfaces of M dwarfs (of around 0.5 M$_\odot$), which changes their opacity and redistributes the stellar flux longwards of 2\,$\mu$m to shorter wavelengths \citep{Linsky1969,Saumon1994}. The kink is an important point along the main sequence used to derive accurate ages for globular clusters \citep[e.g.][]{Correnti2018,Saracino2018}. Figure~\ref{fig:EuclidCMDcomp} highlights the comparison between two sets of CMDs made with the official ERO catalogues and ours. Only 20\% of the stars in common are shown for clarity. Our ePSF-based photometry provides remarkably better defined sequences on CMDs, especially with the NISP instrument, and is a demonstration of the accuracy of our ePSFs.

\begin{figure*}[th!]
    \centering
    \includegraphics[width=\textwidth]{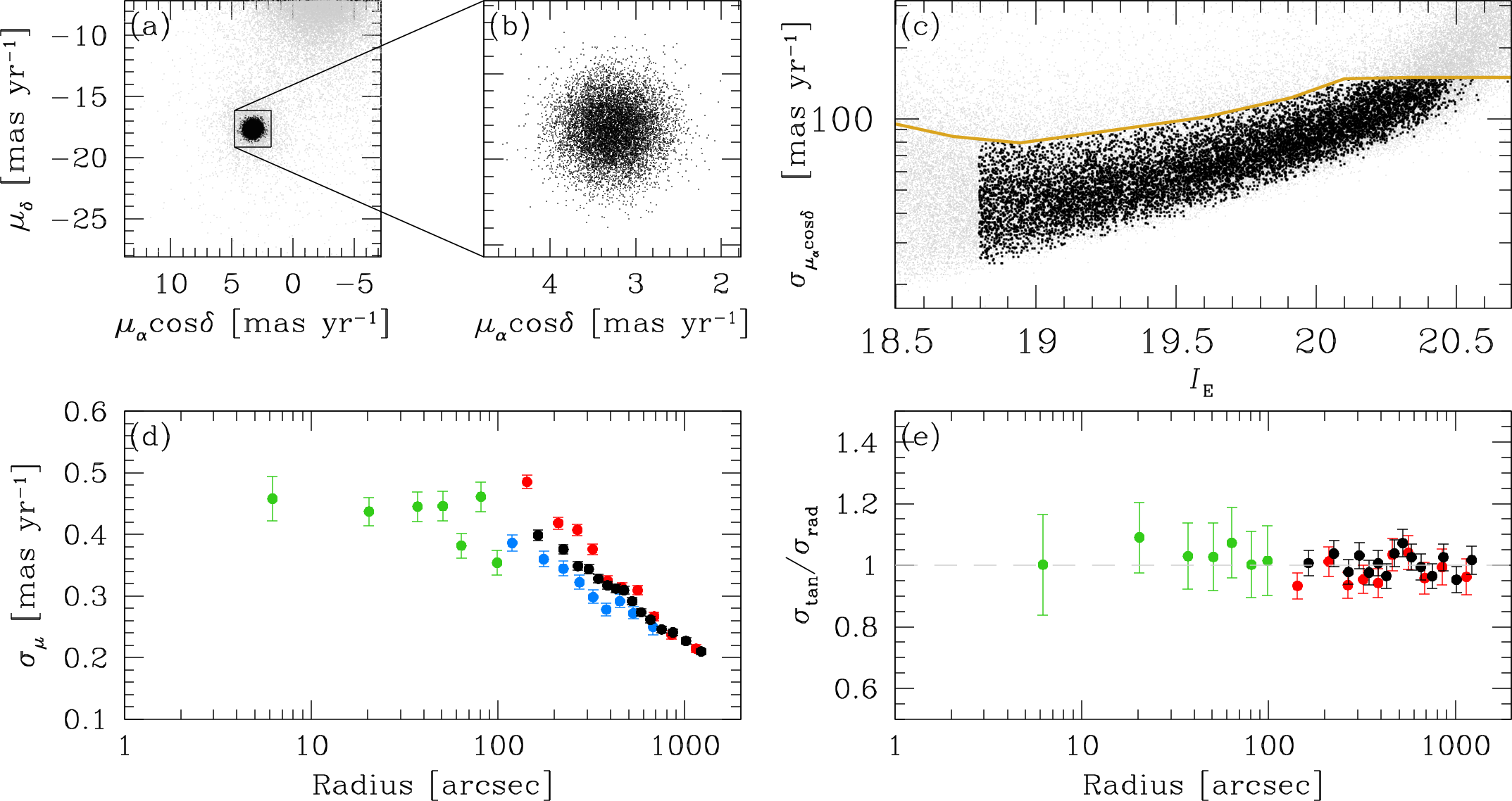}
    \caption{Velocity-dispersion and anisotropy radial profiles for NGC~6397. Panel (a) shows the VPD of the \egpm PMs. Black dots are the cluster members that survived all our selection criteria and were used to compute the velocity dispersion of NGC~6397. All other stars are plotted as grey points. Panel (b) presents a zoom-in around the cluster distribution in the VPD (only black points are included for clarity). The PM error along $\alpha \cos\delta$ as a function of \IE magnitude is shown in panel (c). The gold line represents the PM threshold used to select the best stars for the analysis (the same selection was applied to the PM error along $\delta$; see the text for details). The combined velocity dispersion $\sigma_\mu$ as function of distance from the centre of the cluster (in arcsec) is plotted in the panel (d). Black, filled points are obtained by means of the \egpm PMs; green points are from \citet{2022LibralatoPMcat}; blue points refer to the \gaia-EDR3 measurements of \citet{2021VasilievGCkin}; and the red points represent our cross-check with the \gaia DR3 catalogue. Finally, panel (e) shows the anisotropy ($\sigma_{\rm tan}/\sigma_{\rm rad}$) radial profile. Points are colour-coded as in the previous panel \citep[no anisotropy measurements are provided by][]{2021VasilievGCkin}. The grey, dashed line is set to 1 (i.e. isotropic case) as a reference.}
    \label{fig:kin}
\end{figure*}

Finally, Fig.~\ref{fig:EuclidCMDVPD} shows another example of the potential applications enabled by our astrometry and photometry. We used our \egpm PMs to select members of NGC~6397 and foreground or background stars (all objects, including those saturated, are shown for completeness). Then, we highlighted these two groups of stars in a NISP-based CMD, showing that these groups have different loci in the diagram. As a reference, we plot all other stars in the photometric catalogue that do not have a PM measurement. Using a deeper first-epoch catalogue other than \gaia, we could discriminate the various populations in the CMD according to their kinematics even at fainter magnitudes.

\section{Scientific application: The absolute PM and the internal kinematics of NGC~6397}\label{sec:kin}

The \egpm PMs offer us the opportunity of investigating the absolute PM of NGC~6397 and its internal kinematics out to about a half of its tidal radius ($r_{\rm t} = 52.44$ arcmin; from the GC online database of Holger Baumgardt). This is a pilot project aimed at demonstrating the astrometric and photometric capabilities of \euclid rather than an all-around scientific investigation of NGC~6397, so we simply present here a comparison with literature values.

First, we defined a sample of well-measured, unsaturated (within two magnitudes from saturation in the VIS catalogue) cluster stars with a PM error lower than the minimum between two numbers: a magnitude-dependent threshold set by hand (that increases towards fainter magnitudes) and 0.15 mas yr$^{-1}$. We considered only unsaturated stars that were measured in at least three VIS images and that were corrected for the colour-dependent systematic described in Appendix~\ref{appendix:pmcomp}. Stars included and rejected in our analysis are shown in panels (a), (b), and (c) of Fig.~\ref{fig:kin}.

The absolute PM of NGC~6397, defined as the 3$\sigma$-clipped median PM of well-measured cluster members, is
\begin{equation}
  (\mu_\alpha \cos\delta,\mu_\delta)_{E-G} = (3.271 \pm 0.003,-17.644 \pm 0.003) \phantom{1} \textrm{mas yr}^{-1} \phantom{1} .
\end{equation}
The error bars include only the statistical errors and not systematic ones, which can be an order of magnitude larger. Our estimate of the absolute PM of NGC~6397 is in agreement with that from \gaia DR3 provided in the Baumgardt GC database, $(\mu_\alpha \cos\delta,\mu_\delta)$$=$$(3.251\pm0.005,-17.649\pm0.005)$ mas yr$^{-1}$, at the $\sim$3$\sigma$ level.\looseness=-2

Then, we computed the combined ($\sigma_\mu$), tangential ($\sigma_{\rm tan}$) and radial ($\sigma_{\rm rad}$) velocity dispersions as a function of distance from the centre of the cluster in 15 equally populated bins (663 stars per bin) using a maximum-likelihood approach, as described in section 4 of \citet{2022LibralatoPMcat}, to which we refer the reader for a description of the methodology and the definition of the quantities derived. The values of $\sigma_{\rm tan}$ and $\sigma_{\rm rad}$ were obtained from the PMs along the tangential and radial directions, respectively, with respect to the centre of NGC~6397 in the plane of the sky. The value of $\sigma_\mu$ was obtained by assuming $\sigma_\mu = \sigma_{\rm tan} = \sigma_{\rm rad}$ in the computation \citep{2022LibralatoPMcat}. The result is presented in Fig.~\ref{fig:kin}. Measurements based on the \egpm PMs are shown in black. Blue and green points are taken from \citet[with \gaia DR3]{2021VasilievGCkin} and \citet[with \hst]{2022LibralatoPMcat}, respectively. Our profile suggests that stars are kinematically hotter than what is predicted by the \gaia DR3 PMs. The stars in our sample are main-sequence objects with $G>19$, while those used in the other studies in the literature are brighter and more massive. Thus, the different kinematics could be related to energy equipartition in the cluster. Alternatively, our PM errors could be underestimated, which is not an unreasonable option, given the data set at our disposal. As a cross-check, we independently computed with our tools\footnote{ \citet{2022LibralatoPMcat} has already shown the good agreement between the velocity dispersions computed with the same tools used in our paper and the velocity dispersions in the literature obtained with different tools by \citet{2021VasilievGCkin} and \citet{2015ApJ...803...29W}.} the velocity dispersion of NGC~6397 using the \gaia DR3 PMs in a mass range similar to that of the \egpm sample. However, we could not select the exact same stars as in the \egpm sample because the \gaia DR3 PM errors are too large at $G>19$ for meaningful measurements. The \gaia-based profile is shown in red. Despite the larger error bars, there is a good agreement with our \egpm profile.

Panel (e) of Fig.~\ref{fig:kin} presents the anisotropy profile, defined as the ratio between the tangential and radial components of the velocity dispersion as a function of clustercentric distance. The plot shows that NGC~6397 is isotropic within our FoV. The result is in agreement with what was found by \citet{2021VasilievGCkin}.

\section{Conclusions and future plans}\label{S:CaF}

This work describes the development of the procedures to obtain `state-of-the art' imaging astrometry and photometry from undersampled images collected with the instruments at the focus of the space-based diffraction-limited \euclid telescope. These techniques involve an iterative process aimed at simultaneously solving for: 
(1) the exact shape of the effective PSFs \citep[see details of this concept in][]{2000AKWFPC2};
(2) the position of the sources; and 
(3) the GD correction of the instruments.  
These procedures, applied to ground- and space-based data, have shown in the literature to produce significantly better results for both astrometry and photometry when compared with standard techniques, particularly in crowded environments.

This paper applies these techniques, for the first time, to the \euclid instruments using the ERO data of NGC~6397. Compared to the procedures used in the production of the catalogues of the ERO release, major changes include a chip-by-chip determination of the astrometry using re-evaluated PMs, the computation of photometry in individual unresampled images rather than stacks, optimised PSFs. A comparison with \Euclid Wide Survey pipeline has not been undertaken, as those procedures were developed for less crowded cosmological fields and were not applied to the ERO images. The preliminary results we present, although derived from a limited sample of \euclid ERO images, clearly shows the impact they will have to a broad scientific community. Accurate GD corrections have immediate relevance for optimal stacking of individual images and ePSFs can have a role in improving the main planned science with \euclid, for example, an accurate modelling of the ePSF core would be beneficial for weak lensing investigations. Future efforts will analyse the temporal stability of our ePSFs and GD corrections and, most importantly, will study the dependence of these products with the spectral-energy distribution of sources. These effects can have a non-negligible impact on astrometry and photometry with the relatively broad-band filters of the \euclid instruments, in particular VIS (see Appendix~\ref{appendix:pmcomp}).

We also demonstrated the great potential of the synergy between \euclid and \gaia. For all unsaturated sources in \euclid that are in common with \gaia, we were able to compute PMs with a precision by as much as a factor of 10$\times$ better than in the \gaia DR3 catalogue. In addition, for sources in the \gaia DR3 catalogue with only a 2-parameter astrometric solution (i.e. with positions but not PMs), we were able to obtain a PM estimate, thus de facto extending by about 0.5 mag the \gaia DR3 catalogue. But the astrometric capability of \euclid is much more than this.

\begin{figure}[t!]
    \centering
    \includegraphics[width=\columnwidth]{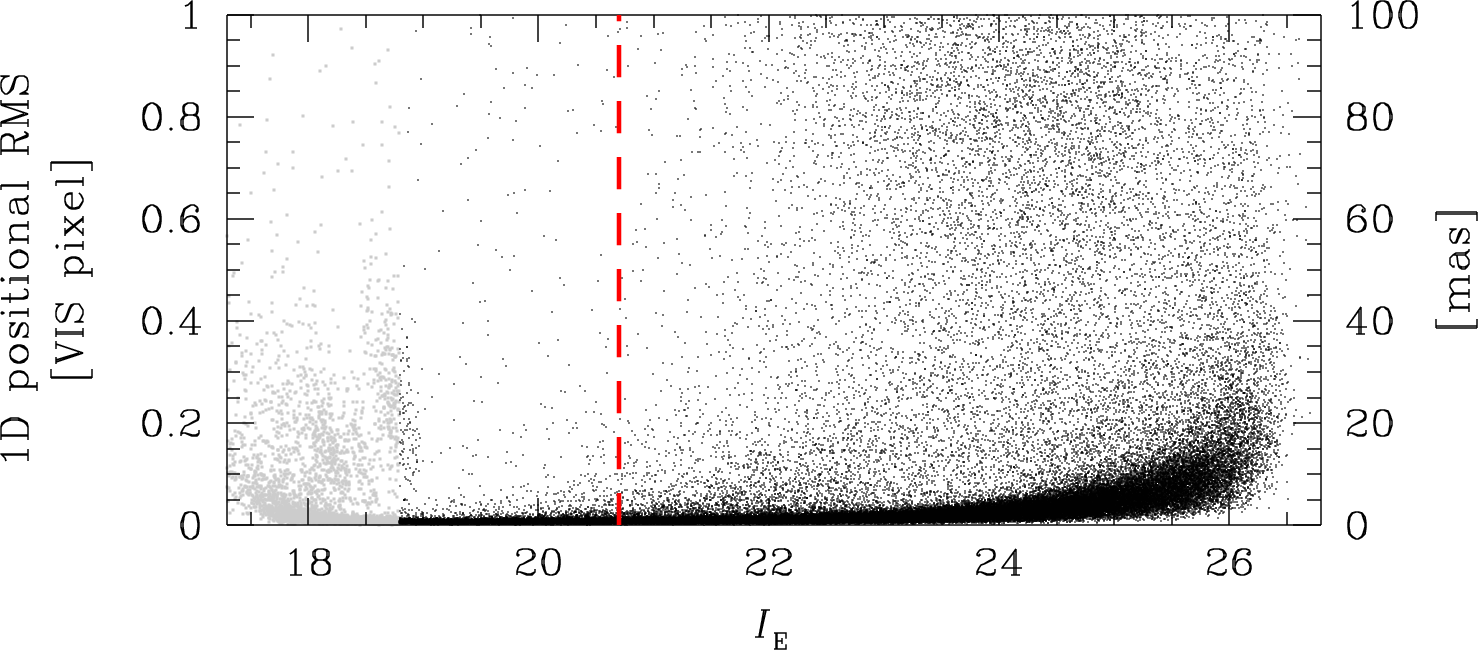}
    \caption{One dimensional (1D) positional RMS as a function of $\IE$ magnitude. Grey points represent saturated stars in the VIS catalogue; all other unsaturated sources are plotted in black. The red, dashed vertical line indicates the \gaia faint limit. Only 20\% of the points are shown for clarity.}
    \label{fig:rmsall}
\end{figure}

Figure~\ref{fig:rmsall} shows the 1D positional RMS as function of $\IE$ magnitude for all sources in our VIS catalogue and it is an extended version of what is presented in Fig.\,\ref{fig:1drms}. \euclid could easily reach positional astrometry with accuracy better than 10 mas down to magnitudes as faint as $\IE \simeq 26$ (about $V \simeq 27$ or $G \simeq 27$ for a main-sequence star). This means that for sources within the planned calibration fields, which will be re-observed a few times over the 6-year duration of the mission, if we assume the same number of images in the ERO data of NGC~6397 (four) per epoch, we could compute \euclid-only PMs with a precision of around 0.1 mas yr$^{-1}$ for bright stars close to saturation and 1.5 mas yr$^{-1}$ at $\IE \simeq 26$. Thus, \euclid can potentially extend the \gaia astrometry in the ICRS absolute reference system up to 6 magnitudes fainter than the \gaia limit. Such an extension would provide a significantly larger number of stars to use as astrometric references. For example, in the deep ERO images of NGC\,6397 analysed here it was possible to recover in the same field over 474\,000 sources, compared to the roughly 100\,000 in the \gaia DR3 catalogue.

As a scientific benchmark, we used the \egpm PMs to study the internal motions of the globular cluster NGC~6397 out to 0.5 $r_{\rm t}$ from the centre of the cluster. This is just one possible application and there can be many others. For example, a \euclid-based astrometric catalogue can be extremely useful for the registration of images onto the ICRS when the number of \gaia stars in the field is too low or the only visible objects are too faint for \gaia \citep[e.g.][Libralato et al., in prep.]{2024AN....34530158B}. To facilitate other applications, we make the astrometric and photometric catalogues and high-resolution stacked images available to the community.

We note that this is just a preliminary demonstration of an alternative reduction of the \euclid data aimed at maximising its astrometric and photometric capabilities, and additional improvements can be implemented. In particular, the photometry presented here is the result of a single pass of finding where stars are measured ignoring the close-by neighbours \citep{2022acs..rept....2A,2022wfc..rept....5A}. We plan to employ the so-called `second-pass' photometry (\citealt{2008AJ....135.2114A}, in prep.; \citealt{2017BelliniwCenI}, \citealt{2018MNRAS.481.3382N}, \citealt{2018LibralatoNGC362}), which has already been successfully exported to other wide-field instruments \citep[see, e.g.][]{2023GriggioCFHT,2024GriggioVVV}. Second-pass photometry uses all images at once to detect sources that are otherwise too faint to be detected in a single exposure, and ePSF-subtracts all neighbours prior to the final ePSF fit of every object. This could allow \euclid to gain in depth (the exact gain will depend on the number of overlapping images covering a given patch of the sky) and improve the quality of the catalogues in crowded regions.

Finally, the current photometry still suffers from uncorrected systematic errors due to the pixel-area correction (related to the fact that due to GD, pixels subtend different areas on the sky, and the flat fields are constructed to preserve surface brightness not flux), along with other systematic effects that can result in local photometric zero-point variations across the field of view. In future efforts, we will also address these issues and provide more accurate photometry.

The very wide field of view, high resolution, and PSF quality of \euclid make this telescope a one-of-a-kind resource for astrometry. The combination of \euclid and \gaia offers the opportunity of extending \gaia-like positions (and PMs) to faint stars outside \gaia's reach. This is beneficial not only for scientific topics such as those discussed above but also for the astrometric calibration and registration of data from future surveys such as those carried out by the \textit{Nancy Grace Roman} Space Telescope \citep{2019WFIRSTAWG}.

\section*{Data availability}As part of this publication, we release the star catalogues (positions, photometry and proper motions) and the astrometrized stacked images through the CDS and our website (\href{https://web.oapd.inaf.it/bedin/files/PAPERs_eMATERIALs/Euclid/Paper01/}{https://web.oapd.inaf.it/bedin/files/PAPERs\_eMATERIALs/\\Euclid/Paper01/}). Astro-photometric catalogues and stacked images are available at the CDS via anonymous ftp to to \href{cdsarc.u-strasbg.fr}{cdsarc.u-strasbg.fr} (130.79.128.5) or via \href{http://cdsweb.u-strasbg.fr/cgi-bin/qcat?J/A+A/}{http://cdsweb.u-strasbg.fr/cgi-bin/qcat?J/A$+$A/}.

\begin{acknowledgement}

\AckERO \AckEC This work has made use of data from the European Space Agency (ESA) mission {\it Gaia} (\url{https://www.cosmos.esa.int/gaia}), processed by the {\it Gaia} Data Processing and Analysis Consortium (DPAC, \url{https://www.cosmos.esa.int/web/gaia/dpac/consortium}). Funding for the DPAC has been provided by national institutions, in particular the institutions participating in the {\it Gaia} Multilateral Agreement. This research made use of \texttt{astropy}, a community-developed core \texttt{python} package for Astronomy \citep{astropy:2013, astropy:2018} and of \texttt{emcee} \citep{2013PASP..125..306F}.

\end{acknowledgement}

\bibliography{Euclid_Gaia_Paper_I_arXiv}{}

\begin{thebibliography}{56}
\expandafter\ifx\csname natexlab\endcsname\relax\def\natexlab#1{#1}\fi

\bibitem[{{Anderson}(2022{\natexlab{a}})}]{2022acs..rept....2A}
{Anderson}, J. 2022{\natexlab{a}}, {One-Pass HST Photometry with hst1pass},
  Instrument Science Report ACS 2022-02

\bibitem[{{Anderson}(2022{\natexlab{b}})}]{2022wfc..rept....5A}
{Anderson}, J. 2022{\natexlab{b}}, {One-Pass HST Photometry with hst1pass},
  Instrument Science Report WFC3 2022-5

\bibitem[{{Anderson} {et~al.}(2006){Anderson}, {Bedin}, {Piotto}, {Yadav}, \&
  {Bellini}}]{2006AndersonWFI}
{Anderson}, J., {Bedin}, L.~R., {Piotto}, G., {Yadav}, R.~S., \& {Bellini}, A.
  2006, \aap, 454, 1029

\bibitem[{{Anderson} \& {King}(2000)}]{2000AKWFPC2}
{Anderson}, J. \& {King}, I.~R. 2000, \pasp, 112, 1360

\bibitem[{{Anderson} \& {King}(2003)}]{2003PASP..115..113A}
{Anderson}, J. \& {King}, I.~R. 2003, \pasp, 115, 113

\bibitem[{{Anderson} {et~al.}(2008){Anderson}, {King}, {Richer}, {Fahlman},
  {Hansen}, {Hurley}, {Kalirai}, {Rich}, \& {Stetson}}]{2008AJ....135.2114A}
{Anderson}, J., {King}, I.~R., {Richer}, H.~B., {et~al.} 2008, \aj, 135, 2114

\bibitem[{{Anderson} \& {van der Marel}(2010)}]{2010JvdMwCen}
{Anderson}, J. \& {van der Marel}, R.~P. 2010, \apj, 710, 1032

\bibitem[{{Astropy Collaboration} {et~al.}(2018){Astropy Collaboration},
  {Price-Whelan}, {Sip{\H{o}}cz}, {G{\"u}nther}, {Lim}, {Crawford}, {Conseil},
  {Shupe}, {Craig}, {Dencheva}, {Ginsburg}, {Vand erPlas}, {Bradley},
  {P{\'e}rez-Su{\'a}rez}, {de Val-Borro}, {Aldcroft}, {Cruz}, {Robitaille},
  {Tollerud}, {Ardelean}, {Babej}, {Bach}, {Bachetti}, {Bakanov}, {Bamford},
  {Barentsen}, {Barmby}, {Baumbach}, {Berry}, {Biscani}, {Boquien}, {Bostroem},
  {Bouma}, {Brammer}, {Bray}, {Breytenbach}, {Buddelmeijer}, {Burke},
  {Calderone}, {Cano Rodr{\'\i}guez}, {Cara}, {Cardoso}, {Cheedella}, {Copin},
  {Corrales}, {Crichton}, {D'Avella}, {Deil}, {Depagne}, {Dietrich}, {Donath},
  {Droettboom}, {Earl}, {Erben}, {Fabbro}, {Ferreira}, {Finethy}, {Fox},
  {Garrison}, {Gibbons}, {Goldstein}, {Gommers}, {Greco}, {Greenfield},
  {Groener}, {Grollier}, {Hagen}, {Hirst}, {Homeier}, {Horton}, {Hosseinzadeh},
  {Hu}, {Hunkeler}, {Ivezi{\'c}}, {Jain}, {Jenness}, {Kanarek}, {Kendrew},
  {Kern}, {Kerzendorf}, {Khvalko}, {King}, {Kirkby}, {Kulkarni}, {Kumar},
  {Lee}, {Lenz}, {Littlefair}, {Ma}, {Macleod}, {Mastropietro}, {McCully},
  {Montagnac}, {Morris}, {Mueller}, {Mumford}, {Muna}, {Murphy}, {Nelson},
  {Nguyen}, {Ninan}, {N{\"o}the}, {Ogaz}, {Oh}, {Parejko}, {Parley}, {Pascual},
  {Patil}, {Patil}, {Plunkett}, {Prochaska}, {Rastogi}, {Reddy Janga},
  {Sabater}, {Sakurikar}, {Seifert}, {Sherbert}, {Sherwood-Taylor}, {Shih},
  {Sick}, {Silbiger}, {Singanamalla}, {Singer}, {Sladen}, {Sooley},
  {Sornarajah}, {Streicher}, {Teuben}, {Thomas}, {Tremblay}, {Turner},
  {Terr{\'o}n}, {van Kerkwijk}, {de la Vega}, {Watkins}, {Weaver}, {Whitmore},
  {Woillez}, {Zabalza}, \& {Astropy Contributors}}]{astropy:2018}
{Astropy Collaboration}, {Price-Whelan}, A.~M., {Sip{\H{o}}cz}, B.~M., {et~al.}
  2018, \aj, 156, 123

\bibitem[{{Astropy Collaboration} {et~al.}(2013){Astropy Collaboration},
  {Robitaille}, {Tollerud}, {Greenfield}, {Droettboom}, {Bray}, {Aldcroft},
  {Davis}, {Ginsburg}, {Price-Whelan}, {Kerzendorf}, {Conley}, {Crighton},
  {Barbary}, {Muna}, {Ferguson}, {Grollier}, {Parikh}, {Nair}, {Unther},
  {Deil}, {Woillez}, {Conseil}, {Kramer}, {Turner}, {Singer}, {Fox}, {Weaver},
  {Zabalza}, {Edwards}, {Azalee Bostroem}, {Burke}, {Casey}, {Crawford},
  {Dencheva}, {Ely}, {Jenness}, {Labrie}, {Lim}, {Pierfederici}, {Pontzen},
  {Ptak}, {Refsdal}, {Servillat}, \& {Streicher}}]{astropy:2013}
{Astropy Collaboration}, {Robitaille}, T.~P., {Tollerud}, E.~J., {et~al.} 2013,
  \aap, 558, A33

\bibitem[{{Bedin} {et~al.}(2024){Bedin}, {Dietrich}, {Burgasser}, {Apai},
  {Libralato}, {Griggio}, {Fontanive}, \& {Pourbaix}}]{2024AN....34530158B}
{Bedin}, L.~R., {Dietrich}, J., {Burgasser}, A.~J., {et~al.} 2024,
  Astronomische Nachrichten, 345, e20230158

\bibitem[{{Bellini} {et~al.}(2011){Bellini}, {Anderson}, \&
  {Bedin}}]{2011BelliniWFC3}
{Bellini}, A., {Anderson}, J., \& {Bedin}, L.~R. 2011, \pasp, 123, 622

\bibitem[{{Bellini} {et~al.}(2017){Bellini}, {Anderson}, {Bedin}, {King}, {van
  der Marel}, {Piotto}, \& {Cool}}]{2017BelliniwCenI}
{Bellini}, A., {Anderson}, J., {Bedin}, L.~R., {et~al.} 2017, \apj, 842, 6

\bibitem[{{Bellini} {et~al.}(2014){Bellini}, {Anderson}, {van der Marel},
  {Watkins}, {King}, {Bianchini}, {Chanam{\'e}}, {Chandar}, {Cool}, {Ferraro},
  {Ford}, \& {Massari}}]{2014BelliniPMcat}
{Bellini}, A., {Anderson}, J., {van der Marel}, R.~P., {et~al.} 2014, \apj,
  797, 115

\bibitem[{{Bellini} \& {Bedin}(2009)}]{2009BelliniWFC3}
{Bellini}, A. \& {Bedin}, L.~R. 2009, \pasp, 121, 1419

\bibitem[{{Bellini} \& {Bedin}(2010)}]{2010BelliniLBT}
{Bellini}, A. \& {Bedin}, L.~R. 2010, \aap, 517, A34

\bibitem[{{Bianchini} {et~al.}(2018){Bianchini}, {van der Marel}, {del Pino},
  {Watkins}, {Bellini}, {Fardal}, {Libralato}, \& {Sills}}]{2018BianchiniRot}
{Bianchini}, P., {van der Marel}, R.~P., {del Pino}, A., {et~al.} 2018, \mnras,
  481, 2125

\bibitem[{{Cantat-Gaudin} \& {Casamiquela}(2024)}]{2024NewAR..9901696C}
{Cantat-Gaudin}, T. \& {Casamiquela}, L. 2024, \nar, 99, 101696

\bibitem[{{Correnti} {et~al.}(2018){Correnti}, {Gennaro}, {Kalirai}, {Cohen},
  \& {Brown}}]{Correnti2018}
{Correnti}, M., {Gennaro}, M., {Kalirai}, J.~S., {Cohen}, R.~E., \& {Brown},
  T.~M. 2018, \apj, 864, 147

\bibitem[{{Cuillandre} {et~al.}(2024){Cuillandre}, {Bertin}, {Bolzonella},
  {et~al.}}]{EROData}
{Cuillandre}, J.-C., {Bertin}, E., {Bolzonella}, M., {et~al.} 2024, \aap,
  submitted, arXiv:2405.13496

\bibitem[{{del Pino} {et~al.}(2022){del Pino}, {Libralato}, {van der Marel},
  {Bennet}, {Fardal}, {Anderson}, {Bellini}, {Tony Sohn}, \&
  {Watkins}}]{2022delPinoGaiaHub}
{del Pino}, A., {Libralato}, M., {van der Marel}, R.~P., {et~al.} 2022, \apj,
  933, 76

\bibitem[{{Euclid Collaboration: Cropper} {et~al.}(2024){Euclid Collaboration:
  Cropper}, {Al Bahlawan}, {Amiaux}, {et~al.}}]{EuclidSkyVIS}
{Euclid Collaboration: Cropper}, M., {Al Bahlawan}, A., {Amiaux}, J., {et~al.}
  2024, \aap, submitted, arXiv:2405.13492

\bibitem[{{Euclid Collaboration: Jahnke} {et~al.}(2024){Euclid Collaboration:
  Jahnke}, {Gillard}, {Schirmer}, {et~al.}}]{EuclidSkyNISP}
{Euclid Collaboration: Jahnke}, K., {Gillard}, W., {Schirmer}, M., {et~al.}
  2024, \aap, submitted, arXiv:2405.13493

\bibitem[{{Euclid Collaboration: Mellier} {et~al.}(2024){Euclid Collaboration:
  Mellier}, {Abdurro'uf}, {Acevedo~Barroso}, {Ach\'ucarro},
  {et~al.}}]{EuclidSkyOverview}
{Euclid Collaboration: Mellier}, Y., {Abdurro'uf}, {Acevedo~Barroso}, J.,
  {Ach\'ucarro}, A., {et~al.} 2024, \aap, submitted, arXiv:2405.13491

\bibitem[{{Euclid Collaboration: Schirmer} {et~al.}(2022){Euclid Collaboration:
  Schirmer}, {Jahnke}, {Seidel}, {Aussel}, {Bodendorf}, {Grupp}, {Hormuth},
  {Wachter}, {Appleton}, {Barbier}, {Brinchmann}, {Carrasco}, {Castander},
  {Coupon}, {De Paolis}, {Franco}, {Ganga}, {Hudelot}, {Jullo}, {Lan{\c{c}}on},
  {Nucita}, {Paltani}, {Smadja}, {Strafella}, {Venancio}, {Weiler}, {Amara},
  {Auphan}, {Auricchio}, {Balestra}, {Bender}, {Bonino}, {Branchini},
  {Brescia}, {Capobianco}, {Carbone}, {Carretero}, {Casas}, {Castellano},
  {Cavuoti}, {Cimatti}, {Cledassou}, {Congedo}, {Conselice}, {Conversi},
  {Copin}, {Corcione}, {Costille}, {Courbin}, {Da Silva}, {Degaudenzi},
  {Douspis}, {Dubath}, {Dupac}, {Dusini}, {Ealet}, {Farrens}, {Ferriol},
  {Fosalba}, {Frailis}, {Franceschi}, {Franzetti}, {Fumana}, {Garilli},
  {Gillard}, {Gillis}, {Giocoli}, {Grazian}, {Guzzo}, {Haugan}, {Hoekstra},
  {Holmes}, {Hornstrup}, {K{\"u}mmel}, {Kermiche}, {Kiessling}, {Kilbinger},
  {Kitching}, {Kohley}, {Kunz}, {Kurki-Suonio}, {Laureijs}, {Ligori}, {Lilje},
  {Lloro}, {Maciaszek}, {Maiorano}, {Mansutti}, {Marggraf}, {Markovic},
  {Marulli}, {Massey}, {Maurogordato}, {Mellier}, {Meneghetti}, {Merlin},
  {Meylan}, {Moresco}, {Moscardini}, {Munari}, {Nakajima}, {Nichol}, {Niemi},
  {Padilla}, {Pasian}, {Pedersen}, {Percival}, {Pettorino}, {Pires}, {Poncet},
  {Popa}, {Pozzetti}, {Prieto}, {Raison}, {Rhodes}, {Rix}, {Roncarelli},
  {Rossetti}, {Saglia}, {Sartoris}, {Scaramella}, {Schneider}, {Secroun},
  {Serrano}, {Sirignano}, {Sirri}, {Stanco}, {Tallada-Cresp{\'\i}}, {Taylor},
  {Teplitz}, {Tereno}, {Toledo-Moreo}, {Torradeflot}, {Trifoglio}, {Valentijn},
  {Valenziano}, {Wang}, {Weller}, {Zamorani}, {Zoubian}, {Andreon}, {Bardelli},
  {Boucaud}, {Camera}, {Farinelli}, {Graci{\'a}-Carpio}, {Maino}, {Medinaceli},
  {Mei}, {Morisset}, {Polenta}, {Renzi}, {Romelli}, {Tenti}, {Vassallo},
  {Zacchei}, {Zucca}, {Baccigalupi}, {Balaguera-Antol{\'\i}nez}, {Biviano},
  {Blanchard}, {Borgani}, {Bozzo}, {Burigana}, {Cabanac}, {Cappi}, {Carvalho},
  {Casas}, {Castignani}, {Colodro-Conde}, {Cooray}, {Courtois}, {Crocce},
  {Cuby}, {Davini}, {de la Torre}, {Di Ferdinando}, {Escartin}, {Farina},
  {Ferreira}, {Finelli}, {Fotopoulou}, {Galeotta}, {Garcia-Bellido},
  {Gaztanaga}, {George}, {Gozaliasl}, {Hook}, {Ili{\'c}}, {Kansal},
  {Kashlinsky}, {Keihanen}, {Kirkpatrick}, {Lindholm}, {Mainetti}, {Maoli},
  {Martinelli}, {Martinet}, {Maturi}, {Mauri}, {McCracken}, {Metcalf},
  {Monaco}, {Morgante}, {Nightingale}, {Patrizii}, {Peel}, {Popa}, {Porciani},
  {Potter}, {Reimberg}, {Riccio}, {S{\'a}nchez}, {Sapone}, {Scottez},
  {Sefusatti}, {Teyssier}, {Tutusaus}, {Valieri}, {Valiviita}, {Viel}, \&
  {Hildebrandt}}]{Schirmer-EP18}
{Euclid Collaboration: Schirmer}, M., {Jahnke}, K., {Seidel}, G., {et~al.}
  2022, \aap, 662, A92

\bibitem[{{Euclid Early Release Observations}(2024)}]{EROcite}
{Euclid Early Release Observations}. 2024,
  \url{https://doi.org/10.57780/esa-qmocze3}

\bibitem[{{Foreman-Mackey} {et~al.}(2013){Foreman-Mackey}, {Hogg}, {Lang}, \&
  {Goodman}}]{2013PASP..125..306F}
{Foreman-Mackey}, D., {Hogg}, D.~W., {Lang}, D., \& {Goodman}, J. 2013, \pasp,
  125, 306

\bibitem[{{Gaia Collaboration} {et~al.}(2023{\natexlab{a}}){Gaia
  Collaboration}, {Drimmel}, {Romero-G{\'o}mez}, {Chemin}, {Ramos}, {Poggio},
  {Ripepi}, {Andrae}, {Blomme}, {Cantat-Gaudin}, {Castro-Ginard}, {Clementini},
  {Figueras}, {Fouesneau}, {Fr{\'e}mat}, {Jardine}, {Khanna}, {Lobel},
  {Marshall}, {Muraveva}, {Brown}, {Vallenari}, {Prusti}, {de Bruijne},
  {Arenou}, {Babusiaux}, {Biermann}, {Creevey}, {Ducourant}, {Evans}, {Eyer},
  {Guerra}, {Hutton}, {Jordi}, {Klioner}, {Lammers}, {Lindegren}, {Luri},
  {Mignard}, {Panem}, {Pourbaix}, {Randich}, {Sartoretti}, {Soubiran}, {Tanga},
  {Walton}, {Bailer-Jones}, {Bastian}, {Jansen}, {Katz}, {Lattanzi}, {van
  Leeuwen}, {Bakker}, {Cacciari}, {Casta{\~n}eda}, {De Angeli}, {Fabricius},
  {Galluccio}, {Guerrier}, {Heiter}, {Masana}, {Messineo}, {Mowlavi},
  {Nicolas}, {Nienartowicz}, {Pailler}, {Panuzzo}, {Riclet}, {Roux},
  {Seabroke}, {Sordo}, {Th{\'e}venin}, {Gracia-Abril}, {Portell}, {Teyssier},
  {Altmann}, {Audard}, {Bellas-Velidis}, {Benson}, {Berthier}, {Burgess},
  {Busonero}, {Busso}, {C{\'a}novas}, {Carry}, {Cellino}, {Cheek}, {Damerdji},
  {Davidson}, {de Teodoro}, {Nu{\~n}ez Campos}, {Delchambre}, {Dell'Oro},
  {Esquej}, {Fern{\'a}ndez-Hern{\'a}ndez}, {Fraile}, {Garabato},
  {Garc{\'\i}a-Lario}, {Gosset}, {Haigron}, {Halbwachs}, {Hambly}, {Harrison},
  {Hern{\'a}ndez}, {Hestroffer}, {Hodgkin}, {Holl}, {Jan{\ss}en}, {Jevardat de
  Fombelle}, {Jordan}, {Krone-Martins}, {Lanzafame}, {L{\"o}ffler}, {Marchal},
  {Marrese}, {Moitinho}, {Muinonen}, {Osborne}, {Pancino}, {Pauwels},
  {Recio-Blanco}, {Reyl{\'e}}, {Riello}, {Rimoldini}, {Roegiers}, {Rybizki},
  {Sarro}, {Siopis}, {Smith}, {Sozzetti}, {Utrilla}, {van Leeuwen}, {Abbas},
  {{\'A}brah{\'a}m}, {Abreu Aramburu}, {Aerts}, {Aguado}, {Ajaj},
  {Aldea-Montero}, {Altavilla}, {{\'A}lvarez}, {Alves}, {Anders}, {Anderson},
  {Anglada Varela}, {Antoja}, {Baines}, {Baker}, {Balaguer-N{\'u}{\~n}ez},
  {Balbinot}, {Balog}, {Barache}, {Barbato}, {Barros}, {Barstow},
  {Bartolom{\'e}}, {Bassilana}, {Bauchet}, {Becciani}, {Bellazzini},
  {Berihuete}, {Bernet}, {Bertone}, {Bianchi}, {Binnenfeld}, {Blanco-Cuaresma},
  {Boch}, {Bombrun}, {Bossini}, {Bouquillon}, {Bragaglia}, {Bramante},
  {Breedt}, {Bressan}, {Brouillet}, {Brugaletta}, {Bucciarelli}, {Burlacu},
  {Butkevich}, {Buzzi}, {Caffau}, {Cancelliere}, {Carballo}, {Carlucci},
  {Carnerero}, {Carrasco}, {Casamiquela}, {Castellani}, {Chaoul}, {Charlot},
  {Chiaramida}, {Chiavassa}, {Chornay}, {Comoretto}, {Contursi}, {Cooper},
  {Cornez}, {Cowell}, {Crifo}, {Cropper}, {Crosta}, {Crowley}, {Dafonte},
  {Dapergolas}, {David}, {de Laverny}, {De Luise}, {De March}, {De Ridder}, {de
  Souza}, {de Torres}, {del Peloso}, {del Pozo}, {Delbo}, {Delgado}, {Delisle},
  {Demouchy}, {Dharmawardena}, {Di Matteo}, {Diakite}, {Diener}, {Distefano},
  {Dolding}, {Enke}, {Fabre}, {Fabrizio}, {Faigler}, {Fedorets}, {Fernique},
  {Fournier}, {Fouron}, {Fragkoudi}, {Gai}, {Garcia-Gutierrez},
  {Garcia-Reinaldos}, {Garc{\'\i}a-Torres}, {Garofalo}, {Gavel}, {Gavras},
  {Gerlach}, {Geyer}, {Giacobbe}, {Gilmore}, {Girona}, {Giuffrida}, {Gomel},
  {Gomez}, {Gonz{\'a}lez-N{\'u}{\~n}ez}, {Gonz{\'a}lez-Santamar{\'\i}a},
  {Gonz{\'a}lez-Vidal}, {Granvik}, {Guillout}, {Guiraud},
  {Guti{\'e}rrez-S{\'a}nchez}, {Guy}, {Hatzidimitriou}, {Hauser}, {Haywood},
  {Helmer}, {Helmi}, {Sarmiento}, {Hidalgo}, {H{\l}adczuk}, {Hobbs}, {Holland},
  {Huckle}, {Jasniewicz}, {Jean-Antoine Piccolo}, {Jim{\'e}nez-Arranz},
  {Juaristi Campillo}, {Julbe}, {Karbevska}, {Kervella}, {Kordopatis}, {Korn},
  {K{\'o}sp{\'a}l}, {Kostrzewa-Rutkowska}, {Kruszy{\'n}ska}, {Kun}, {Laizeau},
  {Lambert}, {Lanza}, {Lasne}, {Le Campion}, {Lebreton}, {Lebzelter}, {Leccia},
  {Leclerc}, {Lecoeur-Taibi}, {Liao}, {Licata}, {Lindstr{\o}m}, {Lister},
  {Livanou}, {Lorca}, {Loup}, {Madrero Pardo}, {Magdaleno Romeo}, {Managau},
  {Mann}, {Manteiga}, {Marchant}, {Marconi}, {Marcos}, {Marcos Santos},
  {Mar{\'\i}n Pina}, {Marinoni}, {Marocco}, {Martin Polo},
  {Mart{\'\i}n-Fleitas}, {Marton}, {Mary}, {Masip}, {Massari},
  {Mastrobuono-Battisti}, {Mazeh}, {McMillan}, {Messina}, {Michalik}, {Millar},
  {Mints}, {Molina}, {Molinaro}, {Moln{\'a}r}, {Monari}, {Mongui{\'o}},
  {Montegriffo}, {Montero}, {Mor}, {Mora}, {Morbidelli}, {Morel}, {Morris},
  {Murphy}, {Musella}, {Nagy}, {Noval}, {Oca{\~n}a}, {Ogden}, {Ordenovic},
  {Osinde}, {Pagani}, {Pagano}, {Palaversa}, {Palicio}, {Pallas-Quintela},
  {Panahi}, {Payne-Wardenaar}, {Pe{\~n}alosa Esteller}, {Penttil{\"a}},
  {Pichon}, {Piersimoni}, {Pineau}, {Plachy}, {Plum}, {Pr{\v{s}}a}, {Pulone},
  {Racero}, {Ragaini}, {Rainer}, {Raiteri}, {Ramos-Lerate}, {Re Fiorentin},
  {Regibo}, {Richards}, {Rios Diaz}, {Riva}, {Rix}, {Rixon}, {Robichon},
  {Robin}, {Robin}, {Roelens}, {Rogues}, {Rohrbasser}, {Rowell}, {Royer}, {Ruz
  Mieres}, {Rybicki}, {Sadowski}, {S{\'a}ez N{\'u}{\~n}ez}, {Sagrist{\`a}
  Sell{\'e}s}, {Sahlmann}, {Salguero}, {Samaras}, {Sanchez Gimenez}, {Sanna},
  {Santove{\~n}a}, {Sarasso}, {Schultheis}, {Sciacca}, {Segol}, {Segovia},
  {S{\'e}gransan}, {Semeux}, {Shahaf}, {Siddiqui}, {Siebert}, {Siltala},
  {Silvelo}, {Slezak}, {Slezak}, {Smart}, {Snaith}, {Solano}, {Solitro},
  {Souami}, {Souchay}, {Spagna}, {Spina}, {Spoto}, {Steele},
  {Steidelm{\"u}ller}, {Stephenson}, {S{\"u}veges}, {Surdej}, {Szabados},
  {Szegedi-Elek}, {Taris}, {Taylor}, {Teixeira}, {Tolomei}, {Tonello}, {Torra},
  {Torra}, {Torralba Elipe}, {Trabucchi}, {Tsounis}, {Turon}, {Ulla}, {Unger},
  {Vaillant}, {van Dillen}, {van Reeven}, {Vanel}, {Vecchiato}, {Viala},
  {Vicente}, {Voutsinas}, {Weiler}, {Wevers}, {Wyrzykowski}, {Yoldas}, {Yvard},
  {Zhao}, {Zorec}, {Zucker}, \& {Zwitter}}]{2023A&A...674A..37G}
{Gaia Collaboration}, {Drimmel}, R., {Romero-G{\'o}mez}, M., {et~al.}
  2023{\natexlab{a}}, \aap, 674, A37

\bibitem[{{Gaia Collaboration} {et~al.}(2016){Gaia Collaboration}, {Prusti},
  {de Bruijne}, {Brown}, {Vallenari}, {Babusiaux}, {Bailer-Jones}, {Bastian},
  {Biermann}, {Evans}, {Eyer}, {Jansen}, {Jordi}, {Klioner}, {Lammers},
  {Lindegren}, {Luri}, {Mignard}, {Milligan}, {Panem}, {Poinsignon},
  {Pourbaix}, {Randich}, {Sarri}, {Sartoretti}, {Siddiqui}, {Soubiran},
  {Valette}, {van Leeuwen}, {Walton}, {Aerts}, {Arenou}, {Cropper}, {Drimmel},
  {H{\o}g}, {Katz}, {Lattanzi}, {O'Mullane}, {Grebel}, {Holland}, {Huc},
  {Passot}, {Bramante}, {Cacciari}, {Casta{\~n}eda}, {Chaoul}, {Cheek}, {De
  Angeli}, {Fabricius}, {Guerra}, {Hern{\'a}ndez}, {Jean-Antoine-Piccolo},
  {Masana}, {Messineo}, {Mowlavi}, {Nienartowicz}, {Ord{\'o}{\~n}ez-Blanco},
  {Panuzzo}, {Portell}, {Richards}, {Riello}, {Seabroke}, {Tanga},
  {Th{\'e}venin}, {Torra}, {Els}, {Gracia-Abril}, {Comoretto},
  {Garcia-Reinaldos}, {Lock}, {Mercier}, {Altmann}, {Andrae}, {Astraatmadja},
  {Bellas-Velidis}, {Benson}, {Berthier}, {Blomme}, {Busso}, {Carry},
  {Cellino}, {Clementini}, {Cowell}, {Creevey}, {Cuypers}, {Davidson}, {De
  Ridder}, {de Torres}, {Delchambre}, {Dell'Oro}, {Ducourant}, {Fr{\'e}mat},
  {Garc{\'\i}a-Torres}, {Gosset}, {Halbwachs}, {Hambly}, {Harrison}, {Hauser},
  {Hestroffer}, {Hodgkin}, {Huckle}, {Hutton}, {Jasniewicz}, {Jordan},
  {Kontizas}, {Korn}, {Lanzafame}, {Manteiga}, {Moitinho}, {Muinonen},
  {Osinde}, {Pancino}, {Pauwels}, {Petit}, {Recio-Blanco}, {Robin}, {Sarro},
  {Siopis}, {Smith}, {Smith}, {Sozzetti}, {Thuillot}, {van Reeven}, {Viala},
  {Abbas}, {Abreu Aramburu}, {Accart}, {Aguado}, {Allan}, {Allasia},
  {Altavilla}, {{\'A}lvarez}, {Alves}, {Anderson}, {Andrei}, {Anglada Varela},
  {Antiche}, {Antoja}, {Ant{\'o}n}, {Arcay}, {Atzei}, {Ayache}, {Bach},
  {Baker}, {Balaguer-N{\'u}{\~n}ez}, {Barache}, {Barata}, {Barbier}, {Barblan},
  {Baroni}, {Barrado y Navascu{\'e}s}, {Barros}, {Barstow}, {Becciani},
  {Bellazzini}, {Bellei}, {Bello Garc{\'\i}a}, {Belokurov}, {Bendjoya},
  {Berihuete}, {Bianchi}, {Bienaym{\'e}}, {Billebaud}, {Blagorodnova},
  {Blanco-Cuaresma}, {Boch}, {Bombrun}, {Borrachero}, {Bouquillon}, {Bourda},
  {Bouy}, {Bragaglia}, {Breddels}, {Brouillet}, {Br{\"u}semeister},
  {Bucciarelli}, {Budnik}, {Burgess}, {Burgon}, {Burlacu}, {Busonero}, {Buzzi},
  {Caffau}, {Cambras}, {Campbell}, {Cancelliere}, {Cantat-Gaudin}, {Carlucci},
  {Carrasco}, {Castellani}, {Charlot}, {Charnas}, {Charvet}, {Chassat},
  {Chiavassa}, {Clotet}, {Cocozza}, {Collins}, {Collins}, {Costigan}, {Crifo},
  {Cross}, {Crosta}, {Crowley}, {Dafonte}, {Damerdji}, {Dapergolas}, {David},
  {David}, {De Cat}, {de Felice}, {de Laverny}, {De Luise}, {De March}, {de
  Martino}, {de Souza}, {Debosscher}, {del Pozo}, {Delbo}, {Delgado},
  {Delgado}, {di Marco}, {Di Matteo}, {Diakite}, {Distefano}, {Dolding}, {Dos
  Anjos}, {Drazinos}, {Dur{\'a}n}, {Dzigan}, {Ecale}, {Edvardsson}, {Enke},
  {Erdmann}, {Escolar}, {Espina}, {Evans}, {Eynard Bontemps}, {Fabre},
  {Fabrizio}, {Faigler}, {Falc{\~a}o}, {Farr{\`a}s Casas}, {Faye}, {Federici},
  {Fedorets}, {Fern{\'a}ndez-Hern{\'a}ndez}, {Fernique}, {Fienga}, {Figueras},
  {Filippi}, {Findeisen}, {Fonti}, {Fouesneau}, {Fraile}, {Fraser}, {Fuchs},
  {Furnell}, {Gai}, {Galleti}, {Galluccio}, {Garabato}, {Garc{\'\i}a-Sedano},
  {Gar{\'e}}, {Garofalo}, {Garralda}, {Gavras}, {Gerssen}, {Geyer}, {Gilmore},
  {Girona}, {Giuffrida}, {Gomes}, {Gonz{\'a}lez-Marcos},
  {Gonz{\'a}lez-N{\'u}{\~n}ez}, {Gonz{\'a}lez-Vidal}, {Granvik}, {Guerrier},
  {Guillout}, {Guiraud}, {G{\'u}rpide}, {Guti{\'e}rrez-S{\'a}nchez}, {Guy},
  {Haigron}, {Hatzidimitriou}, {Haywood}, {Heiter}, {Helmi}, {Hobbs},
  {Hofmann}, {Holl}, {Holland}, {Hunt}, {Hypki}, {Icardi}, {Irwin}, {Jevardat
  de Fombelle}, {Jofr{\'e}}, {Jonker}, {Jorissen}, {Julbe}, {Karampelas},
  {Kochoska}, {Kohley}, {Kolenberg}, {Kontizas}, {Koposov}, {Kordopatis},
  {Koubsky}, {Kowalczyk}, {Krone-Martins}, {Kudryashova}, {Kull}, {Bachchan},
  {Lacoste-Seris}, {Lanza}, {Lavigne}, {Le Poncin-Lafitte}, {Lebreton},
  {Lebzelter}, {Leccia}, {Leclerc}, {Lecoeur-Taibi}, {Lemaitre}, {Lenhardt},
  {Leroux}, {Liao}, {Licata}, {Lindstr{\o}m}, {Lister}, {Livanou}, {Lobel},
  {L{\"o}ffler}, {L{\'o}pez}, {Lopez-Lozano}, {Lorenz}, {Loureiro},
  {MacDonald}, {Magalh{\~a}es Fernandes}, {Managau}, {Mann}, {Mantelet},
  {Marchal}, {Marchant}, {Marconi}, {Marie}, {Marinoni}, {Marrese},
  {Marschalk{\'o}}, {Marshall}, {Mart{\'\i}n-Fleitas}, {Martino}, {Mary},
  {Matijevi{\v{c}}}, {Mazeh}, {McMillan}, {Messina}, {Mestre}, {Michalik},
  {Millar}, {Miranda}, {Molina}, {Molinaro}, {Molinaro}, {Moln{\'a}r},
  {Moniez}, {Montegriffo}, {Monteiro}, {Mor}, {Mora}, {Morbidelli}, {Morel},
  {Morgenthaler}, {Morley}, {Morris}, {Mulone}, {Muraveva}, {Musella},
  {Narbonne}, {Nelemans}, {Nicastro}, {Noval}, {Ord{\'e}novic},
  {Ordieres-Mer{\'e}}, {Osborne}, {Pagani}, {Pagano}, {Pailler}, {Palacin},
  {Palaversa}, {Parsons}, {Paulsen}, {Pecoraro}, {Pedrosa}, {Pentik{\"a}inen},
  {Pereira}, {Pichon}, {Piersimoni}, {Pineau}, {Plachy}, {Plum}, {Poujoulet},
  {Pr{\v{s}}a}, {Pulone}, {Ragaini}, {Rago}, {Rambaux}, {Ramos-Lerate},
  {Ranalli}, {Rauw}, {Read}, {Regibo}, {Renk}, {Reyl{\'e}}, {Ribeiro},
  {Rimoldini}, {Ripepi}, {Riva}, {Rixon}, {Roelens}, {Romero-G{\'o}mez},
  {Rowell}, {Royer}, {Rudolph}, {Ruiz-Dern}, {Sadowski}, {Sagrist{\`a}
  Sell{\'e}s}, {Sahlmann}, {Salgado}, {Salguero}, {Sarasso}, {Savietto},
  {Schnorhk}, {Schultheis}, {Sciacca}, {Segol}, {Segovia}, {Segransan},
  {Serpell}, {Shih}, {Smareglia}, {Smart}, {Smith}, {Solano}, {Solitro},
  {Sordo}, {Soria Nieto}, {Souchay}, {Spagna}, {Spoto}, {Stampa}, {Steele},
  {Steidelm{\"u}ller}, {Stephenson}, {Stoev}, {Suess}, {S{\"u}veges}, {Surdej},
  {Szabados}, {Szegedi-Elek}, {Tapiador}, {Taris}, {Tauran}, {Taylor},
  {Teixeira}, {Terrett}, {Tingley}, {Trager}, {Turon}, {Ulla}, {Utrilla},
  {Valentini}, {van Elteren}, {Van Hemelryck}, {van Leeuwen}, {Varadi},
  {Vecchiato}, {Veljanoski}, {Via}, {Vicente}, {Vogt}, {Voss}, {Votruba},
  {Voutsinas}, {Walmsley}, {Weiler}, {Weingrill}, {Werner}, {Wevers},
  {Whitehead}, {Wyrzykowski}, {Yoldas}, {{\v{Z}}erjal}, {Zucker}, {Zurbach},
  {Zwitter}, {Alecu}, {Allen}, {Allende Prieto}, {Amorim},
  {Anglada-Escud{\'e}}, {Arsenijevic}, {Azaz}, {Balm}, {Beck}, {Bernstein},
  {Bigot}, {Bijaoui}, {Blasco}, {Bonfigli}, {Bono}, {Boudreault}, {Bressan},
  {Brown}, {Brunet}, {Bunclark}, {Buonanno}, {Butkevich}, {Carret}, {Carrion},
  {Chemin}, {Ch{\'e}reau}, {Corcione}, {Darmigny}, {de Boer}, {de Teodoro}, {de
  Zeeuw}, {Delle Luche}, {Domingues}, {Dubath}, {Fodor}, {Fr{\'e}zouls},
  {Fries}, {Fustes}, {Fyfe}, {Gallardo}, {Gallegos}, {Gardiol}, {Gebran},
  {Gomboc}, {G{\'o}mez}, {Grux}, {Gueguen}, {Heyrovsky}, {Hoar}, {Iannicola},
  {Isasi Parache}, {Janotto}, {Joliet}, {Jonckheere}, {Keil}, {Kim},
  {Klagyivik}, {Klar}, {Knude}, {Kochukhov}, {Kolka}, {Kos}, {Kutka}, {Lainey},
  {LeBouquin}, {Liu}, {Loreggia}, {Makarov}, {Marseille}, {Martayan},
  {Martinez-Rubi}, {Massart}, {Meynadier}, {Mignot}, {Munari}, {Nguyen},
  {Nordlander}, {Ocvirk}, {O'Flaherty}, {Olias Sanz}, {Ortiz}, {Osorio},
  {Oszkiewicz}, {Ouzounis}, {Palmer}, {Park}, {Pasquato}, {Peltzer}, {Peralta},
  {P{\'e}turaud}, {Pieniluoma}, {Pigozzi}, {Poels}, {Prat}, {Prod'homme},
  {Raison}, {Rebordao}, {Risquez}, {Rocca-Volmerange}, {Rosen}, {Ruiz-Fuertes},
  {Russo}, {Sembay}, {Serraller Vizcaino}, {Short}, {Siebert}, {Silva},
  {Sinachopoulos}, {Slezak}, {Soffel}, {Sosnowska}, {Strai{\v{z}}ys}, {ter
  Linden}, {Terrell}, {Theil}, {Tiede}, {Troisi}, {Tsalmantza}, {Tur},
  {Vaccari}, {Vachier}, {Valles}, {Van Hamme}, {Veltz}, {Virtanen}, {Wallut},
  {Wichmann}, {Wilkinson}, {Ziaeepour}, \& {Zschocke}}]{2016A&A...595A...1G}
{Gaia Collaboration}, {Prusti}, T., {de Bruijne}, J.~H.~J., {et~al.} 2016,
  \aap, 595, A1

\bibitem[{{Gaia Collaboration} {et~al.}(2023{\natexlab{b}}){Gaia
  Collaboration}, {Vallenari}, {Brown}, {Prusti}, {de Bruijne}, {Arenou},
  {Babusiaux}, {Biermann}, {Creevey}, {Ducourant}, {Evans}, {Eyer}, {Guerra},
  {Hutton}, {Jordi}, {Klioner}, {Lammers}, {Lindegren}, {Luri}, {Mignard},
  {Panem}, {Pourbaix}, {Randich}, {Sartoretti}, {Soubiran}, {Tanga}, {Walton},
  {Bailer-Jones}, {Bastian}, {Drimmel}, {Jansen}, {Katz}, {Lattanzi}, {van
  Leeuwen}, {Bakker}, {Cacciari}, {Casta{\~n}eda}, {De Angeli}, {Fabricius},
  {Fouesneau}, {Fr{\'e}mat}, {Galluccio}, {Guerrier}, {Heiter}, {Masana},
  {Messineo}, {Mowlavi}, {Nicolas}, {Nienartowicz}, {Pailler}, {Panuzzo},
  {Riclet}, {Roux}, {Seabroke}, {Sordo}, {Th{\'e}venin}, {Gracia-Abril},
  {Portell}, {Teyssier}, {Altmann}, {Andrae}, {Audard}, {Bellas-Velidis},
  {Benson}, {Berthier}, {Blomme}, {Burgess}, {Busonero}, {Busso},
  {C{\'a}novas}, {Carry}, {Cellino}, {Cheek}, {Clementini}, {Damerdji},
  {Davidson}, {de Teodoro}, {Nu{\~n}ez Campos}, {Delchambre}, {Dell'Oro},
  {Esquej}, {Fern{\'a}ndez-Hern{\'a}ndez}, {Fraile}, {Garabato},
  {Garc{\'\i}a-Lario}, {Gosset}, {Haigron}, {Halbwachs}, {Hambly}, {Harrison},
  {Hern{\'a}ndez}, {Hestroffer}, {Hodgkin}, {Holl}, {Jan{\ss}en}, {Jevardat de
  Fombelle}, {Jordan}, {Krone-Martins}, {Lanzafame}, {L{\"o}ffler}, {Marchal},
  {Marrese}, {Moitinho}, {Muinonen}, {Osborne}, {Pancino}, {Pauwels},
  {Recio-Blanco}, {Reyl{\'e}}, {Riello}, {Rimoldini}, {Roegiers}, {Rybizki},
  {Sarro}, {Siopis}, {Smith}, {Sozzetti}, {Utrilla}, {van Leeuwen}, {Abbas},
  {{\'A}brah{\'a}m}, {Abreu Aramburu}, {Aerts}, {Aguado}, {Ajaj},
  {Aldea-Montero}, {Altavilla}, {{\'A}lvarez}, {Alves}, {Anders}, {Anderson},
  {Anglada Varela}, {Antoja}, {Baines}, {Baker}, {Balaguer-N{\'u}{\~n}ez},
  {Balbinot}, {Balog}, {Barache}, {Barbato}, {Barros}, {Barstow},
  {Bartolom{\'e}}, {Bassilana}, {Bauchet}, {Becciani}, {Bellazzini},
  {Berihuete}, {Bernet}, {Bertone}, {Bianchi}, {Binnenfeld}, {Blanco-Cuaresma},
  {Blazere}, {Boch}, {Bombrun}, {Bossini}, {Bouquillon}, {Bragaglia},
  {Bramante}, {Breedt}, {Bressan}, {Brouillet}, {Brugaletta}, {Bucciarelli},
  {Burlacu}, {Butkevich}, {Buzzi}, {Caffau}, {Cancelliere}, {Cantat-Gaudin},
  {Carballo}, {Carlucci}, {Carnerero}, {Carrasco}, {Casamiquela}, {Castellani},
  {Castro-Ginard}, {Chaoul}, {Charlot}, {Chemin}, {Chiaramida}, {Chiavassa},
  {Chornay}, {Comoretto}, {Contursi}, {Cooper}, {Cornez}, {Cowell}, {Crifo},
  {Cropper}, {Crosta}, {Crowley}, {Dafonte}, {Dapergolas}, {David}, {David},
  {de Laverny}, {De Luise}, {De March}, {De Ridder}, {de Souza}, {de Torres},
  {del Peloso}, {del Pozo}, {Delbo}, {Delgado}, {Delisle}, {Demouchy},
  {Dharmawardena}, {Di Matteo}, {Diakite}, {Diener}, {Distefano}, {Dolding},
  {Edvardsson}, {Enke}, {Fabre}, {Fabrizio}, {Faigler}, {Fedorets}, {Fernique},
  {Fienga}, {Figueras}, {Fournier}, {Fouron}, {Fragkoudi}, {Gai},
  {Garcia-Gutierrez}, {Garcia-Reinaldos}, {Garc{\'\i}a-Torres}, {Garofalo},
  {Gavel}, {Gavras}, {Gerlach}, {Geyer}, {Giacobbe}, {Gilmore}, {Girona},
  {Giuffrida}, {Gomel}, {Gomez}, {Gonz{\'a}lez-N{\'u}{\~n}ez},
  {Gonz{\'a}lez-Santamar{\'\i}a}, {Gonz{\'a}lez-Vidal}, {Granvik}, {Guillout},
  {Guiraud}, {Guti{\'e}rrez-S{\'a}nchez}, {Guy}, {Hatzidimitriou}, {Hauser},
  {Haywood}, {Helmer}, {Helmi}, {Sarmiento}, {Hidalgo}, {Hilger},
  {H{\l}adczuk}, {Hobbs}, {Holland}, {Huckle}, {Jardine}, {Jasniewicz},
  {Jean-Antoine Piccolo}, {Jim{\'e}nez-Arranz}, {Jorissen}, {Juaristi
  Campillo}, {Julbe}, {Karbevska}, {Kervella}, {Khanna}, {Kontizas},
  {Kordopatis}, {Korn}, {K{\'o}sp{\'a}l}, {Kostrzewa-Rutkowska},
  {Kruszy{\'n}ska}, {Kun}, {Laizeau}, {Lambert}, {Lanza}, {Lasne}, {Le
  Campion}, {Lebreton}, {Lebzelter}, {Leccia}, {Leclerc}, {Lecoeur-Taibi},
  {Liao}, {Licata}, {Lindstr{\o}m}, {Lister}, {Livanou}, {Lobel}, {Lorca},
  {Loup}, {Madrero Pardo}, {Magdaleno Romeo}, {Managau}, {Mann}, {Manteiga},
  {Marchant}, {Marconi}, {Marcos}, {Marcos Santos}, {Mar{\'\i}n Pina},
  {Marinoni}, {Marocco}, {Marshall}, {Martin Polo}, {Mart{\'\i}n-Fleitas},
  {Marton}, {Mary}, {Masip}, {Massari}, {Mastrobuono-Battisti}, {Mazeh},
  {McMillan}, {Messina}, {Michalik}, {Millar}, {Mints}, {Molina}, {Molinaro},
  {Moln{\'a}r}, {Monari}, {Mongui{\'o}}, {Montegriffo}, {Montero}, {Mor},
  {Mora}, {Morbidelli}, {Morel}, {Morris}, {Muraveva}, {Murphy}, {Musella},
  {Nagy}, {Noval}, {Oca{\~n}a}, {Ogden}, {Ordenovic}, {Osinde}, {Pagani},
  {Pagano}, {Palaversa}, {Palicio}, {Pallas-Quintela}, {Panahi},
  {Payne-Wardenaar}, {Pe{\~n}alosa Esteller}, {Penttil{\"a}}, {Pichon},
  {Piersimoni}, {Pineau}, {Plachy}, {Plum}, {Poggio}, {Pr{\v{s}}a}, {Pulone},
  {Racero}, {Ragaini}, {Rainer}, {Raiteri}, {Rambaux}, {Ramos}, {Ramos-Lerate},
  {Re Fiorentin}, {Regibo}, {Richards}, {Rios Diaz}, {Ripepi}, {Riva}, {Rix},
  {Rixon}, {Robichon}, {Robin}, {Robin}, {Roelens}, {Rogues}, {Rohrbasser},
  {Romero-G{\'o}mez}, {Rowell}, {Royer}, {Ruz Mieres}, {Rybicki}, {Sadowski},
  {S{\'a}ez N{\'u}{\~n}ez}, {Sagrist{\`a} Sell{\'e}s}, {Sahlmann}, {Salguero},
  {Samaras}, {Sanchez Gimenez}, {Sanna}, {Santove{\~n}a}, {Sarasso},
  {Schultheis}, {Sciacca}, {Segol}, {Segovia}, {S{\'e}gransan}, {Semeux},
  {Shahaf}, {Siddiqui}, {Siebert}, {Siltala}, {Silvelo}, {Slezak}, {Slezak},
  {Smart}, {Snaith}, {Solano}, {Solitro}, {Souami}, {Souchay}, {Spagna},
  {Spina}, {Spoto}, {Steele}, {Steidelm{\"u}ller}, {Stephenson}, {S{\"u}veges},
  {Surdej}, {Szabados}, {Szegedi-Elek}, {Taris}, {Taylor}, {Teixeira},
  {Tolomei}, {Tonello}, {Torra}, {Torra}, {Torralba Elipe}, {Trabucchi},
  {Tsounis}, {Turon}, {Ulla}, {Unger}, {Vaillant}, {van Dillen}, {van Reeven},
  {Vanel}, {Vecchiato}, {Viala}, {Vicente}, {Voutsinas}, {Weiler}, {Wevers},
  {Wyrzykowski}, {Yoldas}, {Yvard}, {Zhao}, {Zorec}, {Zucker}, \&
  {Zwitter}}]{2023A&A...674A...1G}
{Gaia Collaboration}, {Vallenari}, A., {Brown}, A.~G.~A., {et~al.}
  2023{\natexlab{b}}, \aap, 674, A1

\bibitem[{{Griggio} {et~al.}(2022){Griggio}, {Bedin}, {Raddi}, {Reindl},
  {Tomasella}, {Scalco}, {Salaris}, {Cassisi}, {Ochner}, {Ciroi}, {Rosati},
  {Nardiello}, {Anderson}, {Libralato}, {Bellini}, {Vallenari}, {Spina}, \&
  {Pedani}}]{2022MNRAS.515.1841G}
{Griggio}, M., {Bedin}, L.~R., {Raddi}, R., {et~al.} 2022, \mnras, 515, 1841

\bibitem[{{Griggio} {et~al.}(2024){Griggio}, {Libralato}, {Bellini}, {Bedin},
  {Anderson}, {Smith}, \& {Minniti}}]{2024GriggioVVV}
{Griggio}, M., {Libralato}, M., {Bellini}, A., {et~al.} 2024, \aap, 687, A94

\bibitem[{{Griggio} {et~al.}(2023{\natexlab{a}}){Griggio}, {Nardiello}, \&
  {Bedin}}]{2023GriggioNIRCam}
{Griggio}, M., {Nardiello}, D., \& {Bedin}, L.~R. 2023{\natexlab{a}},
  Astronomische Nachrichten, 344, easna.20230006

\bibitem[{{Griggio} {et~al.}(2023{\natexlab{b}}){Griggio}, {Salaris},
  {Nardiello}, {Bedin}, {Cassisi}, \& {Anderson}}]{2023GriggioCFHT}
{Griggio}, M., {Salaris}, M., {Nardiello}, D., {et~al.} 2023{\natexlab{b}},
  \mnras, 524, 108

\bibitem[{{H{\"a}berle} {et~al.}(2021){H{\"a}berle}, {Libralato}, {Bellini},
  {Watkins}, {Pott}, {Neumayer}, {van der Marel}, {Piotto}, \&
  {Nardiello}}]{2021HaberleNACO}
{H{\"a}berle}, M., {Libralato}, M., {Bellini}, A., {et~al.} 2021, \mnras, 503,
  1490

\bibitem[{{H{\"a}berle} {et~al.}(2024){H{\"a}berle}, {Neumayer}, {Seth},
  {Bellini}, {Libralato}, {Baumgardt}, {Whitaker}, {Dumont}, {Alfaro-Cuello},
  {Anderson}, {Clontz}, {Kacharov}, {Kamann}, {Feldmeier-Krause}, {Milone},
  {Nitschai}, {Pechetti}, \& {van de Ven}}]{2024HaberlewCenIMBH}
{H{\"a}berle}, M., {Neumayer}, N., {Seth}, A., {et~al.} 2024, \nat, 631, 285

\bibitem[{{Ibata} {et~al.}(2024){Ibata}, {Malhan}, {Tenachi},
  {Ardern-Arentsen}, {Bellazzini}, {Bianchini}, {Bonifacio}, {Caffau},
  {Diakogiannis}, {Errani}, {Famaey}, {Ferrone}, {Martin}, {di Matteo},
  {Monari}, {Renaud}, {Starkenburg}, {Thomas}, {Viswanathan}, \&
  {Yuan}}]{2024ApJ...967...89I}
{Ibata}, R., {Malhan}, K., {Tenachi}, W., {et~al.} 2024, \apj, 967, 89

\bibitem[{{Libralato} {et~al.}(2024){Libralato}, {Argyriou}, {Dicken},
  {Garc{\'\i}a Mar{\'\i}n}, {Guillard}, {Hines}, {Kavanagh}, {Kendrew}, {Law},
  {Noriega-Crespo}, \& {{\'A}lvarez-M{\'a}rquez}}]{2024LibralatoMIRI}
{Libralato}, M., {Argyriou}, I., {Dicken}, D., {et~al.} 2024, \pasp, 136,
  034502

\bibitem[{{Libralato} {et~al.}(2016){Libralato}, {Bedin}, {Nardiello}, \&
  {Piotto}}]{2016LibralatoK2i}
{Libralato}, M., {Bedin}, L.~R., {Nardiello}, D., \& {Piotto}, G. 2016, \mnras,
  456, 1137

\bibitem[{{Libralato} {et~al.}(2015){Libralato}, {Bellini}, {Bedin},
  {Anderson}, {Piotto}, {Nascimbeni}, {Platais}, {Minniti}, \&
  {Zoccali}}]{2015LibralatoVIRCAM}
{Libralato}, M., {Bellini}, A., {Bedin}, L.~R., {et~al.} 2015, \mnras, 450,
  1664

\bibitem[{{Libralato} {et~al.}(2014){Libralato}, {Bellini}, {Bedin}, {Piotto},
  {Platais}, {Kissler-Patig}, \& {Milone}}]{2014LibralatoHAWKI}
{Libralato}, M., {Bellini}, A., {Bedin}, L.~R., {et~al.} 2014, \aap, 563, A80

\bibitem[{{Libralato} {et~al.}(2023){Libralato}, {Bellini}, {van der Marel},
  {Anderson}, {Sohn}, {Watkins}, {Alderson}, {Allen}, {Clampin}, {Glidden},
  {Goyal}, {Hoch}, {Huang}, {Kammerer}, {Lewis}, {Lin}, {Long}, {Louie},
  {MacDonald}, {Mountain}, {Pe{\~n}a-Guerrero}, {Perrin}, {Pueyo}, {Rebollido},
  {Rickman}, {Seager}, {Stevenson}, {Valenti}, {Valentine}, \&
  {Wakeford}}]{2023LibralatoNIRISS}
{Libralato}, M., {Bellini}, A., {van der Marel}, R.~P., {et~al.} 2023, \apj,
  950, 101

\bibitem[{{Libralato} {et~al.}(2018){Libralato}, {Bellini}, {van der Marel},
  {Anderson}, {Watkins}, {Piotto}, {Ferraro}, {Nardiello}, \&
  {Vesperini}}]{2018LibralatoNGC362}
{Libralato}, M., {Bellini}, A., {van der Marel}, R.~P., {et~al.} 2018, \apj,
  861, 99

\bibitem[{{Libralato} {et~al.}(2022){Libralato}, {Bellini}, {Vesperini},
  {Piotto}, {Milone}, {van der Marel}, {Anderson}, {Aparicio}, {Barbuy},
  {Bedin}, {Borsato}, {Cassisi}, {Dalessandro}, {Ferraro}, {King}, {Lanzoni},
  {Nardiello}, {Ortolani}, {Sarajedini}, \& {Sohn}}]{2022LibralatoPMcat}
{Libralato}, M., {Bellini}, A., {Vesperini}, E., {et~al.} 2022, \apj, 934, 150

\bibitem[{{Libralato} {et~al.}(2021){Libralato}, {Lennon}, {Bellini}, {van der
  Marel}, {Clark}, {Najarro}, {Patrick}, {Anderson}, {Bedin}, {Crowther}, {de
  Mink}, {Evans}, {Platais}, {Sabbi}, \& {Sohn}}]{2021Libralato2DkinGC}
{Libralato}, M., {Lennon}, D.~J., {Bellini}, A., {et~al.} 2021, \mnras, 500,
  3213

\bibitem[{{Linsky}(1969)}]{Linsky1969}
{Linsky}, J.~L. 1969, \apj, 156, 989

\bibitem[{{Massari} {et~al.}(2024){Massari}, {Dalessandro}, {Erkal},
  {et~al.}}]{EROGalGCs}
{Massari}, D., {Dalessandro}, E., {Erkal}, D., {et~al.} 2024, \aap, submitted,
  arXiv:2405.13498

\bibitem[{{Nardiello} {et~al.}(2022){Nardiello}, {Bedin}, {Burgasser},
  {Salaris}, {Cassisi}, {Griggio}, \& {Scalco}}]{2022NardielloNIRCam}
{Nardiello}, D., {Bedin}, L.~R., {Burgasser}, A., {et~al.} 2022, \mnras, 517,
  484

\bibitem[{{Nardiello} {et~al.}(2018){Nardiello}, {Libralato}, {Piotto},
  {Anderson}, {Bellini}, {Aparicio}, {Bedin}, {Cassisi}, {Granata}, {King},
  {Lucertini}, {Marino}, {Milone}, {Ortolani}, {Platais}, \& {van der
  Marel}}]{2018MNRAS.481.3382N}
{Nardiello}, D., {Libralato}, M., {Piotto}, G., {et~al.} 2018, \mnras, 481,
  3382

\bibitem[{{Saracino} {et~al.}(2018){Saracino}, {Dalessandro}, {Ferraro},
  {Lanzoni}, {Origlia}, {Salaris}, {Pietrinferni}, {Geisler}, {Kalirai},
  {Correnti}, {Cohen}, {Mauro}, {Villanova}, \& {Moni Bidin}}]{Saracino2018}
{Saracino}, S., {Dalessandro}, E., {Ferraro}, F.~R., {et~al.} 2018, \apj, 860,
  95

\bibitem[{{Saumon} {et~al.}(1994){Saumon}, {Bergeron}, {Lunine}, {Hubbard}, \&
  {Burrows}}]{Saumon1994}
{Saumon}, D., {Bergeron}, P., {Lunine}, J.~I., {Hubbard}, W.~B., \& {Burrows},
  A. 1994, \apj, 424, 333

\bibitem[{{Varri} {et~al.}(2018){Varri}, {Cai}, {Concha-Ram{\'\i}rez},
  {Dinnbier}, {L{\"u}tzgendorf}, {Pavl{\'\i}k}, {Rastello}, {Sollima}, {Wang},
  \& {Zocchi}}]{2018ComAC...5....2V}
{Varri}, A.~L., {Cai}, M.~X., {Concha-Ram{\'\i}rez}, F., {et~al.} 2018,
  Computational Astrophysics and Cosmology, 5, 2

\bibitem[{{Vasiliev} \& {Baumgardt}(2021)}]{2021VasilievGCkin}
{Vasiliev}, E. \& {Baumgardt}, H. 2021, \mnras, 505, 5978

\bibitem[{{Watkins} {et~al.}(2015){Watkins}, {van der Marel}, {Bellini}, \&
  {Anderson}}]{2015ApJ...803...29W}
{Watkins}, L.~L., {van der Marel}, R.~P., {Bellini}, A., \& {Anderson}, J.
  2015, \apj, 803, 29

\bibitem[{{WFIRST Astrometry Working Group} {et~al.}(2019{\natexlab{a}}){WFIRST
  Astrometry Working Group}, {Sanderson}, {Bellini}, {Casertano}, {Lu},
  {Melchior}, {Libralato}, {Bennett}, {Shao}, {Rhodes}, {Sohn}, {Malhotra},
  {Gaudi}, {Fall}, {Nelan}, {Guhathakurta}, {Anderson}, \& {Ho}}]{2019WFIRST}
{WFIRST Astrometry Working Group}, {Sanderson}, R.~E., {Bellini}, A., {et~al.}
  2019{\natexlab{a}}, Journal of Astronomical Telescopes, Instruments, and
  Systems, 5, 044005

\bibitem[{{WFIRST Astrometry Working Group} {et~al.}(2019{\natexlab{b}}){WFIRST
  Astrometry Working Group}, {Sanderson}, {Bellini}, {Casertano}, {Lu},
  {Melchior}, {Libralato}, {Bennett}, {Shao}, {Rhodes}, {Sohn}, {Malhotra},
  {Gaudi}, {Fall}, {Nelan}, {Guhathakurta}, {Anderson}, \&
  {Ho}}]{2019WFIRSTAWG}
{WFIRST Astrometry Working Group}, {Sanderson}, R.~E., {Bellini}, A., {et~al.}
  2019{\natexlab{b}}, Journal of Astronomical Telescopes, Instruments, and
  Systems, 5, 044005

\bibitem[{{Zhang} \& {Bloom}(2020)}]{2020ApJ...889...24Z}
{Zhang}, K. \& {Bloom}, J.~S. 2020, \apj, 889, 24

\end{thebibliography}
\bibliographystyle{aa}

\begin{appendix}

\section{The effect of the \texttt{DeepCR} step on the photometry with VIS data}\label{appendix:crphot}

\begin{figure}[t!]
    \centering
    \includegraphics[width=\columnwidth]{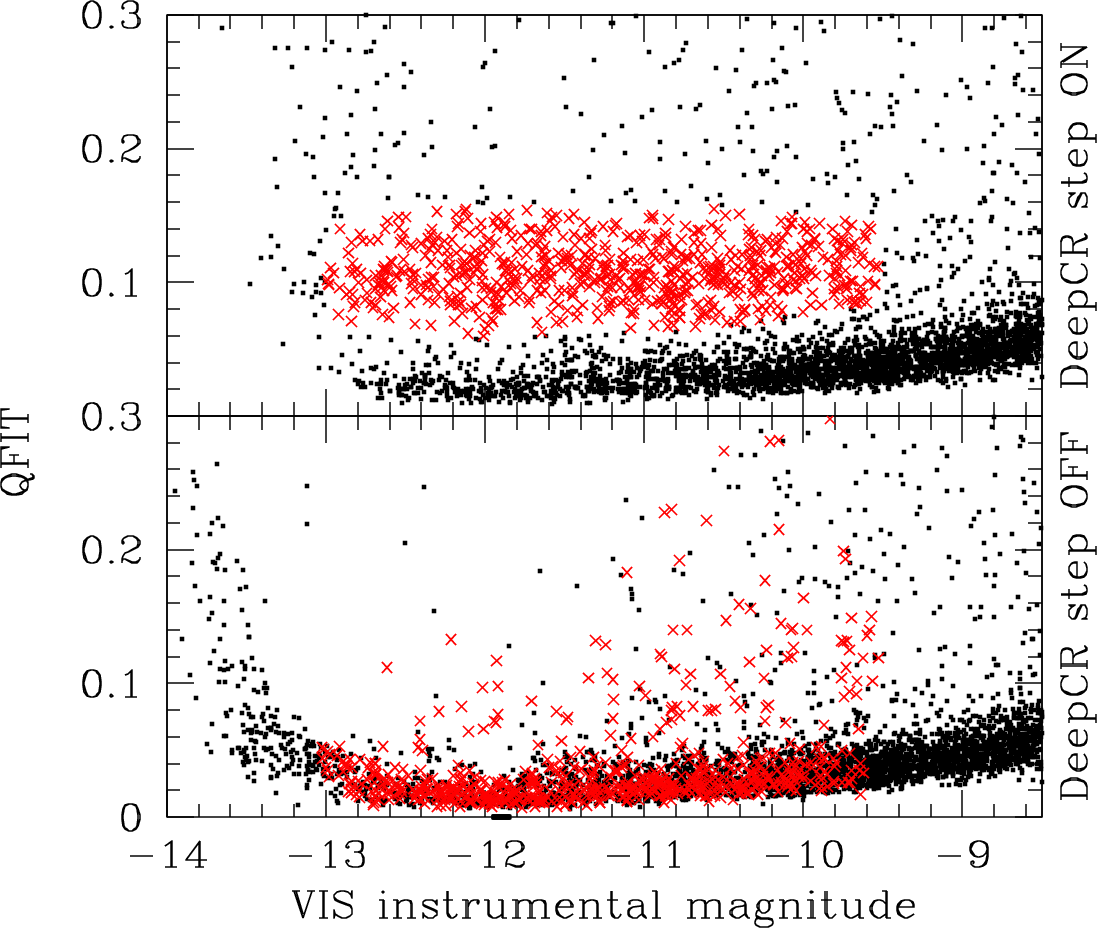}
    \caption{\qfit as a function of VIS instrumental magnitude for stars measured in the VIS quadrant 4-1.F of all four ERO images of NGC~6397. The top panel refers to the result obtained using the images processed with the \texttt{deepCR} tool, while the bottom panel is obtained without the cosmic-ray correction. Red crosses refer to atypical stars with a dimmer centremost pixel in the images processed with the \texttt{deepCR} tool (see the text for details), while all other sources are shown as black dots.}
    \label{fig:deepcr}
\end{figure}

Our initial analysis made use of VIS images processed by the ERO pipeline as described in \citet{EROData}, including the step designed to identify and correct for the effects of the cosmic rays using the tool \texttt{deepCR} \citep{2020ApJ...889...24Z}. We noticed that some bright, unsaturated stars, which should be well measured and thus have a \qfit$<$0.05 (see Sect.~\ref{sec:precision}), instead showed a worse \qfit of $\sim$0.1. The top panel of Fig.~\ref{fig:deepcr} presents the \qfit as a function of VIS instrumental magnitude for stars measured in the VIS quadrant 4-1.F of all four ERO images of NGC~6397 processed using the tool \texttt{deepCR}. The red crosses highlight the `atypical' stars, while all other points are shown in black.

These atypical sources have their centremost pixel dimmer than what predicted by our ePSFs, which results in a poorer fit (and larger \qfit) than what can be obtained for `regular' sources of the same brightness. Upon investigation, we believe that the replacing of pixels hit by cosmic rays performed by the \texttt{deepCR} tool has altered the flux distribution in the centremost pixel of these atypical stars, but it is unclear why not all stars have been impacted in the same way.
We reanalysed the ERO single epoch images without using the \texttt{deepCR} tool and remeasured all sources. The result is presented in the bottom panel of Fig.~\ref{fig:deepcr}. It is clear that the majority of the atypical stars can now be measured well as other objects at the same magnitude level. We also found that the magnitude difference for the atypical stars in the two sets of images can be as large as 0.1 mag. For all these reasons, we chose to use this latter set of images in our paper.

\clearpage

\begin{figure*}[th!]
    \centering
    \includegraphics[width=0.55\textwidth]{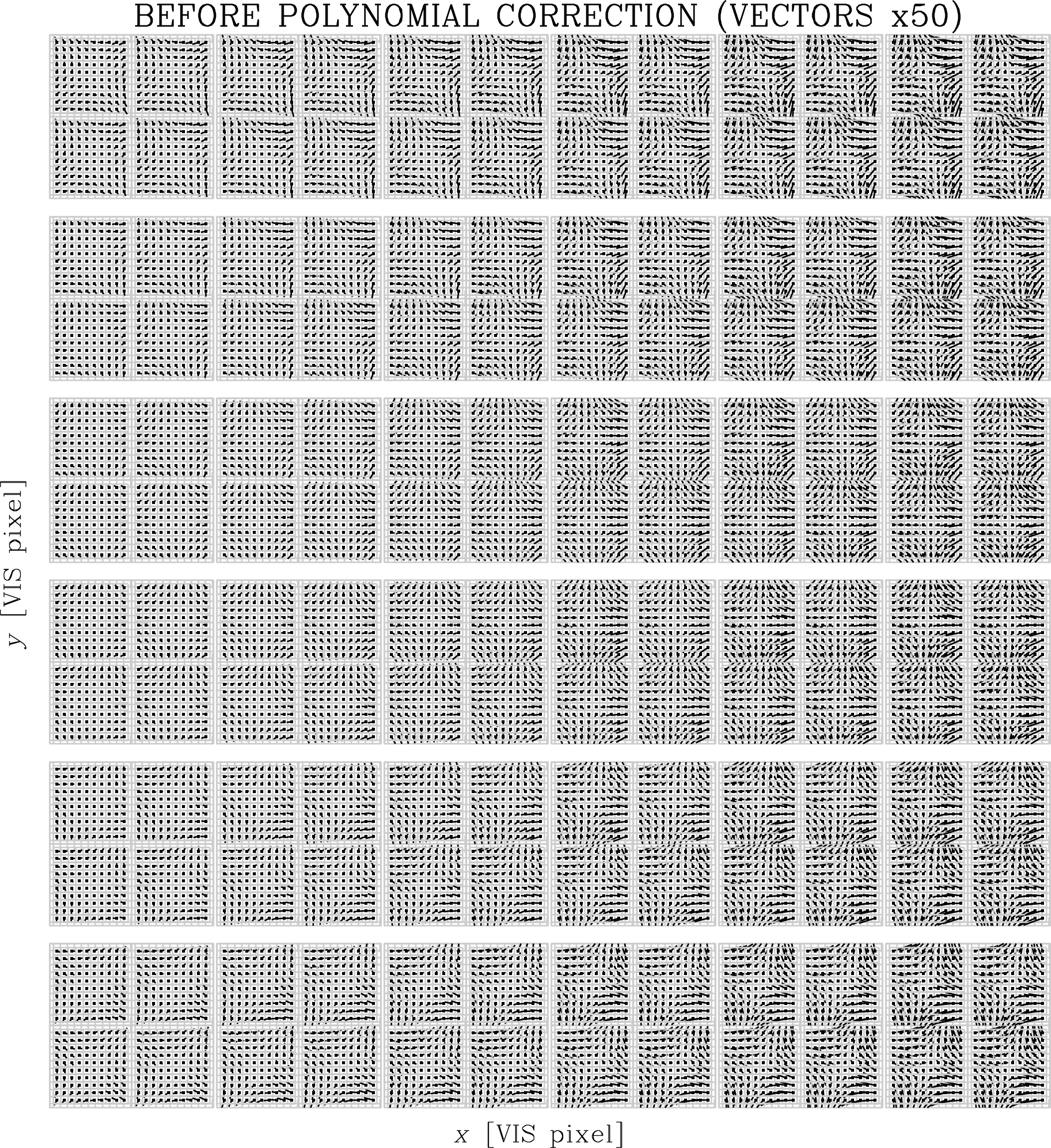}
    \caption{GD maps for all VIS quadrants before applying the polynomial correction. Vectors are magnified by a factor of 50.}
    \label{fig:VISGDall1}
\end{figure*}

\begin{figure*}[th!]
    \centering
    \includegraphics[width=0.55\textwidth]{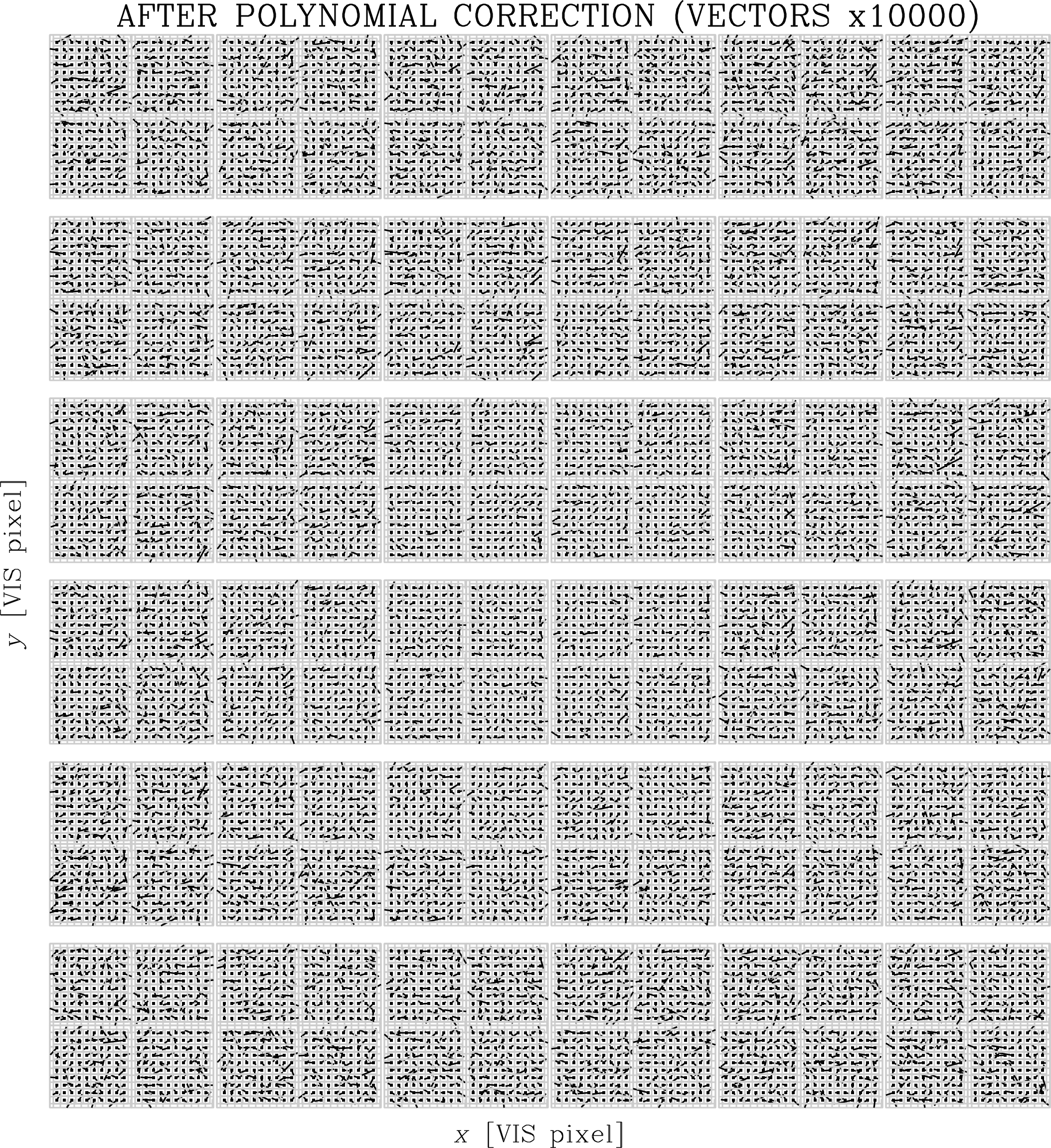}
    \caption{GD maps for all VIS quadrants after applying the polynomial correction. Vectors are magnified by a factor of 10\,000.}
    \label{fig:VISGDall2}
\end{figure*}

\begin{figure*}[th!]
    \centering
    \includegraphics[width=0.55\textwidth]{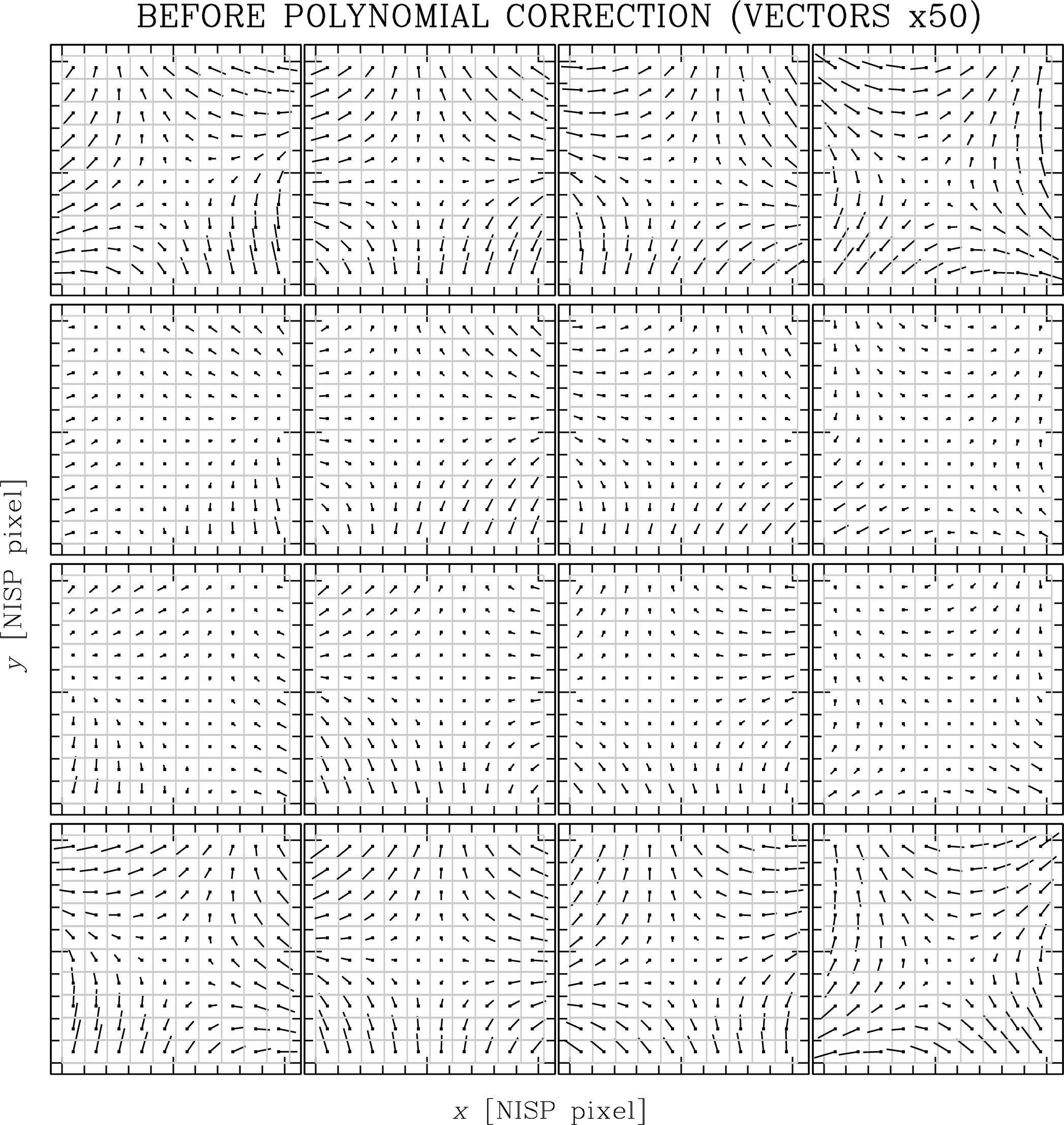}
    \caption{GD maps for all NISP detectors in the \YE filter before applying the polynomial correction. Vectors are magnified by a factor of 50.}
    \label{fig:NISPYall1}
\end{figure*}

\begin{figure*}[th!]
    \centering
    \includegraphics[width=0.55\textwidth]{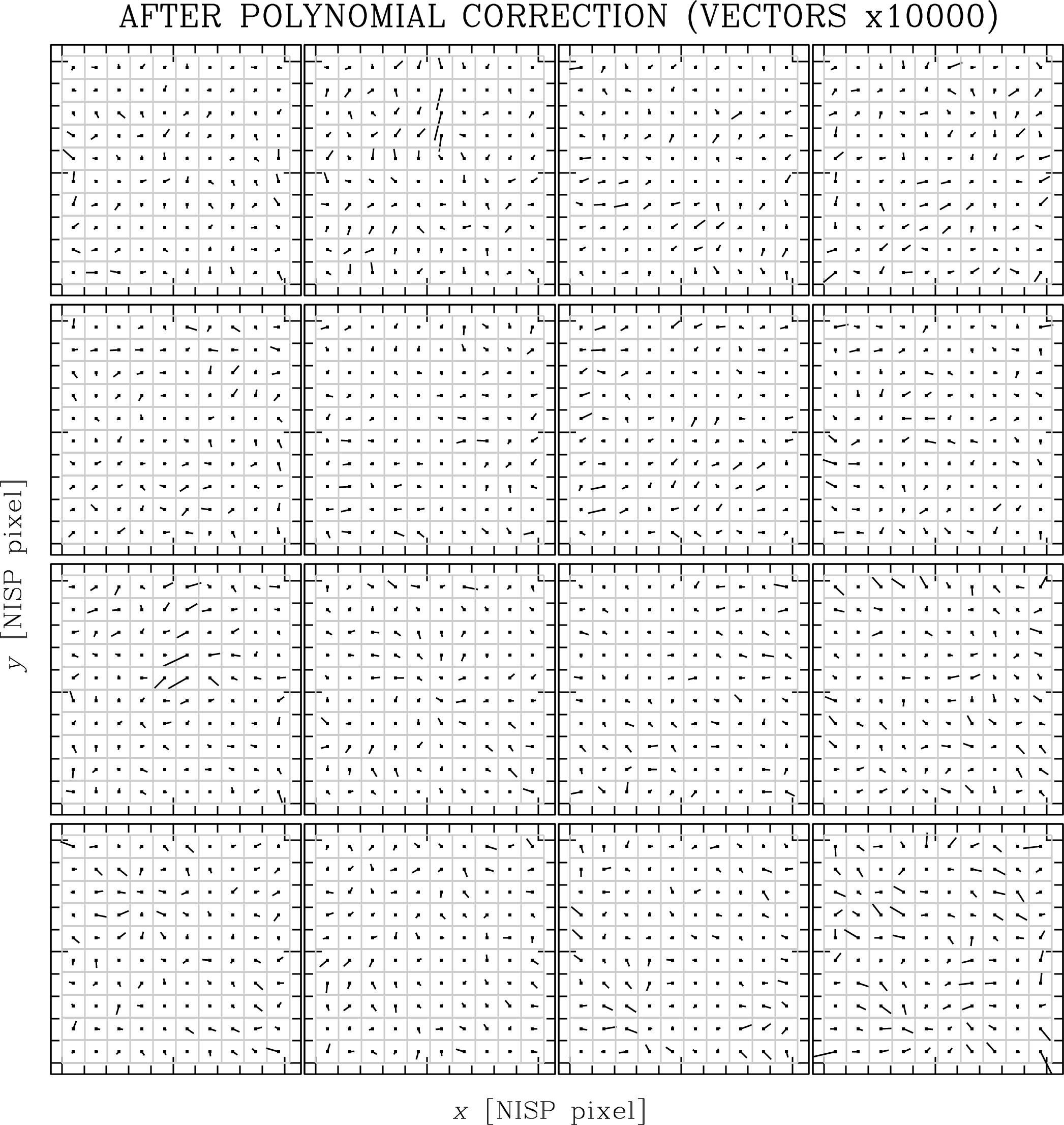}
    \caption{GD maps for all NISP detectors in the \YE filter after applying the polynomial correction. Vectors are magnified by a factor of 10\,000.}
    \label{fig:NISPYall2}
\end{figure*}

\clearpage

\section{GD map for all VIS and NISP detectors}\label{appendix:gdmaps}

We present here the GD maps before and after the correction, for all VIS quadrants (Figs.~\ref{fig:VISGDall1} and \ref{fig:VISGDall2}) and NISP (Figs.~\ref{fig:NISPYall1} and \ref{fig:NISPYall2}) detectors (for the \YE-filter solution). We note that we show here only the non linear terms of the GD.

\clearpage

\begin{figure*}[th!]
    \centering
    \includegraphics[width=0.625\textwidth]{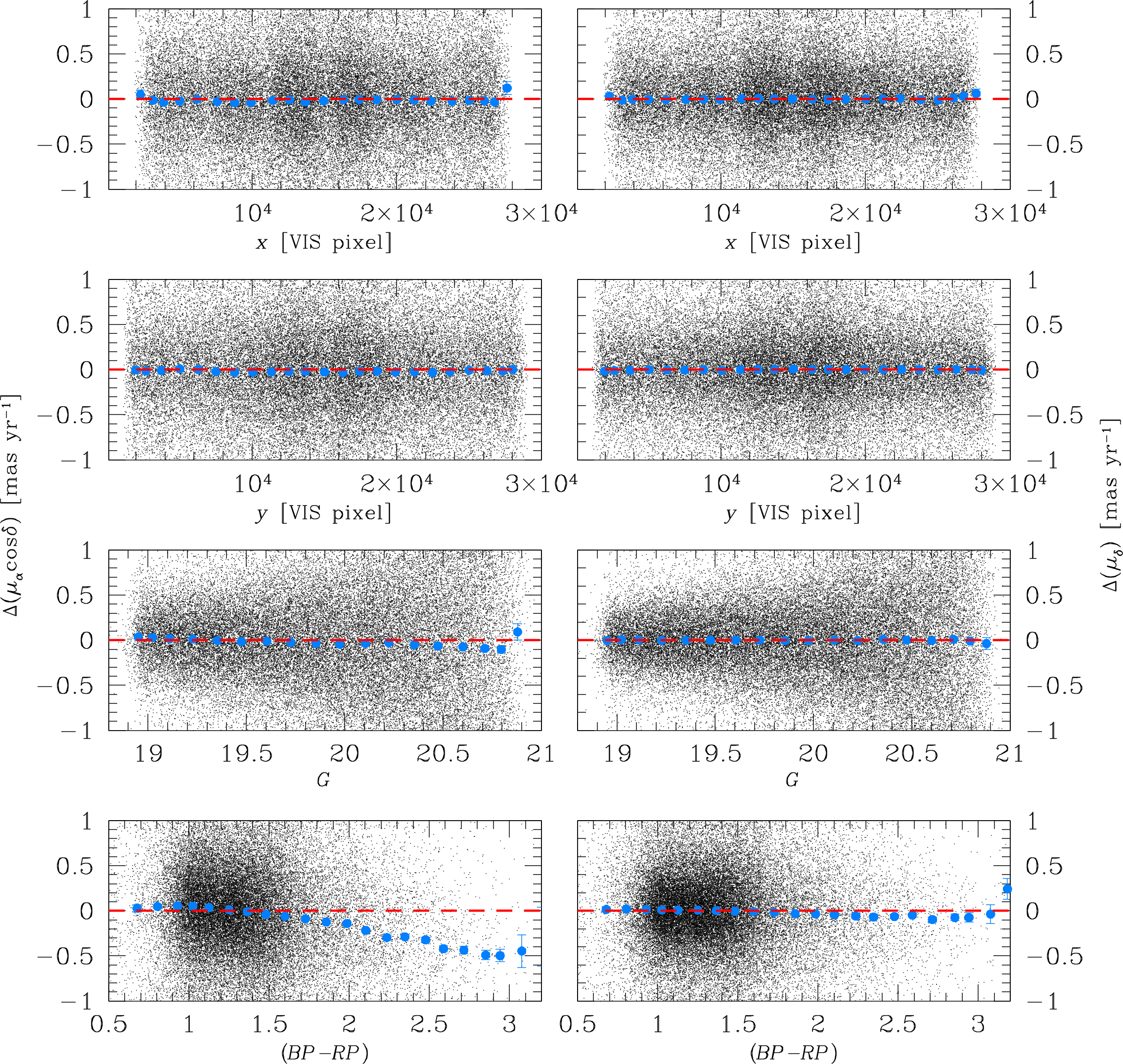}
    \caption{Difference between our \egpm and the \gaia DR3 PMs as a function of (from top to bottom) $x$ position, $y$ position, $G$ magnitude, and ($BP-RP$) colour. The left column refers to the $\mu_\alpha \cos\delta$ component of the PM, whereas the right column focuses on the $\mu_\delta$ component. In all plots, the red dashed line is set to 0 as a reference. Blue points are the median values (and their errors) of the $\Delta$ PM in bins of size 2500 pixels (first two rows from the top; step of 1250 pixels) or 0.25 magnitude (the other panels; step of 0.125 magnitude).}
    \label{fig:gaia1}
\end{figure*}

\begin{figure*}[th!]
    \centering
    \includegraphics[width=0.625\textwidth]{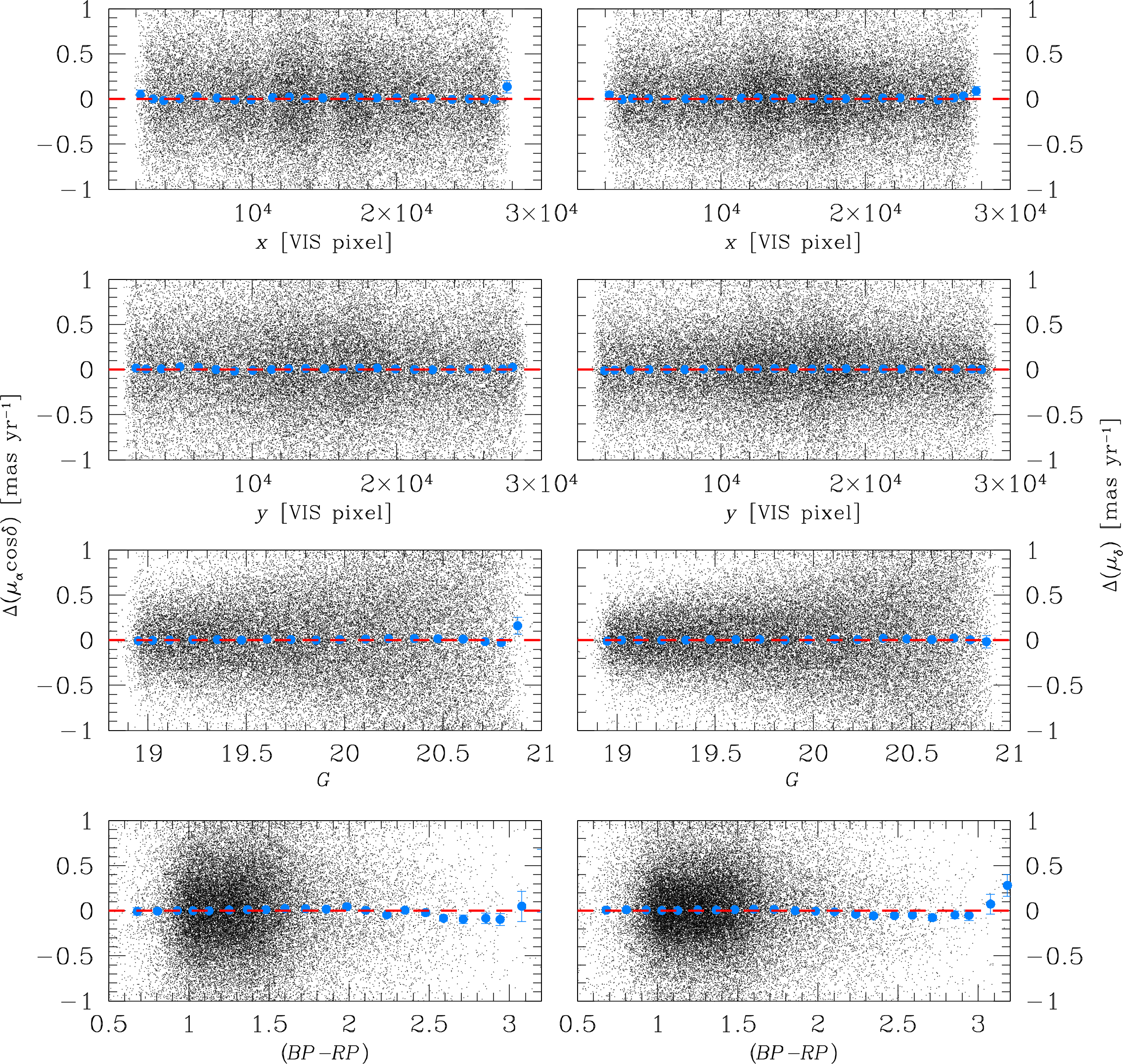}
    \caption{As in Fig.~\ref{fig:gaia2}, but after the colour-dependent systematic has been corrected in the \egpm PMs.}
    \label{fig:gaia2}
\end{figure*}

\clearpage

\section{Comparison with \gaia DR3 PMs}\label{appendix:pmcomp}

Figure~\ref{fig:gaia1} shows a comparison between the \egpm and the \gaia DR3 PMs as a function of $x$ and $y$ VIS master-frame positions (in VIS pixels), $G$ magnitude and ($BP-RP$) colour. The blue points (with error bars) are the median values (and their standard errors) in equally spaced positional, magnitude, or colour bins (depending on the panel). Only unsaturated stars were considered.

There is a clear systematic trend as a function of ($BP-RP$) colour. We corrected for this trend as follows. We divided the sample in 0.25-mag-wide ($BP-RP$) colour bins (with step of 0.125 mag) and, in each bin, we computed the median value (and error) of the difference between the \egpm and \gaia PMs of the stars (red points in the bottom panel of Fig.~\ref{fig:gaia1}). This median difference is the correction to apply to the \egpm PMs. To provide a smooth correction, we linearly interpolated between the various bins. For stars might without both $B$ and $R$ \gaia magnitudes, we computed analogue corrections but using other combinations of VIS and NISP colours (in case a star was not measured in a given filter). We also summed in quadrature the standard error on the median of the correction to the PM error. 

The comparison between the corrected \egpm and the \gaia DR3 PMs is shown in Fig.~\ref{fig:gaia2}. The mild correlation between $\Delta$PM and $G$ magnitude disappears after the colour-dependent correction. Most of the faint stars showing the systematic trend are also very red and thus affected by the aforementioned colour-dependent systematic effect.

\end{appendix}

\label{LastPage}
\end{document}